\documentclass[prc,amsfonts,reprint, superscriptaddress,showpacs,onecolumn, notitlepage,nofootinbib,longbibliography,11pt]{revtex4-1} 
\pdfoutput=1
\usepackage{hyperref}
\hypersetup{colorlinks=true, linkcolor=blue, citecolor=red, urlcolor=blue}
\usepackage{tikz}
\usetikzlibrary{arrows}
\usetikzlibrary{decorations.pathmorphing}
\usepackage{hyperref}
\usepackage{amssymb}
\usepackage{amsmath}
\usepackage{color}
\usepackage{xcolor}
\usepackage{graphicx}
\usepackage{soul}
\usepackage{mathtools}
\usepackage{braket}

\newcommand{\be}{\begin{equation}\begin{aligned}\label}{}
\newcommand{\ee}{\end{aligned}\end{equation}}

\newcommand{\comment}[1]{}

\begin{document}

\title{Transmission spectra of the driven, dissipative Rabi model in the USC regime}

\author{L. Magazz\`u}
\affiliation{Institute for Theoretical Physics, University of
Regensburg, 93040 Regensburg, Germany}

\author{P. Forn-D\'iaz}
\affiliation{Institut de F\'isica d'Altes Energies (IFAE), The Barcelona Institute of Science and Technology (BIST),
Bellaterra (Barcelona) 08193, Spain}
\affiliation{Qilimanjaro Quantum Tech SL, Barcelona, Spain}

\author{M. Grifoni}
\affiliation{Institute for Theoretical Physics, University of Regensburg, 93040 Regensburg, Germany}

\date{\today}

\begin{abstract}
We present theoretical transmission spectra of a strongly driven,  damped flux qubit coupled to a dissipative resonator in the ultrastrong coupling regime. Such a qubit-oscillator system, described within a dissipative Rabi model, constitutes the building block of superconducting circuit QED platforms. The addition of a strong drive allows one to characterize the system properties and study novel phenomena, leading to a better understanding and control of the qubit-oscillator system. In this work, the  calculated transmission of a weak probe field quantifies the response of the qubit, in  frequency domain, under the  influence of the quantized  resonator and of the strong microwave drive. We find distinctive features of the entangled  driven qubit-resonator spectrum, namely resonant features and avoided crossings,   modified by the presence of the dissipative environment. The  magnitude, positions, and broadening of these features are determined by the interplay among qubit-oscillator detuning, the strength of their coupling, the driving amplitude, and the interaction with the heat bath.
This work establishes the theoretical basis for future experiments in the driven ultrastrong coupling regime.
\end{abstract}

\maketitle

\section{Introduction}
Current developments in circuit quantum electrodynamics (QED) are establishing superconducting devices as leading platforms for quantum information and simulations~\cite{You2011, Koch2012, Pekola2015, Wendin2017, Nori2017}. In particular, quantum optics experiments with qubits coupled to superconducting resonators are now performed in (and beyond) the so-called ultrastrong coupling (USC) regime, with the qubit-resonator coupling reaching the same order of magnitude of the qubit splitting and resonator frequency~\cite{Ciuti2005, Bourassa2009,Hausinger2008, Niemczyk2010, Ashhab2010, Hausinger2010PRA, Forn-Diaz2010, Yoshihara2017, Yoshihara2017PRA, Forn-Diaz2018review, Kockum2019}. The strong entanglement between light and matter in the USC regime carries  potential for designing novel quantum hybrid states and for achieving ultrafast information transfer~\cite{Romero2012}.\\
\indent In circuit QED platforms, the qubits are essentially based on superconducting loops interrupted by Josephson junctions, the nonlinear elements that provide the anharmonicity required to single-out the two lowest energy states~\cite{Devoret2004}. In the flux configuration~\cite{Mooij1999}, the qubit states are superpositions of clockwise and anti-clockwise circulating supercurrents, corresponding to the two lowest energy eigenstates of a double-well potential \emph{seen} by the flux coordinate.
The double-well can be biased by applying an external magnetic flux and transitions between states in this qubit basis, where the states are localized in the wells, occur via tunneling through the potential barrier.\\
\indent  The standard theoretical tool to account for the coupling of superconducting qubits to their electromagnetic or phononic environments is provided by the spin-boson model, consisting of a quantum two-level system interacting with a heat bath of harmonic oscillators~\cite{Leggett1987,Weiss2012}. This model has been the subject of extensive studies as an archetype of dissipation in quantum mechanics, and the different coupling regimes of spin-boson systems and the associated dynamical behaviors have been theoretically explored by using a variety of approaches~\cite{Breuer2002, Weiss2012}. 
Only recently though, progress in the design of superconducting circuits has opened the possibility to attain experimental control on the strong qubit-environment coupling regime ~\cite{Peropadre2013PRL,Forn-Diaz2017, Magazzu2018, Ustinov2018, Roch2019, Kuzmin2019}.\\
\indent  In circuit QED, an appropriate description for qubit-resonator systems is provided by the Rabi Hamiltonian, whose interaction part is featured by terms known as rotating and counter-rotating terms. In this context,  USC refers to an interaction regime where the rotating wave approximation, that allows for a description in terms of the Jaynes-Cummings Hamiltonian, appropriate for atom-cavity systems, fails, as the counter-rotating terms cannot be neglected~\cite{Forn-Diaz2018review, Kockum2019}. A refined classification of the different regimes of the Rabi model is provided in~\cite{Rossatto2017}. The USC regime of circuit QED is still the subject of much theoretical work, see for example~\cite{Ashhab2010, Hausinger2010PRA, DiazCamacho2016, Armata2017, DeBernardis2018, DiStefano2019}. However, the intricacies of the driven Rabi problem in the USC regime \cite{Hausinger2011} have been largely unexplored so far. Experiments on strongly driven qubits have  demonstrated the possibility to control properties of engineered quantum two-level systems by intense light  \cite{Oliver2005,Wilson2007, Magazzu2018}. E.g.  complex Landau-Zener patterns of avoided crossings  could be controlled by tuning the driving amplitude,  in agreement with theoretical expectations~\cite{Grifoni1998,Hausinger2010,Nori2017, Kohler2017}. In recent works, spectroscopical signatures of  drive-induced new symmetries~\cite{Engelhardt2021} and nonadiabatic effects~\cite{Reimer2018} in quantum systems have been addressed.\\
\indent Experimentally, transmission spectroscopy has been shown to be a powerful tool  to characterize the complex spectrum of the Rabi problem \cite{Niemczyk2010, Yoshihara2017, Yoshihara2017PRA, Yoshihara2018}.  Usually, the probe couples to the resonator, from which properties of the qubit can be inferred. As shown in this work, a probe coupled to the qubit can also provide precious information. Alternatives to the spectroscopy of the qubit to investigate USC systems exist. For example, spectroscopy of ancillary qubits has been proposed in~\cite{Lolli2015} to probe the ground states of ultrastrongly-coupled systems. Moreover, methods alternative to the analysis of the transmission spectra have been recently devised to probe the USC regime~\cite{Falci2019, Ridolfo2019}.\\
\indent In this work, we consider a dissipative flux qubit ultrastrongly coupled to a superconducting resonator, modeled as a harmonic oscillator, which in turn  interacts with a bosonic heat bath. The qubit is  probed by a weak  incoming field whose transmitted part provides information on the dynamics under the influence of the  resonator and its environment, as demonstrated in a variety of experiments~\cite{Astafiev2010, Hoi2011,Abdumalikov2011, Haeberlein2015, Forn-Diaz2017PRApp}. In addition, the qubit is subject to an intense  microwave field, the drive. Despite the rich literature on the topic, the impact of intense microwave drive on the dissipative Rabi model in the USC regime has not been investigated so far. The setup considered  describes quantum optics experiments in circuit QED  but also the coupling of a qubit to a detector~\cite{Chiorescu2004, Thorwart2004, Johansson2006}, and the qubit-bath coupling mediated by a waveguide resonator in a heat transport platform in the  quantum regime~\cite{Pekola2018}.\\
\indent We address the spectral properties of the driven USC system in a two-fold way. On the one hand, quasi-energy spectra of the driven and closed Rabi model are studied analytically using the Floquet-Van Vleck approach of Ref.~\cite{Hausinger2011}. On the other hand, the transmission spectra of the open Rabi model are first numerically evaluated in the absence of the pump drive in the weak and strong dissipation regimes. This allows us to set up the impact of dissipation on this strongly entangled quantum system. In a last step, the full driven, open USC system is analyzed. We notice that weak dissipation affecting the USC system as a whole can be treated via a master equation approach e.g. along the lines of~\cite{Hausinger2008,Blais2011}.  Since our USC system is probed through the qubit, we conveniently map the dissipative Rabi model to an effective spin-boson model where the spin interacts directly with a bosonic bath characterized by an effective spectral density. The latter function is peaked at the oscillator  frequency~\cite{Garg1985}, and thus describes a so-called structured environment. Using the same approach as the one developed in~\cite{Magazzu2018, Magazzu2019} to analyze the  measured transmission of a probe field in the presence of an Ohmic environment, here we first calculate the transmission spectra of the undriven qubit, considering different qubit-resonator coupling strengths. Specifically, according to the dissipation regime, the qubit response function is evaluated by using the weak USC system-environment approach developed in~\cite{Kohler2018} or within the so-called non-interacting blip approximation (NIBA), which allows one to treat dissipation (and hence the effects of the resonator) in a non-perturbative way \cite{Grifoni1998}. Finally, we look at the impact of a strong pump field within the NIBA.\\
\indent The qualitative difference between the setup of Ref.~\cite{Magazzu2018}, featuring a driven qubit ultrastrongly coupled to an Ohmic environment, and the one in this work, where the driven qubit is ultrastrongly coupled to a dissipative quantum resonator, is reflected in the transmission spectra. 
 In absence of the drive, the spectra show clear signatures of the entanglement between qubit and resonator in the form of avoided crossings and resonances, whose positions depend on the qubit-resonator coupling strength. Furthermore, renormalization effects due to the dissipative environment have to be properly taken into account for a quantitative description. When the drive is added, novel avoided crossings and resonances reveal the interplay between the resonator photons and the driving field.\\
\indent The paper is structured as follows. In Section \ref{setup}, the driven, dissipative Rabi model and its mapping to an effective, driven spin boson model is discussed. In Sec. \ref{section_analytical_USC}, spectral properties of the non-dissipative  Rabi model in the USC regime are analyzed, while a formal expression for the transmission is reported in Sec. \ref{section_transmission}. Numerical results are shown in Sec. \ref{section_results},  and interpreted on the basis of the analytical results of  Sec.  \ref{section_analytical_USC}. Finally, conclusions are drawn in Sec. \ref{Section conclusions}. 
\section{The driven, dissipative Rabi model}
\label{setup}
\begin{figure}[ht!]
\begin{center}
\includegraphics[width=0.85\textwidth,angle=0]{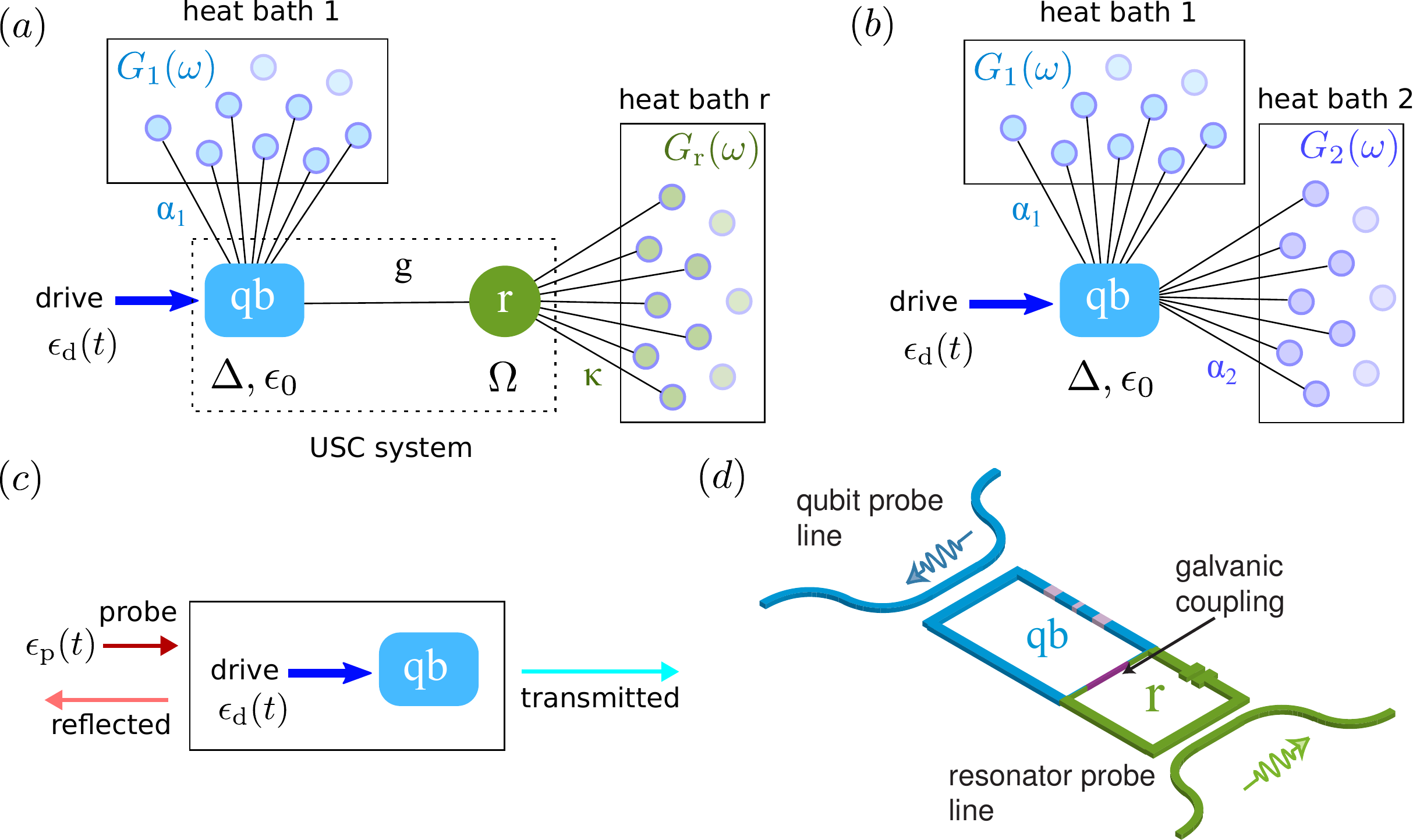}
\caption{\small{Model and experimental design of the driven dissipative Rabi model. (a) - The dissipative Rabi model is realized by coupling a flux qubit  to an Ohmic heat bath, denoted with 1, and to a resonator, which is in turn coupled to an Ohmic environment, denoted with r. (b) - The model is mapped into a two-bath spin-boson model, where the qubit is coupled to two heat baths. The first is the original bath 1 while the second, denoted by 2, is a structured effective bath. (c) - Detail of the two-tone spectroscopy protocol where the transmission line is used to probe the driven qubit, the drive being applied via the transmission line as well. (d) - Simplified circuit implementation of the USC system displaying the qubit (large loop interrupted by three Josephson junctions) and the LC resonator (smaller loop). The qubit and resonator are galvanically connected to enhance the coupling strength into the USC regime. The qubit/resonator probing lines correspond to their respective Ohmic environments.}}
\label{scheme}
\end{center}
\end{figure}

\indent For a realistic description of experiments studying USC systems, the inclusion of decoherence and dissipative effects induced by the electromagnetic environment or by other sources is unavoidable. Previous work used a modified master equation to include the effect of dissipation in the perturbative USC regime \cite{Hausinger2008,DeLiberato2009, Bourassa2011}. In the following, we consider a driven, dissipative flux qubit coupled to a resonator which is in turn subject to dissipation. Specifically, qubit and  resonator interact with two independent Ohmic heat baths, denoted by $1$ and r, respectively. The qubit is characterized by the tunneling matrix element $\Delta$, while the resonator is modeled as a harmonic oscillator of frequency $\Omega$.
 These two systems are coupled with an interaction strength quantified by the frequency $g$. If $g$ is of the order of $\Delta$ and $\Omega$ the system is 
in the ultrastrong coupling regime. The full Hamiltonian of this driven, dissipative Rabi model, which is sketched in Fig.~\ref{scheme}(a), is
\be{H_full}
H(t)=&-\frac{\hbar}{2}\left[\Delta\sigma_x+\epsilon(t)\sigma_z\right]+\hbar \Omega B^\dag B- \hbar\sigma_z g(B^\dag+B)+\sum_{k=1}^{N_1}\hbar\omega_{1 k} b_{1 k}^\dag b_{1 k} -\frac{\hbar}{2}\sigma_z\sum_{k=1}^{N_1} \lambda_{1 k} (b_{1 k}^\dag + b_{1 k})\\
& + \sum_{k=1}^{N_{\rm r}}\hbar\omega_{{\rm r} k} b_{{\rm r} k}^\dag b_{{\rm r} k}-\hbar(B^\dag+B)\sum_{k=1}^{N_{\rm r}}\lambda_{{\rm r} k} (b_{{\rm r} k}^\dag+b_{{\rm r} k})+\hbar(B^\dag + B)^2\sum_{k=1}^{N_{\rm r}}\frac{\lambda_{{\rm r} k}^2}{\omega_{{\rm r} k}}\;,
\ee
see Appendix~\ref{appendix_mapping}, where the qubit operators $\sigma_z=|\downarrow\;\rangle\langle\; \downarrow|-|\uparrow\;\rangle\langle \;\uparrow|$ and $\sigma_x=|\downarrow\;\rangle\langle\; \uparrow|+|\uparrow\;\rangle\langle\;\downarrow|$
are expressed in the so-called qubit basis of localized right- and left-well states $|\downarrow\;\rangle,|\uparrow\;\rangle$.
The qubit is driven by the time-dependent bias
\begin{equation}\begin{aligned}\label{bias_t}
\epsilon(t)=\epsilon_0 + \epsilon_{\rm p}\cos(\omega_{\rm p}t)+ \epsilon_{\rm d}\cos(\omega_{\rm d}t)\;,
\end{aligned}\end{equation}
which is the sum of a static part $\epsilon_0$, a weak probe (p), and a drive (d) with arbitrary amplitude, which we will assume to be of high frequency, see Fig.~\ref{scheme}(c). 
 
The bosonic creation and annihilation operators $B^\dag, b_{1/{\rm r} k}^\dag$ and  $B, b_{1/{\rm r} k}$  create and destroy an excitation in the resonator and in the $k$-th harmonic oscillator of the qubit/resonator bath, respectively. The angular frequencies $\lambda_{1/{\rm r} k}$ correspond to the coupling strengths with the individual modes of the respective baths. Note that, by removing the resonator and its bath from the full Hamiltonian~\eqref{H_full}, we are left with the standard spin-boson Hamiltonian~\cite{Leggett1987}. On the other hand, removing the baths coupled to  qubit and resonator, the standard Rabi Hamiltonian is recovered.\\
\indent Each bath is fully characterized by the spectral density function
\be{}
G(\omega)=\sum_k\lambda_{k}^2\delta(\omega-\omega_k)\;.
\ee
In the continuum limit, we assume an Ohmic spectral density with exponential cutoff for the qubit, i.e. $G_1(\omega)=2\alpha_1\omega e^{-\omega/\omega_{c}}$, with $\omega_{\rm c}$ a high-frequency cutoff, and a strictly Ohmic spectral density $G_{\rm r}(\omega)=\kappa\omega$ for the resonator. The dimensionless dissipation strengths are related to the friction coefficient in the Caldeira-Leggett model~\cite{Caldeira1981, Caldeira1983, Weiss2012}, see Appendix~\ref{appendix_mapping}.\\
\indent  Figure~\ref{scheme}(d) shows a schematic of the implementation of the driven, dissipative Rabi model using a superconducting circuit. A 3-junction flux qubit is galvanically attached to an LC resonator in order to attain ultrastrong coupling strengths, while keeping the interaction linear. In addition, two superconducting waveguides couple to the qubit and resonator, in order to define their own respective ohmic baths. The coupling to the baths determines the decay rate of each system, with an additional loss channel due to  intrinsic microscopic noise present in the neighborhood of the circuit.
In order to experimentally implement the system proposed in this work, coupling strengths in the range $g/\Omega = 0.3-0.5$ are most suitable. Coupling strengths $g/\Omega >1$ have already been achieved in qubit-resonator systems employing shared Josephson junctions as couplers~\cite{Yoshihara2017}. While providing the desired coupling strength, the junctions also contain a nonlinearity that may introduce modifications to the standard Rabi model. Using linear inductors it is also possible to attain ultrastrong couplings~\cite{Forn-Diaz2010}. Thin film aluminum would require a shared wire of significant length due to its low kinetic inductance. An alternative is to employ superinductance material compatible with aluminum such as granular aluminum~\cite{Maleeva2018,Gruenhaupt2019}. This material behaves as a linear inductor up to very large currents with inductances  $0.1-1$~nH per unit area and is therefore very suitable as an ultrastrong qubit-resonator coupler. In our coupling regimes of interest, we can estimate the necessary parameters using $g/\Omega = L_cI_p[(2\Omega (L_c + L_r))/\hbar]^{-1/2}$, which yields the desired range $0.3-0.5$, with qubit persistent current $I_p=100$~nA, shared coupling inductance $L_c = 0.4$~nH, resonator inductance $L_r=3.5$~nH, and resonator frequency $\Omega/2\pi=$  1~GHz.
\\
\indent The full Hamiltonian of the model, Eq.~\eqref{H_full}, can be mapped to that of the two-bath spin-boson model depicted in Fig.~\ref{scheme}(b) which reads
\be{H_SB}
H_{\rm SB}(t)=-\frac{\hbar}{2}\left[\Delta\sigma_x+\epsilon(t)\sigma_z\right]-\frac{\hbar}{2}\sigma_z\sum_{\nu k} \lambda_{\nu k} (a_{\nu k}^{\dagger}+a_{\nu k})+\sum_{\nu k}\hbar\omega_{\nu k} a_{\nu k}^{\dagger}a_{\nu k}\;.
\ee
In this effective Hamiltonian, the qubit is directly coupled to two bosonic baths indexed with $\nu=1,2$: the bath 1 is the same Ohmic bath coupled to the qubit as in Eq.~\eqref{H_full}. The second, $\nu=2$ is a new, \emph{effective} bath whose spectral density, in the continuum limit, reads~\cite{Garg1985, Goorden2004, Thorwart2004, Nesi2007NJP,  Zueco2019}
\be{Geff}
G_2(\omega)=  \frac{2 \alpha_2 \omega
\Omega^4}{(\Omega^2-\omega^2)^2+(\gamma\omega)^2}\;.
\ee
This effective spectral density is structured, meaning that it displays a peak centered at the oscillator frequency $\Omega$  with width $\gamma=2\pi\kappa\Omega$. The latter frequency is the memoryless damping kernel of the resonator's Ohmic bath, see Appendix~\ref{appendix_mapping}. The corresponding effective coupling strength $\alpha_2$ is given by the dimensionless parameter $\alpha_2=8 \kappa g^2/\Omega^2$. In the limit $\Omega\gg \Delta$ the qubit sees a bath with a low-frequency Ohmic behavior. By taking the limit $\kappa\rightarrow 0$ (i.e. disconnecting the Ohmic bath interacting with the resonator), we obtain $\lim_{\kappa\to 0}G_2(\omega)=4g^2\delta(\omega-\Omega)$, which consistently describes the bath $\nu=2$ as comprising a single oscillator coupled to the qubit with strength $\lambda=2g$. The resulting interaction term in Eq.~\eqref{H_SB} reproduces the qubit-resonator coupling term in Eq.~\eqref{H_full}. This problem displays high complexity due to the strong driving on the qubit, the ultrastrong  coupling between qubit and resonator, as well as environmental effects on both qubit and oscillator.
In order to gain physical insight, we first recall some features of the spectrum of the driven Rabi system in the absence of dissipation.

\section{Analytical treatment of the closed Rabi model in the USC regime}
\label{section_analytical_USC}

\subsection{Closed Rabi model  in the USC regime} 
\label{Analytical_Rabi_static}

We start with the non-driven, non-dissipative Rabi model, whose spectrum has been discussed in various works by now, see e.g.~\cite{Irish2005, Irish2007, Ashhab2010, Hausinger2008, Hausinger2010PRA}. 
Because of the coupling to the
qubit, the states of the resonator are  displaced in one of two opposite
directions depending on the persistent-current state of the qubit. In mathematical terms, this gives rise to displaced coherent states of the oscillator. The energy eigenstates of the  Rabi system are then coherent superposition of product states  for the qubit and the displaced oscillator.
Following the polaron approach, an approximate analytical expression for the spectrum of the Rabi model can be derived using Van Vleck perturbation theory in the qubit's tunneling parameter $\Delta$. The approximation is nonperturbative in the qubit-resonator coupling strength $g$ and performs excellently~\cite{Hausinger2010PRA} for negative detuning ($\Omega > \Delta$).\\
\indent For $\Delta=0$, the eigenstates of the system are composed of tensor products of displaced oscillator and qubit eigenstates. The exact spectrum is given by the combination of qubit and resonator energies 
\be{E_Rabi_Delta0}
E_{\mp,j}=\mp\frac{\hbar}{2}\epsilon_0+\hbar j\Omega-\hbar\frac{g^2}{\Omega},\quad j=0, 1, 2,..\;.
\ee
A finite tunneling $\Delta$ mixes the eigenstates nontrivially. To find an approximate analytical solution, we notice that for static bias values $\epsilon_0 = l\Omega$ one can identify two-fold degenerate subspaces in the complete Hilbert space of the problem: $E_{+,j}=E_{-,l}$. A finite tunneling removes the degeneracy and induces  coupling among the dressed states. Using Van Vleck perturbation theory to lowest order in the tunneling one finds the modified energies \cite{Hausinger2010PRA} 
\begin{equation}\label{RabiVVPT}
 E_{\mp,j}^l \simeq \hbar \left[ \left(j+ \frac{l}{2}\right) \Omega - \frac{g^2}{\Omega} + \frac{1}{8} \left(\varepsilon_{\downarrow,j}^{(2),l}-\varepsilon_{\uparrow,j+l}^{(2),l}\right) \mp \frac{1}{2} \Omega_j^l\right]\;,
\end{equation} 
where $l$ identifies the degeneracy points $\epsilon_0=l\Omega$. For example, around $\epsilon_0=0$ the energy levels are, in ascending order, $ E_{-,0}^0,E_{+,0}^0,E_{-,1}^0,\dots $ and around $\epsilon_0=\Omega$ they are given by $E_{-,0}^0,E_{-,0}^1,E_{+,0}^1,\dots$, while for $\epsilon_0=-\Omega$ we have $E_{-,0}^0,E_{-,1}^{-1},E_{+,1}^{-1},\dots$.
For the level splitting in Eq.~\eqref{RabiVVPT} it is found
\begin{equation} \label{DressedOmega_static}
 \Omega_j^l =\sqrt{\left[ \epsilon_0 - l \Omega + \frac{1}{4} \left(\varepsilon_{\downarrow,j}^{(2),l}+\varepsilon_{\uparrow,j+l}^{(2),l}\right) \right]^2+ \left(\Delta_j^{j+|l|}\right)^2}\;,
\end{equation} 
where the so-called diagonal corrections are given by
\begin{equation}\label{2nd-order_corr}
\varepsilon^{(2),l}_{\downarrow/\uparrow,j} = \sum_{\substack{k=-j \\k \neq \pm l}}^\infty \frac{ \left(\Delta_{j}^{k+j}\right)^2}{\epsilon_0\mp k \Omega}\;.
\end{equation}
The dressed tunneling elements carry information on the overlap of the displaced oscillator states. They read
\begin{align} \label{DressedDelta}
\Delta^{j^\prime}_j 
=& \Delta \, [{\rm sgn}(j'-j)]^{|j'-j|} \,D^{|j^\prime -j|}_{{\rm min}\{j,j'\}} (\tilde{\alpha})\;,
\end{align}
with $j$ and $j'$ the number of oscillator quanta involved in the dressing,
\begin{equation}\label{Dkj}
 D^{k}_{j} (\tilde{\alpha})=  \tilde{\alpha}^{k/2} \sqrt{\frac{j!}{(j+k)!}}  \mathsf{L}^k_j (\tilde{\alpha}) e^{-\frac{\tilde{\alpha}}{2}},
\end{equation} 
and $\tilde{\alpha} = (2 g / \Omega)^2$.
Here, $\mathsf{L}^k_j(\tilde{\alpha})$ are the generalized Laguerre polynomials defined by the recurrence relation
\be{}
\mathsf{L}^k_{j+1}(\tilde{\alpha})=\frac{(2j+1+k-\tilde{\alpha})\mathsf{L}^k_j(\tilde{\alpha})-(j+k)\mathsf{L}^k_{j-1}(\tilde{\alpha})}{j+1}\;,
\ee
with $\mathsf{L}^k_0(\tilde{\alpha})=1$ and $\mathsf{L}^k_1(\tilde{\alpha})=1+k-\tilde{\alpha}$. For $|\epsilon_0|<\Omega$ we fix $l=0$.\\
\indent For $\Omega = \epsilon_0$, corresponding to $l=1$,  one finds for the avoided crossing involving the first and second excited state, $\Delta_0^1=\Delta\sqrt{\tilde \alpha}e^{-\tilde{\alpha}/2}$. 
In the limit of small $\tilde{\alpha}$, one can expand the dressed tunneling splittings in order to obtain the famous Rabi splitting of the Jaynes-Cummings model which, at resonance, assumes the value $2\hbar g\sqrt{j+1}$.
Noticeably, in the high-photon limit, $j,j' \to  \infty$, and for finite $j-j'$, this dressing by Laguerre polynomials becomes a dressing by Bessel functions known for quantum systems under intense electromagnetic fields \cite{Grifoni1998}. The interplay between quantum and classical radiation is the topic of the next subsection. 

\subsection{Driven Rabi model} 
\label{Analytical_Rabi_driven}

We include now a drive on the qubit. The picture is  enriched, with respect to the static case, by the presence of new resonances and by the modulation induced by the Bessel functions, which stems from the classical drive, on top of the Laguerre dressing given by the quantum oscillator, i.e. the resonator. The spectrum can now be calculated within a dressed Floquet picture, with quasi-energies known exactly for the case $\Delta=0$. Similar to the static case, two-fold  degeneracies now occur when $\epsilon_0=l\Omega -m \omega_{\rm d}$.
Within leading order Floquet-Van Vleck perturbation theory in $\Delta$,  the quasi-energy spectrum of the driven, nondissipative Rabi model now reads \footnote{We keep the same convention for the indexes as in~\cite{Hausinger2010PRA}.}~\cite{Hausinger2011}
\be{Quasienergies}
E^{m,l}_{\mp,n,j} =  \hbar \left[ - \left( n + \frac{m}{2}\right) \omega_{\rm d} + \left(j + \frac{l}{2}\right) \Omega - \frac{g^2}{\Omega} + \frac{1}{8}(\varepsilon^{(2),m,l}_{\downarrow, n, j} - \varepsilon^{(2),m,l}_{\uparrow, n+m, j+l}) \mp \frac{1}{2}\Omega_{n,j}^{m,l} \right],
\ee 
where the indexes $n$ and $j$ denote the Floquet mode and oscillator quantum number, respectively, while $m$ and $l$ give the resonance condition. The doublets' amplitudes  are now given by, cf. Eq.~\eqref{DressedOmega_static},
\be{DressedOmega_driven}
\Omega_{n,j}^{m,l} \simeq  \sqrt{\left[\epsilon_0 + m \omega_{\rm d} - l \Omega +\frac{1}{4}\left(\varepsilon^{(2),m,l}_{\downarrow, n, j} + \varepsilon^{(2),m,l}_{\uparrow, n+m, j+l}\right) \right]^2 + \left(\Delta^{n+m,j+|l|}_{n,j}\right)^2}\;,
\ee
where the diagonal corrections read 
\be{}
\varepsilon^{(2),m,l}_{\downarrow/\uparrow,n,j} = \sum_{\substack{p=-\infty \\p \neq -m}}^\infty \sum_{\substack{k=-j \\k \neq \pm l}}^\infty \frac{ \left(\Delta_{n,j}^{n+p,j+k}\right)^2}{\epsilon_0+p\omega_{\rm d}\mp k \Omega}\;.
\ee
The  tunneling elements are further dressed by  Bessel functions as
\be{DressedDelta_driven}
\Delta^{n',j'}_{n,j}=J_{n'-n}(\epsilon_{\rm d}/\omega_{\rm d})\Delta_j^{j'} \;,
\ee
with $\Delta_j^{j'} $ defined in Eq.~\eqref{DressedDelta}. Noticeably, the bare tunneling splitting is now dressed by {\em both} quanta of the resonator and of the driving microwave radiation.\\
\indent At the symmetry point, $\epsilon_0=0$, the resonances still occur when $m\omega_{\rm d}\simeq l\Omega$.
For example, in the case $m=1$, $l=2$, one finds  avoided crossings with tunneling splitting 
$\Delta_{n,j}^{n+1,j+2}=J_1(\epsilon_{\rm d}/\omega_{\rm d})\Delta_j^{j+2}$. 
 As we shall see in the Sec.~\ref{section_results}, these resonances dominate the low-energy transmission of the driven Rabi model for the chosen parameter set. Before this, we illustrate in the coming section how to relate the qubit transmission spectra, namely the  spectral properties of the Rabi model as probed in the experimental setup, see Fig.~\ref{scheme}(c)-(d), to the steady-state response of the qubit. 

\section{Transmission}
\label{section_transmission}

In actual experiments, see Fig.~\ref{scheme}(c), the probe field  $V_{\rm in}^{\rm p}(t)=f_{\rm Z}\epsilon_{\rm p}\cos(\omega_{\rm p} t)$ is applied to the qubit via an external transmission line. 
Following~\cite{Vool2017,Magazzu2018}, the corresponding transmitted field is  
$V_{\rm transm}(t)=V_{\rm in}^{\rm p}(t)-f\dot{P}(t)/2$, where $P(t):=\langle\sigma_z(t)\rangle$ is the so-called qubit population difference in the localized qubit basis. The proportionality constants $f_{\rm Z}$ and $f$ depend on the  details of the experimental setup and have  dimensions of a magnetic flux.
In terms of the Fourier-transformed probe and transmitted fields calculated at the probe frequency, the transmission is defined as the square modulus of the complex coefficient
\begin{equation}\begin{aligned}\label{T}
\mathcal{T}(\omega_{\rm p})=\tilde{V}_{\rm transm}(\omega_{\rm p})/\tilde{V}_{\rm in}^{\rm p}(\omega_{\rm p})\;.
\end{aligned}\end{equation}
At the steady state, the population difference has the period of the probe, also in the presence of a high-frequency drive, provided that we average over its period $2\pi/\omega_{\rm d}$ the kernels of the exact generalized master equation (GME) for $P
(t)$, see Appendix~\ref{appendix_driven_NIBA}. Expanding in Fourier series the time-periodic asymptotic population difference  
$P_{\rm as}(t)=\lim_{t\to\infty}P(t)$ as~\footnote{Note the different convention used for the signs with respect to Ref.~\cite{Magazzu2018}.}   
\begin{eqnarray}\label{Pasdot}
\dot{P}_{\rm as}(t)&=&\sum_m -{\rm i}m\omega_{\rm p} p_m e^{-{\rm i}m\omega_{\rm p} t}\;,
\end{eqnarray}
where 
\begin{eqnarray}\label{Scoeff}
p_m=\frac{\omega_{\rm p}}{2\pi}\int_{-\mathcal{\pi}/\omega_{\rm p}}^{\mathcal{\pi}/\omega_{\rm p}}dt\; P_{\rm as}(t)e^{{\rm i}m\omega_{\rm p} t}\;,
\end{eqnarray}
we find for the transmission at the probe frequency $\omega_{\rm p}$, within linear response to the probe field,
\be{transmission}
\mathcal T(\omega_{\rm p})=1+{\rm i}\mathcal{N}\hbar\omega_{\rm p}  \chi(\omega_{\rm p})\;,
\ee
where $\mathcal{N}= f/f_{\rm Z}$ and $\chi(\omega_{\rm p})=p_1
/\hbar\epsilon_{\rm p}$. 
Hence, the theoretical quantity of interest is the linear susceptibility evaluated at the probe frequency. Its form depends on the considered dissipation regime. In the next subsection we provide an explicit approximate expression for the response function.
\subsection{Linear susceptibility within the NIBA}
\label{subsec_chi_NIBA}
Within the NIBA~\cite{Leggett1987, Dekker1987, Weiss2012}, the GME that describes the driven, dissipative qubit dynamics, yields an analytical expression for the linear susceptibility $\chi(\omega_{\rm p})$, see Appendix~\ref{appendix_chi}, which is nonperturbative in the qubit-baths coupling. It reads 
\be{chi}
\chi(\omega_{\rm p})=\frac{(\hbar\omega_{\rm p})^{-1}}{-{\rm i}\omega_{\rm p}+\hat{k}_{0}^+(-{\rm i}\omega_{\rm p})}\left[  \hat{\kappa}_{+1}^-(0)-\hat{\kappa}_{+1}^+(0)\frac{ \hat{k}_{0}^-(0)}{\hat{k}_{0}^+(0)}\right]\;,
\ee
where the GME kernels are defined as
\be{kernels}
\hat{k}^\pm_0(\lambda)=&\Delta^2\int_0^\infty d\tau\; e^{-\lambda\tau} e^{-Q'(\tau)}c^\pm[Q''(\tau)]J_0\left[\frac{2\epsilon_{\rm d}}{\omega_{\rm d}}\sin\left(\frac{\omega_{\rm d} \tau}{2}\right) \right] c^\pm(\epsilon_0\tau)\;,\\
\hat{\kappa}^\pm_1(0)=&\mp \Delta^2\int_0^\infty d\tau\;e^{{\rm i}\omega_{\rm p}\tau/2}e^{-Q'(\tau)}c^\pm[Q''(\tau)]J_0\left[\frac{2\epsilon_{\rm d}}{\omega_{\rm d}}\sin\left(\frac{\omega_{\rm d} \tau}{2}\right) \right]\sin(\omega_{\rm p}\tau/2) c^\mp(\epsilon_0\tau)\;,
\ee
with $c^+(x)=\cos(x)$ and $ c^-(x)=\sin(x)$, see also~\cite{Grifoni1998, Magazzu2018, Magazzu2019}. Here, $J_0(x)$ is the Bessel function of the first kind~\cite{Gradshteyn1980}, which stems from averaging the kernels over the drive period, see Appendix~\ref{appendix_k_tau}. The baths' correlation function is the sum $Q(t)=\sum_\nu Q_\nu(t)$, where
\begin{equation}\label{Qdef}
Q_{\nu}(t)=\sum_{\nu}\int_{0}^{\infty}d\omega \frac{G_{\nu}(\omega)}{\omega^2}\left[ \coth{\left(\frac{\hbar\omega\beta_{\nu}}{2}\right)}(1-\cos{\omega t})+{\rm i}\sin{\omega t} \right]\;,
\end{equation}
with  $\beta_\nu:=(k_{\rm B} T_\nu)^{-1}$ the inverse temperature of bath $\nu$. Explicit expressions for $Q_\nu(t)$ are provided in Appendix~\ref{appendix_driven_NIBA}.\\
\indent The NIBA is perturbative in the qubit splitting $\Delta$ and provides accurate results in the case of zero bias, $\epsilon_0=0$, and in the presence of a finite bias for sufficiently strong dissipative coupling and/or high temperatures. This is due to the enhanced downward-renormalization of $\Delta$ by increasing $\alpha$ and to the fact that the real parts of the correlation functions $Q(t)$ suppress effectively the time-nonlocal correlations in the two-state path integral, rendering the present NIBA treatment appropriate~\cite{Weiss2012}. These considerations hold for both the Ohmic and the effective structured bath acting on the qubit, see Eqs.~\eqref{Qsl1-a}-\eqref{Qmats}. In the presence of a high-frequency drive, the additional drive-induced renormalization of the qubit parameter $\Delta$, Eq.~\eqref{DressedDelta_driven},  extends the reach of this approximation scheme to regimes of lower dissipation. The NIBA has been applied in the presence of multiple baths, notably in the context of heat transport, see e.g.~\cite{Nicolin2011,Boudjada2014, Segal2014, Yamamoto2018}.\\
\indent In the following, for the case of a biased qubit and weak dissipation/low temperatures, we use a weak damping master equation approach. The NIBA is used for strong dissipation and in the presence of a drive in its range of applicability, which includes the unbiased case at weak dissipation. Transmission spectra of the static and driven setup are shown in the following section for both dissipation regimes.

\section{Transmission spectra}
\label{section_results}

\begin{figure}[ht!]
\begin{center}
\includegraphics[width=1\textwidth,angle=0]{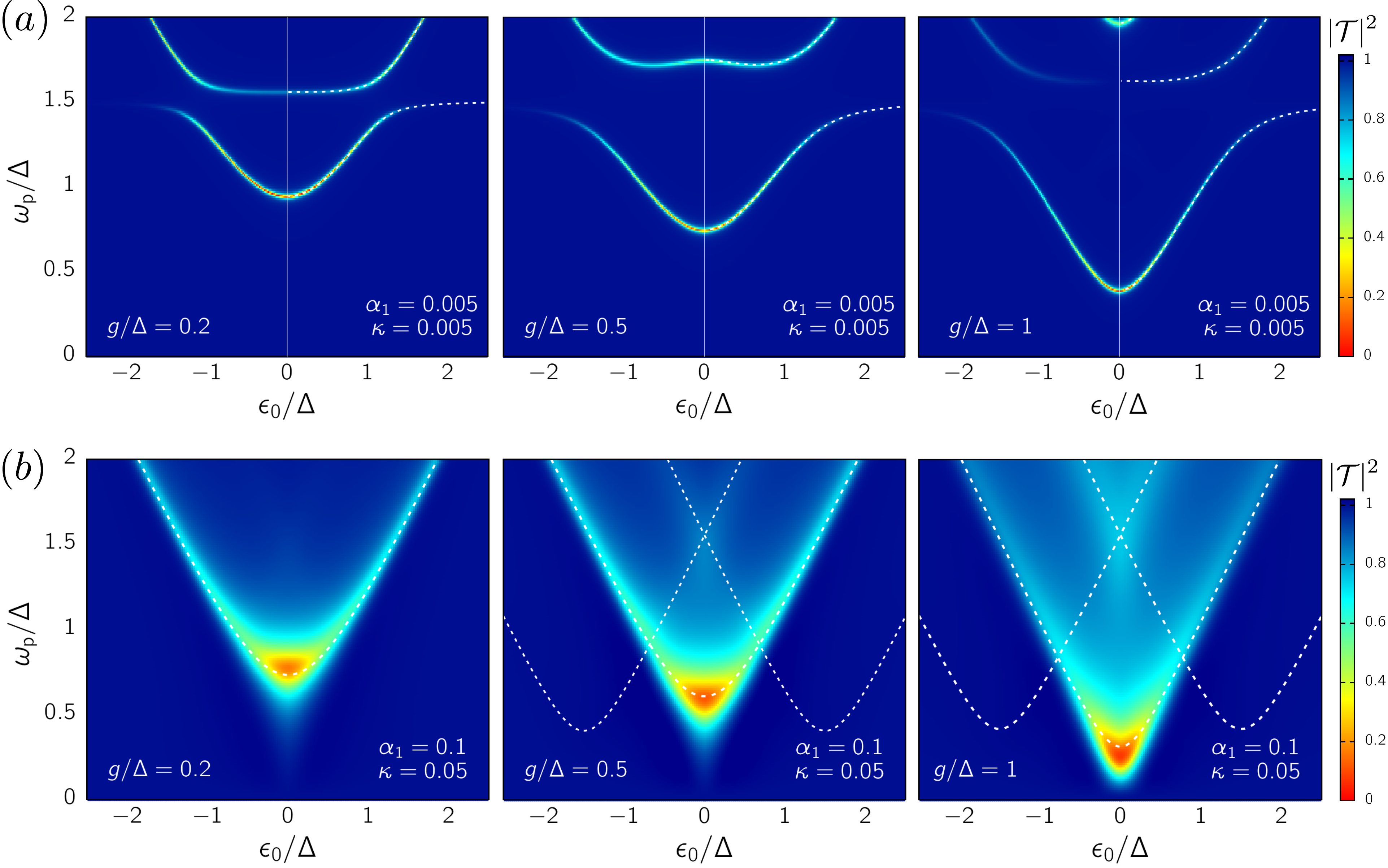}
\caption{Transmission spectra of the static system with resonator frequency $\Omega=1.5~\Delta$ and different values of the qubit-resonator coupling $g$. The transmission is evaluated using Eq.~\eqref{transmission} with $\mathcal{N}$  chosen differently for each plot in order to have the minimum of the transmission assuming the value $|\mathcal{T}|^2\simeq 0$.
(a) - Weak dissipation:  The susceptibility $\chi(\omega_{\rm p})$ is calculated using the approach of~\cite{Kohler2018} with dephasing rates from the Bloch-Redfield master equation in the \emph{dressed} basis of the USC system. The dashed white lines are the transition energies between the ground and the first excited states of the closed Rabi model obtained by numerical diagonalization. The resonator's Hilbert space is truncated to the first 10 energy levels. (b) - Strong dissipation: The susceptibility is calculated with the path integral approach within the NIBA, Eq.~\eqref{chi}, starting from the effective bath mapping in Eq.~\eqref{H_SB}. The dashed lines are given by the frequency gaps $\tilde{\Omega}_0^l$, Eq.~\eqref{omega_r}, with $l=0,\pm1$. In both panels the cutoff frequency of the qubit bath is $\omega_{\rm c}=10~\Delta$ and temperature is $k_{\rm B}T = 0.1~\hbar\Delta$ for both baths. }
\label{fig_T_static}
\end{center}
\end{figure}

In the following, we show the results for the transmission with the probe on the qubit, Eq.~\eqref{transmission}, with the resonator frequency set to $\Omega=1.5~\Delta$, both in the static case and in the presence of the drive on the qubit. In the latter setting, we fix the drive frequency to the value $\omega_{\rm d}=2.7~\Delta$. 
In order to see how the picture of the static Rabi model is impacted by a classical drive on the qubit, we start by showing  in Fig.~\ref{fig_T_static} the static case in two dissipation regimes, which in turn provides insight on the effect of dissipation on the spectra. In the first, where the USC system is weakly coupled to the environment, we calculate the susceptibility $\chi(\omega_{\rm p})$ using the approach developed in~\cite{Kohler2018} with dephasing rates from the Bloch-Redfield master equation calculated in the dressed basis of the USC system. In the other regime considered, which is of strong dissipation, we use the path-integral approach within the NIBA, Eqs.~\eqref{chi}-\eqref{Qdef}, for the spin-boson model with the effective bath mapping, Eq.~\eqref{H_SB}. For both cases, we consider three values of the qubit-resonator coupling, namely $g/\Delta=0.2,~0.5$ and $1$, the latter two being well into the USC regime. 
The picture that emerges from the spectra differs, especially at strong dissipation, from the standard spectroscopy of USC systems, where the transmission of a weakly dissipative USC system is recorded by probing the resonator~\cite{Yoshihara2017}, see Appendix~\ref{probe_resonator} for a comparison. 
In our transmission measurement protocol, which essentially measures the qubit operator $\sigma_z$, the transmission is  given by the difference in populations of the states $|\uparrow\;\rangle, |\downarrow\;\rangle$ of the qubit basis.  As such, the resonator is traced out from the dynamical response of the system, its presence being reflected in the pattern of resonances involving qubit and resonator. 
Qualitatively, this leads to a spectrum resembling the one of the qubit alone, 
with a principal feature, a pronounced dip centered at $\epsilon_0=0$, which is faithfully reproduced by the transition frequency $\tilde{\Omega}_{0}^0$ that is renormalized by the Ohmic bath  acting on the qubit.
There are, however, two major modifications that are peculiar of the Rabi model. i) Emission and absorption of $l$ oscillator quanta produce {\em sidebands} of the main spectral features. ii) Avoided crossings are visible at weak dissipation, where the environment-induced renormalization of the bare qubit splitting $\Delta$ and oscillator frequency $\Omega$ is negligible, which signal the strong resonator-qubit entanglement, according to Eqs.~\eqref{RabiVVPT} and~\eqref{DressedOmega_static}.\\
\indent For our analysis, we introduce the \emph{bath-renormalized} transition frequencies
\be{omega_r}
\tilde\Omega_{j}^l=(\tilde{E}_{+,j}^l-\tilde{E}_{-,j}^l)/\hbar\;,
\ee
see Eqs~\eqref{RabiVVPT}-\eqref{DressedOmega_static}, where $\tilde{E}_{\pm,j}^l$ give the eigenenergies [or quasienergies, with appropriate additional indexes, Eq.~\eqref{Quasienergies}] of the closed Rabi model around the bias point $\epsilon_0 = l\Omega$. When comparing with the NIBA results, these renormalized energies are calculated by substituting the bare qubit splitting $\Delta$ with its dissipation-renormalized version: $\Delta\rightarrow\Delta_T$, where 
$\Delta_T=\Delta_r(2\pi k_{\rm B}T/\hbar\Delta_r)^{\alpha_1}$ and $\Delta_r=\Delta(\Delta/\omega_c)^{[\alpha_1/(1-\alpha_1)]}$~\cite{Weiss2012}.\\
\indent In the absence of the resonator, the principal dip at zero static bias would occur at $\omega_{\rm p}\sim\Delta$ for weak dissipation, with a downwards renormalization for increased system-bath coupling, see~\cite[Fig.~2]{Magazzu2018}. In Fig.~\ref{fig_T_static}, the presence of the (dissipative) resonator moves this principal feature towards lower frequencies upon  increasing $g$. This effect can be understood in terms of the renormalization of $\Delta$ by Laguerre polynomials discussed in Sec.~\ref{Analytical_Rabi_static}, see Eq.~\eqref{DressedDelta}, which yields the spectrum of the Rabi model. Thus, now the main dip is centered at $\omega_{\rm p}=\tilde{\Omega}_0^0|_{\epsilon_0=0}\simeq \Delta_T \exp(-\tilde\alpha/2)$, where $\tilde\alpha=(2g/\Omega)^2$. Here and in what follows, we neglect for simplicity the second-order corrections $\varepsilon^{(2),l}_{\downarrow/\uparrow,j}$, Eq.~\eqref{2nd-order_corr}. \\
\begin{figure}[ht!]
\begin{center}
\includegraphics[width=1\textwidth,angle=0]{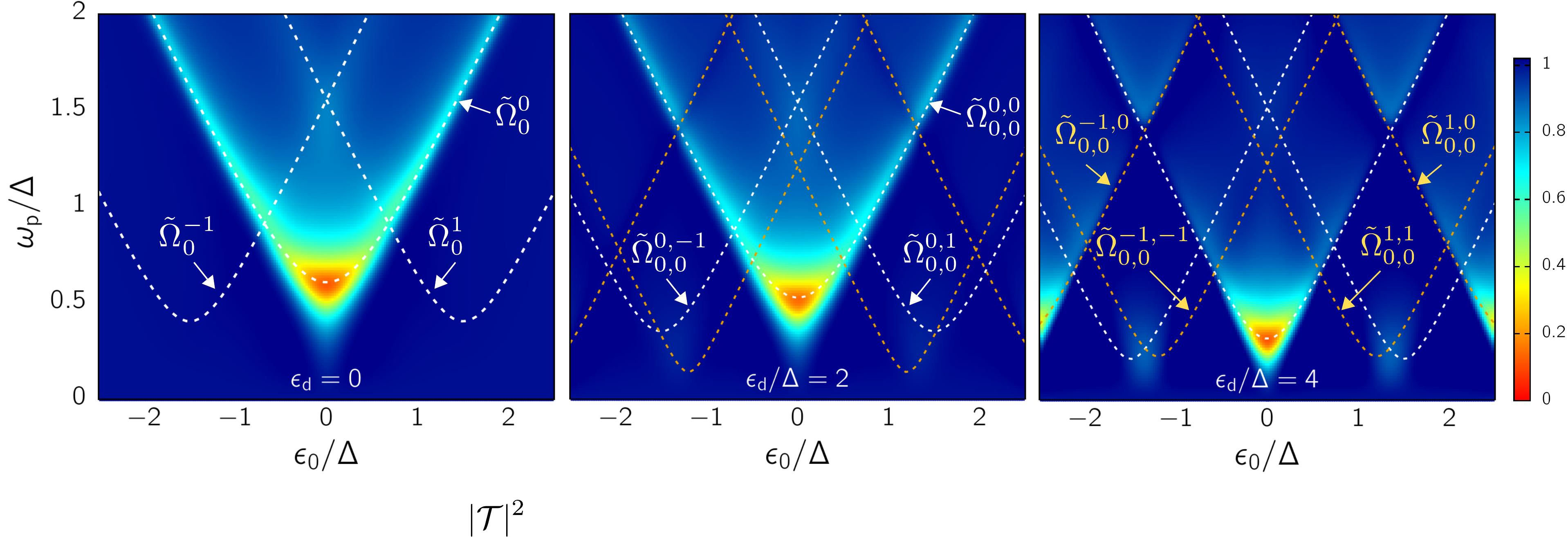}
\caption{Transmission spectra of the driven system at strong dissipation. The qubit-resonator coupling is fixed to $g=0.5~\Delta$ and three values of the drive amplitude are considered. From left to right:  $\epsilon_{\rm d}/\Delta=0,~2,$ and $4$. The remaining parameters are as in Fig.~\ref{fig_T_static}(b). Note that the panel on the left (static case) coincides with the central panel of Fig.~\ref{fig_T_static}(b) with the frequency gaps $\tilde{\Omega}_0^l$ given by Eq.~\eqref{omega_r}. The drive frequency is fixed to the value $\omega_{\rm d}/\Delta=2.7$. We note that increasing the driving strength has the two-fold effect of downward-renormalizing the frequency of the principal dip a zero bias, similarly to what happens as $g$ is increased in the static system (see Fig.~\ref{fig_T_static}(b) ), and of creating a pattern of multi-photon resonances described by $\tilde{\Omega}_{0,0}^{m,l}$ (dashed lines, Eqs.~\eqref{DressedOmega_driven} and~\eqref{omega_r}). The golden dashed lines,  $m\neq 0$, involve the drive frequency $\omega_{\rm d}$.}
\label{fig_T_driven}
\end{center}
\end{figure}
\indent The avoided crossings present at weak dissipation, Fig.~\ref{fig_T_static}(a), are given by the difference between the transition energies $E_2-E_0$ and $E_1-E_0$ at $\epsilon_0=\Omega$, namely by the dressed tunneling element $E_2-E_1=E_{+,0}^1-E_{-,0}^1\simeq |\Delta^1_0|=\Delta \tilde{\alpha}^{1/2}\exp(-\tilde{\alpha}/2)/\sqrt{2}$, see Eqs.~\eqref{RabiVVPT} and~\eqref{DressedDelta}, showing a non-monotonic behavior with respect to $g$. This non-monotonicity is more evident at resonance, see Fig.~\ref{fig_T_static_Om1}.
The crossing pattern at strong dissipation, Fig.~\ref{fig_T_static}(b), in the region $\epsilon_0 \sim 0$ is well reproduced by the frequency gaps $\tilde{\Omega}_0^0$ and $\tilde{\Omega}_0^{\pm 1}$ which render the transitions $|\uparrow\rangle \leftrightarrow |\downarrow\rangle$ with the oscillator in a fixed state $l$, adiabatically following the qubit transitions~\cite{Ashhab2010}, which is consistent with a strongly suppressed qubit splitting. We note that the avoided crossings are suppressed by dissipation. Figure~\ref{fig_T_static} shows that, in the static case, there is no response from the qubit outside the region $|\epsilon_0|\leq \omega_{\rm p}$. This is because, contrary to the driven case, see Fig.~\ref{fig_T_driven}, the weak probe does not induce qubit transitions as its frequency cannot match the qubit frequency gap $\sqrt{\Delta_T^2+\epsilon_0^2}$. In Appendix~\ref{intermediate_regime}, we compare, in an intermediate dissipation regime, the results from the two different approaches used here. Both treatments well reproduce the main resonance involving the ground/first excited state transition. However, in agreement with the findings in Fig.~\ref{fig_T_static}, strong differences appear for the higher lying excitations. \\
\indent A similar, though richer, picture is revealed by Fig.~\ref{fig_T_driven}, where we consider the spectra in the same strongly dissipative setting of Fig.~\ref{fig_T_static}(b) in the presence of the drive, for three values of the drive amplitude $\epsilon_{\rm d}$, the first being $\epsilon_{\rm d}=0$ for reference. While varying  $\epsilon_{\rm d}$, we fix the coupling to $g=0.5~\Delta$.
Here, additional renormalization of $\Delta$ with appropriate Bessel functions is expected, see Eqs.~\eqref{DressedOmega_driven} and~\eqref{DressedDelta_driven}. For the main transmission dip, the zero order Bessel function is involved, and its location is found to be at
\be{dip_driven}
\omega_{\rm p}=\tilde{\Omega}_{0,0}^{0,0}|_{\epsilon_0=0}\simeq |\Delta_T \exp(-\tilde \alpha/2)J_0(\epsilon_{\rm d}/\omega_{\rm d})|\;.
\ee
 In the resulting set of spectra, the drive amplitude $\epsilon_{\rm d}$ takes the role played by $g$ in the static case, in that $\epsilon_{\rm d}$ tunes the frequency of the main transmission dip and the size of the drive-induced avoided crossings. A detailed account of this effect for zero static bias is provided in Fig.~\ref{fig_T_Bessel}(b).
Additional features emerge in the driven case displayed in Fig.~\ref{fig_T_driven} which are  due to multiple resonances of the bias with the drive frequency.
 These resonances yield replicas of the Rabi pattern, in the form of sidebands, reproduced by the transition frequencies $\tilde{\Omega}^{00}_{0,0},\tilde{\Omega}^{0,\pm 1}_{0,0}$,$\tilde{\Omega}^{\pm 1, 0}_{0,0}$, $\tilde{\Omega}^{1, 1}_{0,0}$, and ,$\tilde{\Omega}^{- 1,- 1}_{0,0}$, the last three involving the drive frequency $\omega_{\rm d}$, see Eqs.~\eqref{Quasienergies} and~\eqref{DressedOmega_driven}. An interesting feature emerging by comparison of the central and the right panels of Fig.~\ref{fig_T_driven} is that, while the central pattern fades for increased drive amplitude, the side replicas are enhanced, as the strong drive is able to induce these multi-photon transitions (which are absent in the probe-only spectra). 
\begin{figure}[h!]
\begin{center}
\includegraphics[width=1\textwidth,angle=0]{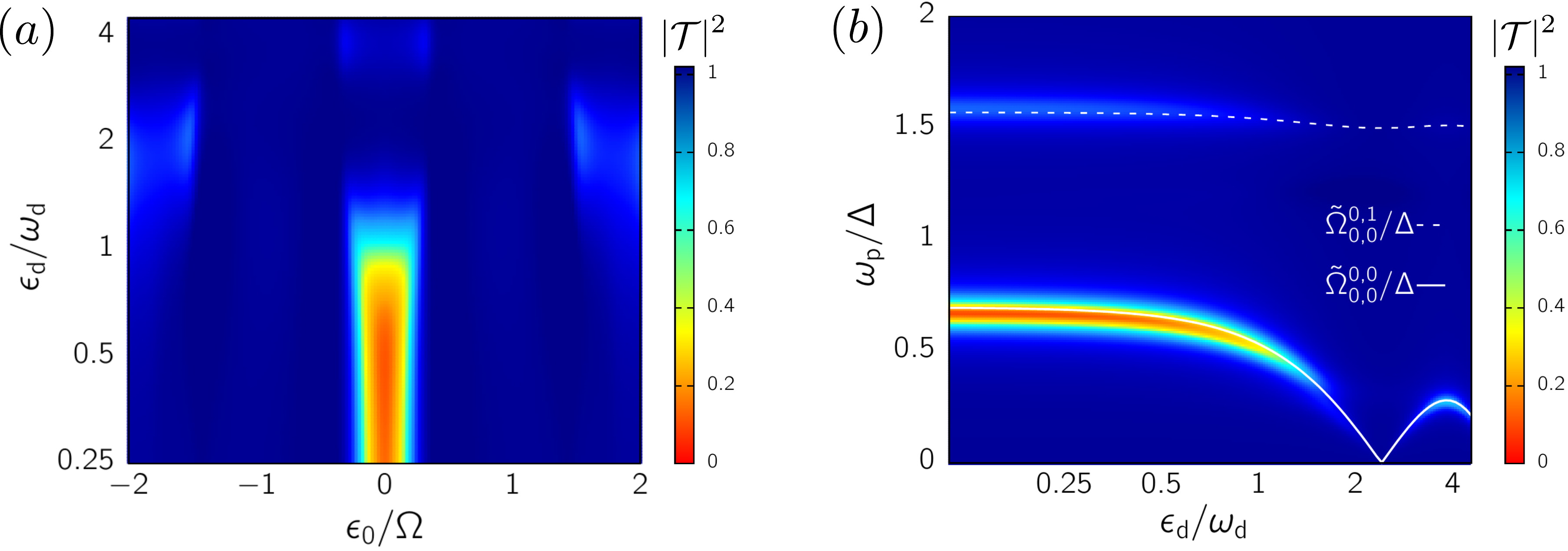}
\caption{Bessel pattern from drive-induced renormalization of the resonances. (a) - Transmission as a function of the static bias $\epsilon_0$ and of the drive amplitude $\epsilon_{\rm d}$, at fixed probe frequency $\omega_{\rm p}/\Delta=0.55$ and with $\alpha_1=0.1$ and $\kappa=0.05$. (b) - Transmission spectrum, as a function of the drive amplitude, for $\epsilon_0=0$, at lower dissipation strength, $\alpha_1=0.05$ and $\kappa=0.005$. For both panels, the remaining parameters are as in Fig.~\ref{fig_T_driven}. The solid and dashed curves in panel (b) are given by the renormalized transition frequencies $\tilde\Omega_{0,0}^{0,l}$, see Eqs.~\eqref{DressedOmega_driven} and~\eqref{omega_r}, where in the present unbiased case, $\tilde\Omega_{0,0}^{0,+1}=\tilde\Omega_{0,0}^{0,-1}$. In the transition frequency $\tilde{\Omega}_{0,0}^{0,0}$ it is clearly visible the Bessel pattern induced by the function $J_0(\epsilon_{\rm d}/\omega_{\rm d})$ due to the drive, see Eq.~\eqref{dip_driven}. In particular, the dip at  $\epsilon_d/\omega_d\approx 2.4$ corresponds to the first zero of such Bessel function.}
\label{fig_T_Bessel}
\end{center}
\end{figure}
\begin{figure}[ht!]
\begin{center}
\includegraphics[width=1\textwidth,angle=0]{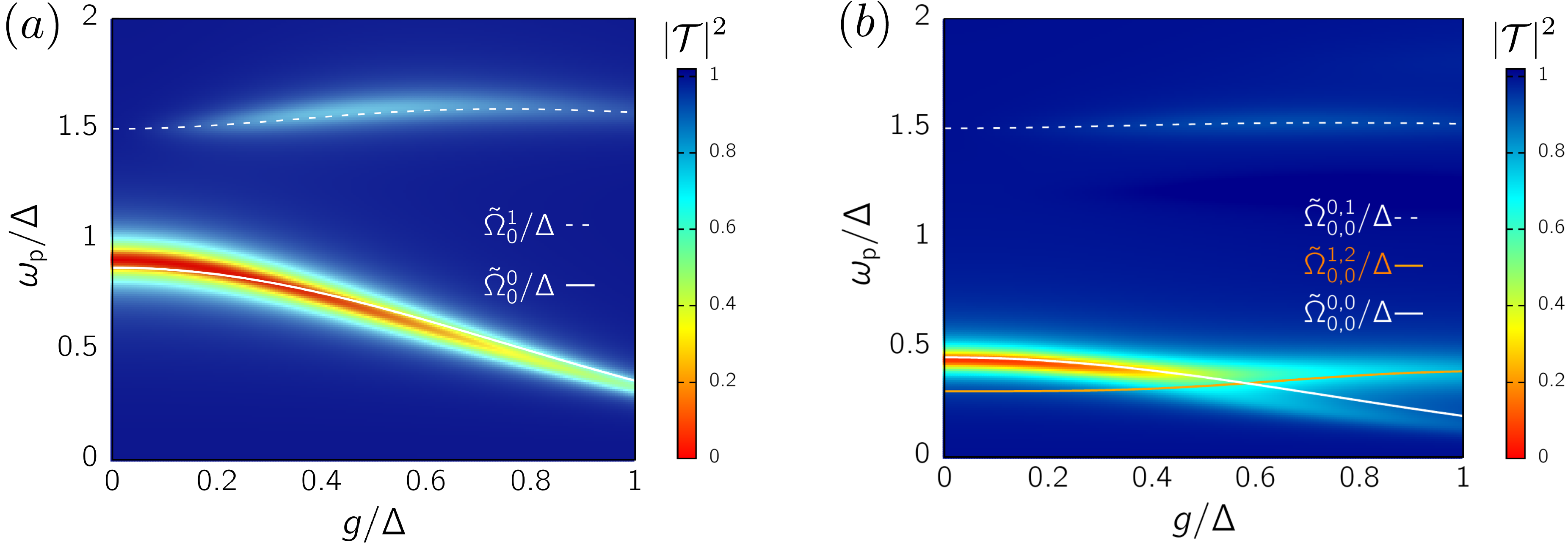}
\caption{Laguerre renormalization at $\epsilon_0=0$. Transmission at zero bias \emph{vs.} the qubit-resonator coupling $g$. (a) Static case, $\epsilon_{\rm d}=0$. (b) Driven case with $\epsilon_{\rm d}=4~\Delta$.
For both panels the other parameters are as in Fig.~\ref{fig_T_Bessel}(b). 
Panel (a) displays, in the transition $\tilde{\Omega}_{0}^0$, the Laguerre pattern from $\mathsf{L}_0$. In panel (b), the splitting of the resonance is given by higher order Laguerre polynomials stemming from the interplay of the resonator with the drive~\cite{Hausinger2011}. The frequencies $\tilde{\Omega}_{n,j}^{m,l}$ are given by Eq.~\eqref{DressedOmega_driven} with the substitution $\Delta\rightarrow\Delta_T$.}
\label{fig_T_Laguerre}
\end{center}
\end{figure}
Mathematically, this is due to the fact that, depending on the bias, the dressed tunneling element is modulated by Bessel functions $J_n(x)$ with different index $n$. In the case of  
Fig.~\ref{fig_T_driven}, the central pattern, $\epsilon_0\sim0$ is modulated by $J_0(\epsilon_{\rm d}/\omega_{\rm d})$ and the side replicas by  $J_1(\epsilon_{\rm d}/\omega_{\rm d})$.\\
\indent Such Bessel pattern is highlighted in Fig.~\ref{fig_T_Bessel}. In panel (a), the qubit transmission in the driven case is shown as a function of the static bias and of the drive amplitude, displaying the ``v"-shaped trace centered at zero bias. The modulation by $J_0(\epsilon_{\rm d}/\omega_{\rm d})$ causes the suppression of the qubit response at zero bias when the first zero of $J_0$ is reached. The plot shows the system's response at a specific probe frequency $\omega_{\rm p}$. 
By setting the static bias to zero, we study in Fig.~\ref{fig_T_Bessel}(b) the full spectrum \emph{vs.} the drive amplitude at weak dissipation, a regime where the NIBA is reliable for $\epsilon_0=0$~\cite{Nesi2007}. Such spectrum reveals that the renormalized transition frequency $\tilde{\Omega}_{0}^0$ follows the Bessel pattern induced by~$J_0(\epsilon_{\rm d}/\omega_{\rm d})$ as suggested by Eq.~\eqref{DressedDelta_driven}.\\
\indent The same is true for both panels of Fig.~\ref{fig_T_Laguerre}, where we show the spectrum at zero static bias as a function of the qubit-resonator coupling strength, for $\epsilon_{\rm d}=0$ and $\epsilon_{\rm d}=4~\Delta$. In panel (a), the features in the transmission are reproduced by the transition frequencies in Eq.~\eqref{DressedOmega_static} with $\Delta\rightarrow\Delta_T$.
Panel (b) of Fig.~\ref{fig_T_Laguerre} shows that the condition $l=0$ at zero bias, used for the spectrum of the static system, is no more generally true. Indeed, as can be seen from Eq.~\eqref{DressedOmega_driven}, the drive introduces novel resonance conditions with $l\neq 0$, i.e. when $l\Omega \simeq m \omega_{\rm d}$. In turn, this allows the contribution of dressed tunneling elements of the type $\Delta_{n,j}^{n+m,j+l}$. By inspection of Eqs.~\eqref{DressedDelta} and~\eqref{Quasienergies}-\eqref{DressedDelta_driven}, we see that, if  $m=1$ and $l=2$, the dressing involves $J_1(\epsilon_{\rm d}/\omega_{\rm d})\tilde{\alpha}\mathsf{L}_j^2(\tilde{\alpha})\exp(-\tilde{\alpha}/2)$, which, already for $j=0$, yields a nonmonotonic behavior of the corresponding resonance with respect to $\tilde{\alpha}=(2g/\Omega)^2$. The combined effect of the different resonances in the presence of the driving at $\epsilon_0=0$ is the splitting of the main resonance in Fig.~\ref{fig_T_Laguerre}(b). This feature is not visible in the absence of driving,  Fig.~\ref{fig_T_Laguerre}(a), because $l=0$ and the resulting dressed tunneling element $\Delta^j_j$ in Eq.~\eqref{DressedOmega_static} yields a simple exponential suppression.

\section{Conclusions}
\label{Section conclusions}

We have theoretically investigated the transmission spectra of the driven, dissipative Rabi model in the USC regime.
While transmission spectra generically carry information about spectral properties of the underlying quantum system being probed, the intensity and even the presence of resonance features crucially depend on which part of the system is coupled to the probe signal. Recent  experiments by Yoshihara et al. \cite{Yoshihara2017} have provided  spectroscopic data  of the (undriven) Rabi model in the deep USC regime, with the probe on the resonator. In the setup considered in this work, the probe couples to the qubit. Hence, the relevant observed quantity is the population difference between the qubit's supercurrent states. This in turn implies, especially at strong dissipation, a different spectral response than so far reported in the literature~\cite{Yoshihara2017,Yoshihara2017PRA}. 
To highlight the differences between the two probe settings, a comparison is provided in Appendix~\ref{probe_resonator}, see also the experimental spectra in Figs. 2 and 3 of Ref.~\cite{Yoshihara2017}.\\
\indent When probing the qubit, the strong coupling to the quantized resonator leads to sidebands in the spectrum, reflecting multiple absorption or emission of resonator's quanta. Furthermore, the  doublet structure of the Rabi system  is reflected in  avoided crossings  between subbands.  When the drive is present, additional photon sidebands appear which also display avoided crossings. An advantage of this setup is the possibility to tune the light-matter coupling in a continuous way. Indeed, the size and position of the avoided crossings depend on both the driving parameters as well as on the qubit-resonator  coupling strength. Characteristic Bessel/Laguerre evolutions upon varying the driving/coupling strength witness the interplay between the classical drive and the resonator in the nonequilibrium steady-state response of the qubit.\\
\indent The platform studied in this work is also suited for investigating the phenomenon of photon blockade~\cite{Imamoglu1997,Birnbaum2005}, where a coherent drive on a cavity coupled to an artificial atom, the resonator-qubit system in our case, produces an output of single photons. Such a nonclassical resonator output can be detected by measuring the photon-photon correlation function $g^{(2)}(t)$~\cite{Hoffman2011,Ridolfo2012}. This phenomenon has been investigated in standard cavity-QED models described by the Jaynes-Cummings Hamiltonian~\cite{Carmichael2015,Curtis2021} and the effects of the ultrastrong qubit-resonator coupling have been studied in~\cite{Ridolfo2012,LeBoite2016}. Since the phenomenon is ultimately due to the nonlinearity of the artificial atom, it is reasonable to expect that a similar emission would be observed by coherently driving (and probing) the qubit. In support of this expectation, the experiment carried out in~\cite{Hoi2012} already demonstrated photon blockade by a single qubit. 
Although the method used here is not suited for capturing the correlation function of the qubit output, an extension of the formalism in this direction could allow for studying how the phenomenon of photon blockade in a dissipative Rabi system is affected by driving and probing the qubit.\\
\indent In summary, we presented theoretical predictions for the spectroscopy of the driven, dissipative Rabi model.  Our results provide new insight and tools to investigate the physics of USC systems. Furthermore, they can be used to optimize the design of future experiments and for the interpretation of spectroscopic results.

\section{Acknowledgments}
The authors thank G. Falci and A. Ridolfo for fruitful discussions on the spectroscopy of the  Rabi model.
L. M. and M. G. acknowledge support by the BMBF (German Ministry for Education and Research), project 13N15208, QuantERA  SiUCs. P. F.-D. acknowledges support from  ``la Caixa" Foundation - Junior leader fellowship (ID100010434-LCF/BQ/PR19/11700009), Ministry of Science and Innovation and Agencia Estatal de Investigaci\'{o}n (FIS2017-89860-P; SEV-2016-0588; PCI2019-111838-2), European Commission (FET-Open AVaQus GA 899561; QuantERA SiUCs), and program  ``Doctorat Industrial" of the Agency for Management of University and Research Grants (2020 DI 41; 2020 DI 42). IFAE is partially funded by the CERCA program of the Generalitat de Catalunya.

\appendix

\section{Caldeira-Leggett Hamiltonian and spectral density function}
\label{appendix_mapping}

The Caldeira-Leggett model~\cite{Caldeira1981, Caldeira1983} describes an open system bi-linearly interacting via the operator $\hat{X}$ with a heat bath of harmonic oscillators
\be{H_Caldeira_Leggett}
H_{\rm CL}
=\frac{\hat{P}^2}{2M}+V(\hat{X})+\frac{1}{2}\sum_{j=1}^{N} \left[
\frac{\hat{p}^{2}_{j}}{m_{j}}+m_{j}\omega^{2}_{j}\left ( \hat{x}_{j}
-\frac{c_{j}} {m_{j}\omega^{2}_{j} } \hat{X} \right )^{2}\right].
\ee
The interaction term is $\hat{X} \sum_jc_j \hat{x}_j$ and the spectral density function $J(\omega)$ is defined as
\begin{equation}\label{J}
J(\omega)=\frac{\pi}{2}\sum_{j=1}^{N}\frac{c_{j}^{2}}{m_{j}\omega_{j}}\delta(\omega-\omega_{j})\;.
\end{equation}
For a bosonic bath we have $\hat{x}_j=\sqrt{\hbar/2m_j\omega_j}(a_j+a_j^\dag)$. Introducing the dimensionless system position operator $\hat{Q}:=\hat{X}/X_0$ we can write
\be{H_Caldeira_Leggett2}
H_{\rm CL}
=&H_{\rm S}+\sum_l\hbar\omega_j a_j^\dag a_j -\hat{Q}\sum_j \hbar\lambda_j(a_j+a_j^\dag)+A^2\;,
\ee
where $A^2$ is proportional to the square of $\hat{X}$. If the open system is a qubit, this term is a constant since $\sigma_z^2=\mathbf{1}$ while for a harmonic oscillator it constitutes a nonlinear term which ensures position-independent friction~\cite{Weiss2012}. In Eq.~\eqref{H_Caldeira_Leggett2}, we introduced the coupling with dimension of an angular  frequency 
\[
\lambda_j=\frac{X_0}{\sqrt{2\hbar m_j \omega_j}}c_j\;.
\]
In the context of the spin-boson model, it is also customary to define the modified spectral density function 
\be{G_appendix}
G(\omega):=\sum_{j=1}^{N}\lambda_j^2\delta(\omega-\omega_{j})=\frac{X_0^2}{\pi\hbar}J(\omega)\;.
\ee
For an Ohmic bath, in the continuum limit, $J(\omega)=M\gamma\omega$, where the friction coefficient $\gamma$ coincides with the memoryless friction kernel in the corresponding generalized Langevin equation for the operator $\hat{X}$.

\begin{itemize}

\item If the system coupled to the bath is a harmonic oscillator then $\hat{X}=X_0(B^\dag+B)$, with $X_0=\sqrt{\hbar/2M\Omega}$. Setting  $G(\omega)=\kappa\omega$, Eq.~\eqref{G_appendix} yields $\kappa=M\gamma X_0^2/(\pi/\hbar)$
 and  we can identify $\gamma=2\pi\kappa\Omega$.

\item For a qubit coupled to the Ohmic bath the coupling coordinate is $\hat{X}=(X_0/2) \sigma_z$, where  $X_0$ is the interwell distance~\cite{Weiss2012}. The spin-boson spectral density function is defined as $G(\omega)=2\alpha\omega$. From Eq.~\eqref{G_appendix} we have $\alpha=M\gamma X_0^2/(2\pi/\hbar)$. 

\end{itemize}

\section{The driven, dissipative Rabi model within NIBA}
\label{appendix_driven_NIBA}

As shown in Sec.~\ref{section_transmission}, the transmission is related to the response of the qubit to the probe field via the population difference $P(t)=\langle\sigma_z(t)\rangle$, i.e. the expectation value of $\sigma_z$, expressed in the localized (flux) states of the qubit. 
An exact formal expression for $P(t)$ in the presence of external heat baths, the Ohmic bath and the dissipative resonator in our case, and of a classical time-dependent drive is found  within the path integral representation of the qubit reduced dynamics~\cite{Carrega2015, Carrega2016}.

The time evolution of the population difference can be given in terms of an exact generalized master equation (GME) which reads~\cite{Grifoni1998, Weiss2012}
\be{GME}
\dot{P}(t)=\int_{t_0}^tdt'\left[ {K}^{-}(t,t')-{K}^{+}(t,t')P(t')\right]\;.
\ee
In the presence of the time-dependent bias in Eq.~\eqref{bias_t},  a closed form for the kernels of the GME~\eqref{GME} is obtained within the non-interacting blip approximation (NIBA)~\cite{Leggett1987, Dekker1987}. The NIBA kernels are nonperturbative in the dissipation strength and the drive and read~\cite{Grifoni1998}
\begin{equation}\begin{aligned}\label{K}
{K}^{+}(t,t')=&\; h^{+}(t-t')\cos\left[\zeta_{\rm full}(t,t')\right]\;,\\
{K}^{-}(t,t')=&\; h^{-}(t-t')\sin\left[\zeta_{\rm full}(t,t')\right]\;,
\end{aligned}\end{equation}
with the dynamical phase reading
\begin{equation}\begin{aligned}\label{zeta_full}
\zeta_{\rm full}(t,t')=&\int_{t'}^{t}dt''\;\epsilon(t'')\\
=&\epsilon_0 (t-t')+\frac{\epsilon_{\rm p}}{\omega_{\rm p}}\left[\sin(\omega_{\rm p} t)-\sin\left(\omega_{\rm p} t' \right)\right]+\frac{\epsilon_{\rm d}}{\omega_{\rm d}}\left[\sin(\omega_{\rm d} t)-\sin\left(\omega_{\rm d} t' \right)\right]
\;,
\end{aligned}\end{equation}
and where 
\begin{equation}\begin{aligned}\label{hp_full}
h^{+}(t)=&\; \Delta^2 e^{-Q'(t)}\cos[Q''(t)]\;,\\
h^{-}(t)=&\; \Delta^2 e^{-Q'(t)}\sin[Q''(t)]\;.
\end{aligned}\end{equation}
The baths correlation function $Q(t)$ is the sum of the contributions from the different baths, $Q(t)=\sum_{\nu}Q_{\nu}(t)$~\cite{Nicolin2011}. In our model, $Q(t)=Q_1(t)+Q_2(t)$, where  the contribution from the Ohmic bath with exponential cutoff, acting directly on the qubit ($\nu=1$) is, in the so-called scaling limit~\cite{Weiss2012},
\begin{eqnarray}
Q_1'(t)&=&2\alpha_1\ln\left[\sqrt{1+\omega_{c}^{2} t^{2}}\frac{\sinh[\pi t/(\hbar\beta_1)]}{\pi t/(\hbar\beta_1)}\right]\;,\label{Qsl1-a}\\
Q_1''(t)&=&2\alpha_1\arctan(\omega_{c}t)\;.\label{Qsl1-b}
\end{eqnarray}
Applying Eq.~\eqref{Qdef} to the spectral desity function $G_2(\omega)$, Eq.~\eqref{Geff}, one obtains for the effective bath of the dissipative resonator ($\nu=2$)~\cite{Garg1985,Nesi2007NJP,Magazzu2019} 
\begin{eqnarray}
\label{Q}
Q_2'(t)&=& X t + L \left(e^{-\gamma t/2}\cos{\bar{\Omega}\tau}-1\right) - Z e^{-\gamma t/2}\sin{\bar{\Omega}t}+Q'_{\rm Mats}(t)\;, \\
Q_2''(t)&=& \pi\alpha_2-e^{-\gamma t/2} \pi\alpha_2 \left(\cos{\bar{\Omega}t}+N\sin{\bar{\Omega}t}\right),
\end{eqnarray}
with $X=2\pi \alpha_2 k_B T/\hbar$ and $\bar{\Omega}=\sqrt{\Omega^2-\gamma^2/4}$ and where
\begin{equation}\begin{aligned}\label{}
N&=\frac{\gamma^2/2-\Omega^2}{\gamma\bar\Omega}\;,\\
L&=\pi \alpha_2\frac{N\sinh{(\beta\hbar\bar{\Omega})}+\sin{\left(\beta\hbar\gamma/2\right)}}{\cosh{(\beta\hbar\bar{\Omega})}-\cos{\left(\beta\hbar\gamma/2\right)}}\;,  \\
Z&=\pi \alpha_2\frac{\sinh{(\beta\hbar\bar{\Omega})}-N\sin{\left(\beta\hbar\gamma/2\right)}}{\cosh{(\beta\hbar\bar{\Omega})}-\cos{\left(\beta\hbar\gamma/2\right)}}\;.
\end{aligned}\end{equation}
The term $Q'_{\rm Mats}(t)$ is the following series over the Matsubara frequencies $\nu_n := n\;2\pi k_B T/\hbar$
\begin{equation}\label{Qmats}
Q'_{\rm Mats}(t)=4 \pi\alpha_2 \frac{\Omega^4}{\hbar\beta}\sum_{n=1}^{+\infty} \dfrac{1}{(\Omega^2+\nu_n^2)^2-\gamma^2 \nu_n^2} \left[\dfrac{1-e^{-\nu_n t}}{\nu_n}\right]\;.
\end{equation}

\section{linear susceptibility in the presence of a high-frequency drive}
\label{appendix_chi}

Averaging the kernels in Eq.~\eqref{K} over a drive period $T_{\rm d}=2\pi/\omega_{\rm d}$, we make the substitution $K^\pm(t,t') \rightarrow K_{\rm d}^\pm(t,t')$ in the GME, where
\begin{equation}\begin{aligned}\label{K_hf}
{K}_{\rm d}^{+}(t,t')=&\; h_{\rm d}^{+}(t-t')\cos\left[\zeta(t,t')\right]\;,\\
{K}_{\rm d}^{-}(t,t')=&\; h_{\rm d}^{-}(t-t')\sin\left[\zeta(t,t')\right]\;,
\end{aligned}\end{equation}
%
with the dynamical phase $\zeta(t,t')$ which is now independent of the drive, as it contains exclusively the static bias and the time-dependent probe
\begin{equation}\begin{aligned}
\label{zeta}
\zeta(t,t')=\epsilon_0 (t-t')+\frac{\epsilon_{\rm p}}{\omega_{\rm p}}\left[\sin(\omega_{\rm p} t)-\sin\left(\omega_{\rm p} t' \right)\right]
\;,
\end{aligned}\end{equation}
cf. Eq.~\eqref{zeta_full}. The drive is taken into account effectively by the functions
\be{hpm}
h_{\rm d}^{+}(t)&=\; \Delta^2 e^{-Q'(t)}\cos[Q''(t)]J_0\left[\frac{2\epsilon_{\rm d}}{\omega_{\rm d}}\sin\left(\frac{\omega_{\rm d} t}{2}\right) \right]\;,\\
h_{\rm d}^{-}(t)&=\; \Delta^2 e^{-Q'(t)}\sin[Q''(t)]J_0\left[\frac{2\epsilon_{\rm d}}{\omega_{\rm d}}\sin\left(\frac{\omega_{\rm d} t}{2}\right) \right]\;.
\ee
The Bessel function can be expanded in Fourier series as 
\be{J0_Fourier}
J_0\left[\frac{2\epsilon_{\rm d}}{\omega_{\rm d}}\sin\left(\frac{\omega_{\rm d} t}{2}\right) \right]=\sum_n J_n^2(\epsilon_{\rm d}/\omega_{\rm d})e^{-{\rm i}n\omega_{\rm d}t}\;.
\ee

\indent Due to the effect of the monochromatic probe, we assume the asymptotic population difference $P_{\rm as}(t)$ to be  periodic with the period of the probe. The function  $P_{\rm as}(t)$ is the solution of the GME for $t_0\rightarrow -\infty$, namely 
\be{GME_as}
\dot{P}_{\rm as}(t)=&\int_{-\infty}^tdt'\left[ {K}_{\rm d}^{-}(t,t')-{K}_{\rm d}^{+}(t,t')P_{\rm as}(t')\right]\\
=&\int_{0}^\infty d\tau\;\left[ {K}_{\rm d}^{-}(t,t-\tau)-{K}_{\rm d}^{+}(t,t-\tau)P_{\rm as}(t-\tau)\right]\;.
\ee
The NIBA kernels are periodical in $t$ with the periodicity of the probe, and can be expanded in Fourier series as
\be{K_fourier}
{K}_{\rm d}^{\pm}(t,t-\tau)=\sum_m k_m^\pm(\tau)e^{-{\rm i}\omega_{\rm p}t}\;, 
\ee
where
\be{}
 k_m^\pm(\tau)=\frac{\omega_{\rm p}}{2\pi}\int_{-\mathcal{\pi}/\omega_{\rm p}}^{\mathcal{\pi}/\omega_{\rm p}}dt\; K_{\rm d}^\pm(t,t-\tau)e^{{\rm i}m\omega_{\rm p} t}\;.
\ee
Defining $ c^+(x)=\cos(x)$ and $ c^-(x)=\sin(x)$, from Appendix~\ref{appendix_k_tau}, we can write for $m=0,1$
\be{k_tau}
k^\pm_0(\tau)=& h_{\rm d}^\pm(\tau) c^\pm(\epsilon_0\tau)\;,\\
k^\pm_1(\tau)=&\mp\frac{\epsilon_{\rm p}}{\omega_{\rm p}} e^{{\rm i}\omega_{\rm p}\tau/2}h_{\rm d}^\pm(\tau)\sin(\omega_{\rm p}\tau/2) c^\mp(\epsilon_0\tau)\;,
\ee
which are of order 0 and 1, respectively, in the ratio $\epsilon_{\rm p}/\omega_{\rm p}$.\\
\indent Under the assumption that the memory time of the kernels is finite, $t_{\rm memory}<\infty$, so that when the kernels are different from zero the asymptotic population different is already at the steady state, $P(t)=P_{\rm as}(t)$, we Fourier-expand $P_{\rm as}(t)$ on the LHS and inside the integral in the GME~\eqref{GME_as}. The latter adopts the form
\be{GME_fourier}
\sum_m -{\rm i}m\omega_{\rm p} p_m e^{-{\rm i}m\omega_{\rm p} t}
=
\int_{0}^\infty d\tau\;\sum_m k_m^-(\tau)e^{-{\rm i}m\omega_{\rm p}t}\nonumber-\int_{0}^\infty d\tau\;\sum_{m,n} k_m^+(\tau)e^{-{\rm i}(m+n)\omega_{\rm p}t}  p_n e^{{\rm i}n\omega_{\rm p} \tau}\;.
\ee
Defining
\be{k_lambda}
\hat{k}_{m}^\pm(\lambda)=\int_0^\infty d\tau\;e^{-\lambda \tau} k_m^\pm(\tau)\;,
\ee
Eq.~\eqref{GME_fourier} reads
\be{GME_Fourier_2}
\sum_m -{\rm i}m\omega_{\rm p} p_m e^{-{\rm i}m\omega_{\rm p} t}=\sum_m \hat{k}_m^-(0)e^{-{\rm i}m\omega_{\rm p}t}-\sum_{m,n} \hat{k}_m^+(-{\rm i} n \omega_{\rm p})p_ne^{-{\rm i}(m+n)\omega_{\rm p}t}
\ee
Taking the component  at frequency $\omega_{\rm p}$ 
\be{GME_omega_p}
-{\rm i}\omega_{\rm p} p_1 =& \;\hat{k}_{+1}^-(0)-\sum_{m+n=1} \hat{k}_{m}^+(-{\rm i} n \omega_{\rm p})p_n\\
\ee
%
%
%
%
Since $\hat{k}_m(\lambda)\propto (\epsilon_{\rm p}/\omega_{\rm p})^{|m|}$, see 
Eq.~\eqref{k_tau}, we have
\begin{eqnarray}\label{}
p_1(\omega_{\rm p})&=&\frac{1}{-{\rm i}\omega_{\rm p}+\hat{k}_{0}^+(-{\rm i}\omega_{\rm p})}\left[  \hat{k}_{+1}^-(0)-\hat{k}_{+1}^+(0)p_0\right]+\mathcal{O}[(\epsilon_{\rm p}/\omega_{\rm p})^2]\;.
\end{eqnarray}
Taking the zero-frequency component of Eq.~\eqref{GME_fourier} 
$$
0 = \hat{k}_{0}^-(0)- \hat{k}_{0}^+(0)p_0+\mathcal{O}(\epsilon_{\rm p}/\omega_{\rm p})\;.
$$
Thus, to the lowest order in the probe amplitude~\cite{Grifoni1995,Grifoni1998}
\begin{eqnarray}\label{p1}
p_1(\omega_{\rm p})&=&\frac{1}{-{\rm i}\omega_{\rm p}+\hat{k}_{0}^+(-{\rm i}\omega_{\rm p})}\left[  \hat{k}_{+1}^-(0)-\hat{k}_{+1}^+(0)\frac{ \hat{k}_{0}^-(0)}{\hat{k}_{0}^+(0)}\right]\;,
\end{eqnarray}
where, from Eqs.~\eqref{k_tau} and~\eqref{k_lambda}, 
\be{}
\hat{k}^\pm_0(\lambda)=&\int_0^\infty d\tau\; e^{-\lambda\tau}h_{\rm d}^\pm(\tau) c^\pm(\epsilon_0\tau)\\
\hat{k}^\pm_1(0)=&\mp \frac{\epsilon_{\rm p}}{\omega_{\rm p}}\int_0^\infty d\tau\;e^{{\rm i}\omega_{\rm p}\tau/2}h_{\rm d}^\pm(\tau)\sin(\omega_{\rm p}\tau/2) c^\mp(\epsilon_0\tau)\;.
\ee
The corresponding expression for the linear susceptibility $\chi(\omega_{\rm p})=p_1(\omega_{\rm p})/\hbar\epsilon_{\rm p}$ is given in Eq.~\eqref{chi}. This expression is the one used throughout the present work. 
\subsection*{Markovian limit}
For completeness we also give the susceptibility in the Markovian limit, namely when the decay time of the kernels is much shorter than the relevant time scales of variation of $P(t)$. In this case
\be{GME_Markov}
\dot{P}_{\rm as}(t)
=&\int_{0}^\infty d\tau\; {K}_{\rm d}^{-}(t,-\tau)-\int_{0}^\infty d\tau\;{K}_{\rm d}^{+}(t,t-\tau)P_{\rm as}(t)\;.
\ee
By expanding the kernels in Fourier series
we obtain the equation
\be{GME_Markov_Fourier}
\sum_m -{\rm i}m\omega_{\rm p} p_m e^{-{\rm i}m\omega_{\rm p} t}=\sum_m \hat{k}_m^-(0)e^{-{\rm i}m\omega_{\rm p}t}-\sum_{m,n} \hat{k}_m^+(0)p_ne^{-{\rm i}(m+n)\omega_{\rm p}t}\;
\ee
whose component at frequency $\omega_{\rm p}$ is now
\be{}
-{\rm i}\omega_{\rm p} p_1 =&\; \hat{k}_{+1}^-(0)
-\sum_{m+n=1} \hat{k}_m^+(0)p_ne^{-{\rm i}(m+n)\omega_{\rm p}t}\\
\simeq &\; \hat{k}_{+1}^-(0)- \hat{k}_{+1}^+(0)p_0-\hat{k}_{0}^+(0)p_{+1}-\hat{k}_{+2}^+(0)p_{-1}\;.
\ee
As a result, in the Markovian limit and  to the lowest order in $\epsilon_{\rm p}/\omega_{\rm p}$, we have
\be{p1_Markov}
p_1(\omega_{\rm p})&=\;\frac{1}{-{\rm i}\omega_{\rm p}+\hat{k}_{0}^+(0)}\left[  \hat{k}_{+1}^-(0)-\hat{k}_{+1}^+(0)\frac{ \hat{k}_{0}^-(0)}{\hat{k}_{0}^+(0)}\right]\;.
\ee

\section{Calculation of $k^\pm_m(\tau)$}
\label{appendix_k_tau}

We express the dynamical phase in Eq.~\eqref{zeta} as
\be{}
\zeta(t,t-\tau)=&\epsilon_0 \tau+\frac{\epsilon_{\rm p}}{\omega_{\rm p}}\left\{\sin(\omega_{\rm p} t)-\sin\left[\omega_{\rm p} (t-\tau) \right]\right\}\\
=&\epsilon_0 \tau+2\frac{\epsilon_{\rm p}}{\omega_{\rm p}}\cos(\omega_{\rm p}t-\phi_\tau)\sin(\phi_\tau)\\
=&\epsilon_0 \tau+E(t,\tau)
\;,
\ee
where $\phi_\tau:=\omega_{\rm p}\tau/2$. Using the notation $ c^+(x):=\cos(x)$ and $c^-(x):=\sin(x)$, the  functions of the dynamical phase in Eq.~\eqref{K_hf} read
\be{}
 c^\pm[\zeta(t,t-\tau)]= c^\pm(\epsilon_0\tau) c^+[E(t,\tau)]\mp c^\mp(\epsilon_0\tau) c^-[E(t,\tau)]\;.
\ee
As a result,
\be{k_pm_tau_A}
 k_m^\pm(\tau)=&\;\frac{\omega_{\rm p}}{2\pi}\int_{-\mathcal{\pi}/\omega_{\rm p}}^{\mathcal{\pi}/\omega_{\rm p}}dt\; K_{\rm d}^\pm(t,t-\tau)e^{{\rm i}m\omega_{\rm p} t}\\
 =& \;h_{\rm d}^\pm(\tau)\;\frac{\omega_{\rm p}}{2\pi}\int_{-\mathcal{\pi}/\omega_{\rm p}}^{\mathcal{\pi}/\omega_{\rm p}}dt\; c^\pm[\zeta(t,t-\tau)]e^{{\rm i}m\omega_{\rm p} t}\\
 =& \;h_{\rm d}^\pm(\tau)\left[ c^\pm(\epsilon_0\tau)F^+_m(\tau)\mp c^\mp(\epsilon_0\tau)F^-_m(\tau)
 \right]\;,
\ee
where
\be{Fpm}
F^\pm_m(\tau)=&\frac{\omega_{\rm p}}{2\pi}\int_{-\mathcal{\pi}/\omega_{\rm p}}^{\mathcal{\pi}/\omega_{\rm p}}dt\;e^{{\rm i}m\omega_{\rm p}t} c^\pm[E(t,\tau)]\\
=&\;\frac{e^{{\rm i}m\phi_\tau}}{2\pi}\int_{-\pi-\phi_\tau}^{\pi-\phi_{\tau}}dx\; c^\pm\left[\frac{2\epsilon_{\rm p}}{\omega_{\rm p}}\cos(x)\sin(\phi_\tau)\right]e^{{\rm i}mx}\\
=&\;\frac{e^{{\rm i}m\phi_\tau}}{2\pi}\int_{-\pi}^{\pi}dx\; c^\pm\left[z\cos(x)\right]e^{{\rm i}mx}\;,
\ee
with $x:=\omega_{\rm p} t-\phi_\tau$ and $z:=2\epsilon_{\rm p}\sin(\phi_\tau)/\omega_{\rm p}$. In the last line we have shifted the integration domain (of length one period) exploiting the periodicity of the integrand.
Since only the even part of the integrand contributes to the integral in Eq.~\eqref{Fpm}, we can write the latter as~\cite{Gradshteyn1980}
\be{Fpm2}
F^\pm_m(\tau)
=&\;\frac{e^{{\rm i}m\phi_\tau}}{\pi}\int_{0}^{\pi}dx\; c^\pm\left[z\cos(x)\right]{\rm i}^{(1\mp 1)/2} c^\pm(mx)\\
=&\;\frac{e^{{\rm i}m\phi_\tau}}{2\pi}\int_{0}^{\pi}dx\;\left[e^{{\rm i}z\cos(x)}\pm e^{-{\rm i}z\cos(x)}\right] c^\pm(mx)\\
=&\; e^{{\rm i}m\phi_\tau}{\rm i}^{m-(1\mp 1)/2}\frac{J_m(z)\pm J_m(-z)}{2}\;,
\ee
where $J_m$ is the Bessel function of order $m$ which has the parity $J_m(-z)=(-1)^{m}J_m(z)$. As a result
\be{}
F^+_m(\tau)=&\begin{cases}
{\rm i}^{m} e^{{\rm i}m\omega_{\rm p}\tau/2}J_m\left[\frac{2\epsilon_{\rm p}}{\omega_{\rm p}}\sin\left(\frac{\omega_{\rm p}\tau}{2}\right)\right]\;,\qquad m~{\rm even}\\
0\;,\qquad\qquad\qquad\qquad\qquad\qquad\qquad m~{\rm odd}
\end{cases}
\ee
and
\be{}
F^-_m(\tau)=&\begin{cases}
0\;,\qquad\qquad\qquad\qquad\qquad\qquad\qquad\quad m~{\rm even}\\
{\rm i}^{m-1} e^{{\rm i}m\omega_{\rm p}\tau/2}J_m\left[\frac{2\epsilon_{\rm p}}{\omega_{\rm p}}\sin\left(\frac{\omega_{\rm p}\tau}{2}\right)\right]\;,\qquad m~{\rm odd}
\end{cases}
\;.
\ee
To lowest order in $z$ we have
$J_m(z)\simeq \left(z/2\right)^m$.
Then, substituting the above expressions for $F_m^\pm(\tau)$ into Eq.~\eqref{k_pm_tau_A}, we obtain
\be{}
 k_m^\pm(\tau)\simeq h_{\rm d}^\pm(\tau)\begin{cases}
{\rm i}^{m} e^{{\rm i}m\omega_{\rm p}\tau/2}\left[\frac{\epsilon_{\rm p}}{\omega_{\rm p}}\sin\left(\frac{\omega_{\rm p}\tau}{2}\right)\right]^m c^\pm(\epsilon_0\tau)\;,\qquad m~{\rm even}\\
\mp{\rm i}^{m-1} e^{{\rm i}m\omega_{\rm p}\tau/2}\left[\frac{\epsilon_{\rm p}}{\omega_{\rm p}}\sin\left(\frac{\omega_{\rm p}\tau}{2}\right)\right]^m c^\mp(\epsilon_0\tau)\;,\qquad m~{\rm odd}
\end{cases}\;,
\ee
to lowest order in $\epsilon_{\rm p}/\omega_{\rm p}$.

\section{Comparison with the  probe on the resonator}
\label{probe_resonator}
\begin{figure}[ht!]
\begin{center}
\includegraphics[width=1\textwidth,angle=0]{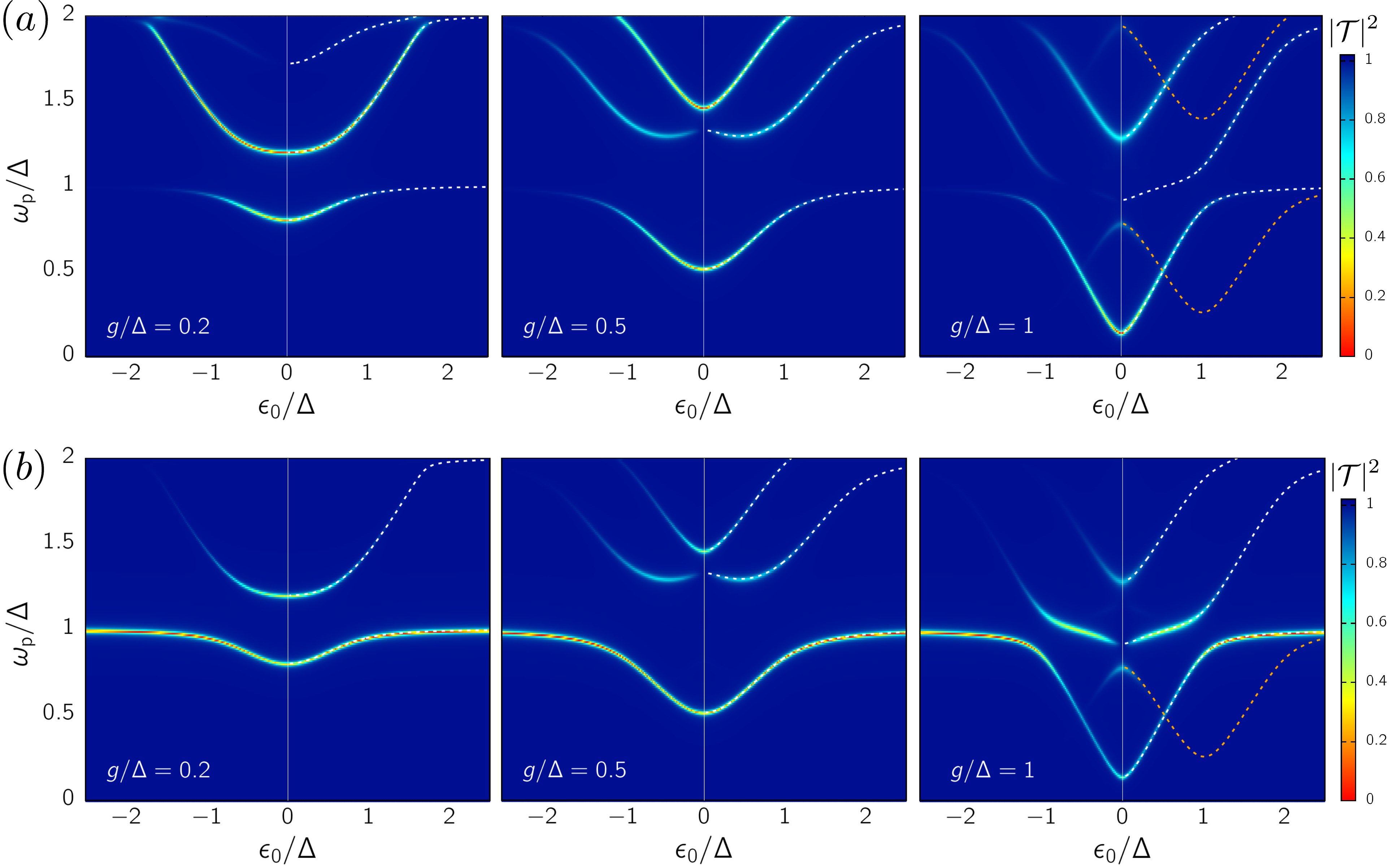}
\caption{Transmission spectra of the static system in the weak dissipation regime ($\alpha_1=\kappa=0.005$) at resonance, $\Omega=\Delta$, for different values of the qubit-resonator coupling $g$. (a) - Probe on the qubit. (b) - Probe on the resonator. For both panels, the transmission is evaluated using Eq.~\eqref{transmission} and the susceptibility is calculated using the approach of~\cite{Kohler2018} with dephasing rates within a Bloch-Redfield master equation approach. The white (orange) dashed lines mark the numerically evaluated transition energies between the ground (first excited) and the higher excited states.
The remaining parameters are the same as in Fig.~\ref{fig_T_static}(a).}
\label{fig_T_static_Om1}
\end{center}
\end{figure}

\indent In contrast to the situation addressed in the present work, where the qubit is probed, standard spectroscopy on USC systems~\cite{Yoshihara2017} is performed by probing the resonator, namely the operator $\hat{X}$ that couples to the transmission line is proportional to resonator position operator, $\hat{X}\propto B^\dag+B$. In order to highlight the differences in the spectra, in Fig.~\ref{fig_T_static_Om1} we compare the transmission for the two probe settings in the weak dissipation regime. As done in Fig.~\ref{fig_T_static}(a), the qubit susceptibility is calculated according to the approach in~\cite{Kohler2018} with dephasing rates from the Bloch-Redfield master equation in the dressed basis of the USC system. The spectra are shown for the same set of values of the qubit-resonator coupling as in Fig.~\ref{fig_T_static}, but this time  at resonance, $\Omega=\Delta$. In both probe settings, the spectra  display resonances corresponding to the transition energies of the Rabi model, with a linewidth given by the decoherence rates, and no environment-induced renormalization of the resonances' positions, which is neglected by the theory and assumed to be small. 
The results for the probe on the qubit, Fig.~\ref{fig_T_static_Om1}(a), display complementary features with respect to the ones obtained with the probe on the resonator, Fig.~\ref{fig_T_static_Om1}(b). In the first setting, similarly to Fig.~\ref{fig_T_static}, resonances are suppressed outside the region $\omega_{\rm p}>|\epsilon_0|$. On the contrary, when the probe is on the resonator the horizontal features, which are insensitive to changes in the static bias, are the most pronounced.
We also note that the qubit spectrum with the largest qubit-resonator coupling, $g=\Delta$ in Fig.~\ref{fig_T_static_Om1}(a), resembles the ones at strong dissipation in Fig.~\ref{fig_T_static}(b). The reason is the strong renormalization effect on the bare qubit splitting $\Delta$, in this case exerted by the resonator at resonance, which depends on the ratio $g/\Omega$, see Eq.~\eqref{Dkj}.

\section{Intermediate dissipation regime}
\label{intermediate_regime}
\begin{figure}[ht!]
\begin{center}
\includegraphics[width=1\textwidth,angle=0]{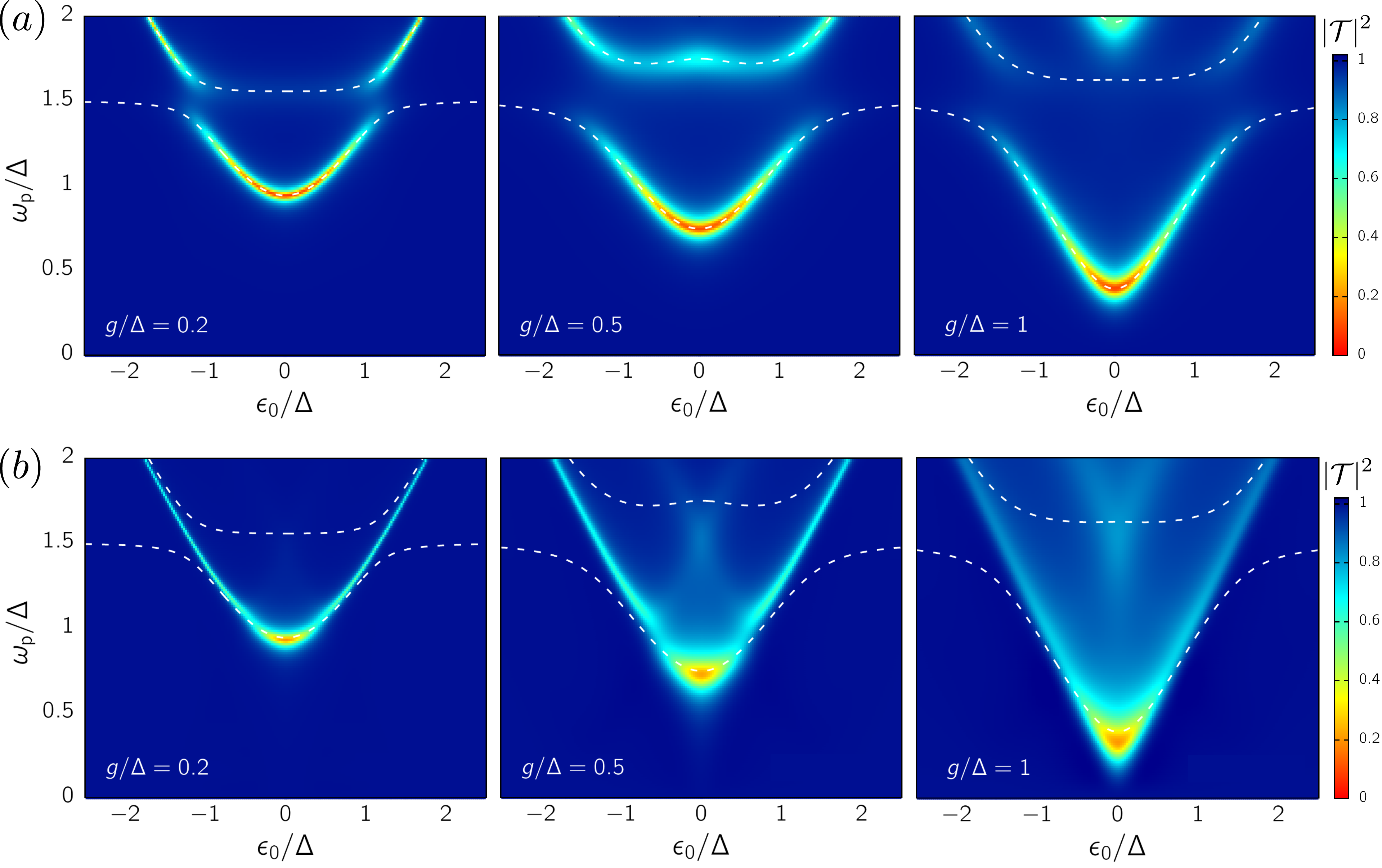}
\caption{Transmission spectra of the static system with probe on the qubit in the intermediate dissipation regime, $\alpha_1=0.01$ and $\kappa=0.05$, for different values of the qubit-resonator coupling $g$. For both panels, the transmission is evaluated using Eq.~\eqref{transmission} and the susceptibility is calculated using  (a) the approach of~\cite{Kohler2018} with dephasing rates within a Bloch-Redfield master equation approach and (b) the path integral approach within NIBA. In both panels, the white dashed lines show, for reference, the numerically evaluated transition energies between the ground and higher excited states of the closed Rabi model.
The remaining parameters are the same as in Fig.~\ref{fig_T_static}.}
\label{fig_intermediate}
\end{center}
\end{figure}
In Fig.~\ref{fig_intermediate}, we compare the results from the weak dissipation approach used in Figs.~\ref{fig_T_static}(a) and~\ref{fig_T_static_Om1} with the ones from the path integral approach within NIBA. In both panels (a) and (b) of Fig.~\ref{fig_intermediate} we show, for reference, the numerically evaluated transition energies of the closed Rabi model. Based on the validity  of the NIBA at zero static bias we can make two observations. First, both approaches reproduce the main resonance involving the ground/first excited state, as given by the numerical diagonalization  of the closed Rabi Hamiltonian, meaning that the bath-induced renormalization of the quit frequency is not an important effect in the intermediate regime considered here, $\alpha=0.01$ and $\kappa=0.05$. Second, the weak coupling approach fails to capture the renormalization of the resonances involving excited states higher than the first, see panel (b) of Fig.~\ref{fig_intermediate}. On the other hand,  one would expect to still be able to observe, even if reduced to some degree, the avoided crossings present in Fig.~\ref{fig_intermediate}(a) which are completely suppressed in the NIBA results of panel (b).\\
\indent In the intermediate dissipation regime  considered here, signatures of the bath-induced renormalization of the system's bare frequencies and the transition to the strong dissipation regime analyzed in Sec.~\ref{section_results} are expected to be revealed in the spectra. We note however, that in this intermediate regime the use of  accurate numerical treatments to check the validity of both approximation schemes might be appropriate. These are, for example, the quasi-adiabatic path integral approach~\cite{Thorwart2004} applied to the spin-boson model with the effective bath mapping used here for the NIBA, or the matrix-product state ansatz based on a chain mapping of the full Hamiltonian~\cite{Zueco2019}.


\begin{thebibliography}{75}%
\makeatletter
\providecommand \@ifxundefined [1]{%
 \@ifx{#1\undefined}
}%
\providecommand \@ifnum [1]{%
 \ifnum #1\expandafter \@firstoftwo
 \else \expandafter \@secondoftwo
 \fi
}%
\providecommand \@ifx [1]{%
 \ifx #1\expandafter \@firstoftwo
 \else \expandafter \@secondoftwo
 \fi
}%
\providecommand \natexlab [1]{#1}%
\providecommand \enquote  [1]{``#1''}%
\providecommand \bibnamefont  [1]{#1}%
\providecommand \bibfnamefont [1]{#1}%
\providecommand \citenamefont [1]{#1}%
\providecommand \href@noop [0]{\@secondoftwo}%
\providecommand \href [0]{\begingroup \@sanitize@url \@href}%
\providecommand \@href[1]{\@@startlink{#1}\@@href}%
\providecommand \@@href[1]{\endgroup#1\@@endlink}%
\providecommand \@sanitize@url [0]{\catcode `\\12\catcode `\$12\catcode
  `\&12\catcode `\#12\catcode `\^12\catcode `\_12\catcode `\%12\relax}%
\providecommand \@@startlink[1]{}%
\providecommand \@@endlink[0]{}%
\providecommand \url  [0]{\begingroup\@sanitize@url \@url }%
\providecommand \@url [1]{\endgroup\@href {#1}{\urlprefix }}%
\providecommand \urlprefix  [0]{URL }%
\providecommand \Eprint [0]{\href }%
\providecommand \doibase [0]{http://dx.doi.org/}%
\providecommand \selectlanguage [0]{\@gobble}%
\providecommand \bibinfo  [0]{\@secondoftwo}%
\providecommand \bibfield  [0]{\@secondoftwo}%
\providecommand \translation [1]{[#1]}%
\providecommand \BibitemOpen [0]{}%
\providecommand \bibitemStop [0]{}%
\providecommand \bibitemNoStop [0]{.\EOS\space}%
\providecommand \EOS [0]{\spacefactor3000\relax}%
\providecommand \BibitemShut  [1]{\csname bibitem#1\endcsname}%
\let\auto@bib@innerbib\@empty
\bibitem [{\citenamefont {You}\ and\ \citenamefont {Nori}(2011)}]{You2011}%
  \BibitemOpen
  \bibfield  {author} {\bibinfo {author} {\bibfnamefont {J.~Q.}\ \bibnamefont
  {You}}\ and\ \bibinfo {author} {\bibfnamefont {F.}~\bibnamefont {Nori}},\
  }\bibfield  {title} {\enquote {\bibinfo {title} {{Atomic physics and quantum
  optics using superconducting circuits}},}\ }\href {\doibase
  10.1038/nature10122} {\bibfield  {journal} {\bibinfo  {journal} {Nature}\
  }\textbf {\bibinfo {volume} {474}},\ \bibinfo {pages} {589--597} (\bibinfo
  {year} {2011})}\BibitemShut {NoStop}%
\bibitem [{\citenamefont {Houck}\ \emph {et~al.}(2012)\citenamefont {Houck},
  \citenamefont {T{\"u}reci},\ and\ \citenamefont {Koch}}]{Koch2012}%
  \BibitemOpen
  \bibfield  {author} {\bibinfo {author} {\bibfnamefont {A.~A.}\ \bibnamefont
  {Houck}}, \bibinfo {author} {\bibfnamefont {H.~E.}\ \bibnamefont
  {T{\"u}reci}}, \ and\ \bibinfo {author} {\bibfnamefont {J.}~\bibnamefont
  {Koch}},\ }\bibfield  {title} {\enquote {\bibinfo {title} {{On-chip quantum
  simulation with superconducting circuits}},}\ }\href {\doibase
  10.1038/nphys2251} {\bibfield  {journal} {\bibinfo  {journal} {Nat. Phys.}\
  }\textbf {\bibinfo {volume} {8}},\ \bibinfo {pages} {292} (\bibinfo {year}
  {2012})}\BibitemShut {NoStop}%
\bibitem [{\citenamefont {Pekola}(2015)}]{Pekola2015}%
  \BibitemOpen
  \bibfield  {author} {\bibinfo {author} {\bibfnamefont {J.~P.}\ \bibnamefont
  {Pekola}},\ }\bibfield  {title} {\enquote {\bibinfo {title} {{Towards quantum
  thermodynamics in electronic circuits}},}\ }\href {\doibase
  10.1038/nphys3169} {\bibfield  {journal} {\bibinfo  {journal} {Nat. Phys.}\
  }\textbf {\bibinfo {volume} {11}},\ \bibinfo {pages} {118--123} (\bibinfo
  {year} {2015})}\BibitemShut {NoStop}%
\bibitem [{\citenamefont {Wendin}(2017)}]{Wendin2017}%
  \BibitemOpen
  \bibfield  {author} {\bibinfo {author} {\bibfnamefont {G.}~\bibnamefont
  {Wendin}},\ }\bibfield  {title} {\enquote {\bibinfo {title} {{Quantum
  information processing with superconducting circuits: a review}},}\ }\href
  {\doibase 10.1088/1361-6633/aa7e1a} {\bibfield  {journal} {\bibinfo
  {journal} {Rep. Prog. Phys.}\ }\textbf {\bibinfo {volume} {80}},\ \bibinfo
  {pages} {106001} (\bibinfo {year} {2017})}\BibitemShut {NoStop}%
\bibitem [{\citenamefont {Gu}\ \emph {et~al.}(2017)\citenamefont {Gu},
  \citenamefont {Kockum}, \citenamefont {Miranowicz}, \citenamefont {Liu},\
  and\ \citenamefont {Nori}}]{Nori2017}%
  \BibitemOpen
  \bibfield  {author} {\bibinfo {author} {\bibfnamefont {X.}~\bibnamefont
  {Gu}}, \bibinfo {author} {\bibfnamefont {A.~F.}\ \bibnamefont {Kockum}},
  \bibinfo {author} {\bibfnamefont {A.}~\bibnamefont {Miranowicz}}, \bibinfo
  {author} {\bibfnamefont {Y.-X.}\ \bibnamefont {Liu}}, \ and\ \bibinfo
  {author} {\bibfnamefont {F.}~\bibnamefont {Nori}},\ }\bibfield  {title}
  {\enquote {\bibinfo {title} {{Microwave photonics with superconducting
  quantum circuits}},}\ }\href {\doibase 10.1016/j.physrep.2017.10.002}
  {\bibfield  {journal} {\bibinfo  {journal} {Phys. Rep.}\ }\textbf {\bibinfo
  {volume} {718-719}},\ \bibinfo {pages} {1--102} (\bibinfo {year}
  {2017})}\BibitemShut {NoStop}%
\bibitem [{\citenamefont {Ciuti}\ \emph {et~al.}(2005)\citenamefont {Ciuti},
  \citenamefont {Bastard},\ and\ \citenamefont {Carusotto}}]{Ciuti2005}%
  \BibitemOpen
  \bibfield  {author} {\bibinfo {author} {\bibfnamefont {C.}~\bibnamefont
  {Ciuti}}, \bibinfo {author} {\bibfnamefont {G.}~\bibnamefont {Bastard}}, \
  and\ \bibinfo {author} {\bibfnamefont {I.}~\bibnamefont {Carusotto}},\
  }\bibfield  {title} {\enquote {\bibinfo {title} {{Quantum vacuum properties
  of the intersubband cavity polariton field}},}\ }\href {\doibase
  10.1103/PhysRevB.72.115303} {\bibfield  {journal} {\bibinfo  {journal} {Phys.
  Rev. B}\ }\textbf {\bibinfo {volume} {72}},\ \bibinfo {pages} {115303}
  (\bibinfo {year} {2005})}\BibitemShut {NoStop}%
\bibitem [{\citenamefont {J.}\ \emph {et~al.}(2009)\citenamefont {J.},
  \citenamefont {M.}, \citenamefont {A.}, \citenamefont {O.}, \citenamefont
  {Y.},\ and\ \citenamefont {A.}}]{Bourassa2009}%
  \BibitemOpen
  \bibfield  {author} {\bibinfo {author} {\bibfnamefont {J. Bourassa}}, \bibinfo {author} {\bibfnamefont {J. M. Gambetta}}, \bibinfo {author} {\bibfnamefont {A. A. Abdumalikov}}, \bibinfo {author} {\bibfnamefont {O. Astafiev}}, \bibinfo {author} {\bibfnamefont {Y. Nakamura}}, \
  and\ \bibinfo {author} {\bibfnamefont {Blais}\ \bibnamefont {A.}},\
  }\bibfield  {title} {\enquote {\bibinfo {title} {{Ultrastrong coupling regime
  of cavity QED with phase-biased flux qubits}},}\ }\href {\doibase
  10.1103/PhysRevA.80.032109} {\bibfield  {journal} {\bibinfo  {journal} {Phys.
  Rev. A}\ }\textbf {\bibinfo {volume} {80}},\ \bibinfo {pages} {032109}
  (\bibinfo {year} {2009})}\BibitemShut {NoStop}%
\bibitem [{\citenamefont {Hausinger}\ and\ \citenamefont
  {Grifoni}(2008)}]{Hausinger2008}%
  \BibitemOpen
  \bibfield  {author} {\bibinfo {author} {\bibfnamefont {J.}~\bibnamefont
  {Hausinger}}\ and\ \bibinfo {author} {\bibfnamefont {M.}~\bibnamefont
  {Grifoni}},\ }\bibfield  {title} {\enquote {\bibinfo {title} {{Dissipative
  dynamics of a biased qubit coupled to a harmonic oscillator: analytical
  results beyond the rotating wave approximation}},}\ }\href {\doibase
  10.1088/1367-2630/10/11/115015} {\bibfield  {journal} {\bibinfo  {journal}
  {New J. Phys.}\ }\textbf {\bibinfo {volume} {10}},\ \bibinfo {pages} {115015}
  (\bibinfo {year} {2008})}\BibitemShut {NoStop}%
\bibitem [{\citenamefont {{\emph{et al.}}}(2010)}]{Niemczyk2010}%
  \BibitemOpen
  \bibfield  {author} {\bibinfo {author} {\bibfnamefont {{T. Niemczyk}}\
  \bibnamefont {{\emph{et al.}}}},\ }\bibfield  {title} {\enquote {\bibinfo
  {title} {{Circuit quantum electrodynamics in the ultrastrong-coupling
  regime}},}\ }\href {\doibase 10.1038/nphys1730} {\bibfield  {journal}
  {\bibinfo  {journal} {Nat. Phys.}\ }\textbf {\bibinfo {volume} {6}},\
  \bibinfo {pages} {772--776} (\bibinfo {year} {2010})}\BibitemShut {NoStop}%
\bibitem [{\citenamefont {Ashhab}\ and\ \citenamefont
  {Nori}(2010)}]{Ashhab2010}%
  \BibitemOpen
  \bibfield  {author} {\bibinfo {author} {\bibfnamefont {S.}~\bibnamefont
  {Ashhab}}\ and\ \bibinfo {author} {\bibfnamefont {F.}~\bibnamefont {Nori}},\
  }\bibfield  {title} {\enquote {\bibinfo {title} {{Qubit-oscillator systems in
  the ultrastrong-coupling regime and their potential for preparing
  nonclassical states}},}\ }\href {\doibase 10.1103/PhysRevA.81.042311}
  {\bibfield  {journal} {\bibinfo  {journal} {Phys. Rev. A}\ }\textbf {\bibinfo
  {volume} {81}},\ \bibinfo {pages} {042311} (\bibinfo {year}
  {2010})}\BibitemShut {NoStop}%
\bibitem [{\citenamefont {Hausinger}\ and\ \citenamefont
  {Grifoni}(2010)}]{Hausinger2010PRA}%
  \BibitemOpen
  \bibfield  {author} {\bibinfo {author} {\bibfnamefont {J.}~\bibnamefont
  {Hausinger}}\ and\ \bibinfo {author} {\bibfnamefont {M.}~\bibnamefont
  {Grifoni}},\ }\bibfield  {title} {\enquote {\bibinfo {title}
  {{Qubit-oscillator system: An analytical treatment of the ultrastrong
  coupling regime}},}\ }\href {\doibase 10.1103/PhysRevA.82.062320} {\bibfield
  {journal} {\bibinfo  {journal} {Phys. Rev. A}\ }\textbf {\bibinfo {volume}
  {82}},\ \bibinfo {pages} {062320} (\bibinfo {year} {2010})}\BibitemShut
  {NoStop}%
\bibitem [{\citenamefont {Forn-D{\'i}az}\ \emph {et~al.}(2010)\citenamefont
  {Forn-D{\'i}az}, \citenamefont {Lisenfeld}, \citenamefont {Marcos},
  \citenamefont {Garc{\'i}a-Ripoll}, \citenamefont {Solano}, \citenamefont
  {Harmans},\ and\ \citenamefont {Mooij}}]{Forn-Diaz2010}%
  \BibitemOpen
  \bibfield  {author} {\bibinfo {author} {\bibfnamefont {P.}~\bibnamefont
  {Forn-D{\'i}az}}, \bibinfo {author} {\bibfnamefont {J.}~\bibnamefont
  {Lisenfeld}}, \bibinfo {author} {\bibfnamefont {D.}~\bibnamefont {Marcos}},
  \bibinfo {author} {\bibfnamefont {J.~J.}\ \bibnamefont {Garc{\'i}a-Ripoll}},
  \bibinfo {author} {\bibfnamefont {E.}~\bibnamefont {Solano}}, \bibinfo
  {author} {\bibfnamefont {C.~J. P.~M.}\ \bibnamefont {Harmans}}, \ and\
  \bibinfo {author} {\bibfnamefont {J.~E.}\ \bibnamefont {Mooij}},\ }\bibfield
  {title} {\enquote {\bibinfo {title} {{Observation of the Bloch-Siegert Shift
  in a Qubit-Oscillator System in the Ultrastrong Coupling Regime}},}\ }\href
  {\doibase 10.1103/PhysRevLett.105.237001} {\bibfield  {journal} {\bibinfo
  {journal} {Phys. Rev. Lett.}\ }\textbf {\bibinfo {volume} {105}},\ \bibinfo
  {pages} {237001} (\bibinfo {year} {2010})}\BibitemShut {NoStop}%
\bibitem [{\citenamefont {Yoshihara}\ \emph {et~al.}(2016)\citenamefont
  {Yoshihara}, \citenamefont {Fuse}, \citenamefont {Ashhab}, \citenamefont
  {Kakuyanagi}, \citenamefont {Saito},\ and\ \citenamefont
  {Semba}}]{Yoshihara2017}%
  \BibitemOpen
  \bibfield  {author} {\bibinfo {author} {\bibfnamefont {F.}~\bibnamefont
  {Yoshihara}}, \bibinfo {author} {\bibfnamefont {T.}~\bibnamefont {Fuse}},
  \bibinfo {author} {\bibfnamefont {S.}~\bibnamefont {Ashhab}}, \bibinfo
  {author} {\bibfnamefont {K.}~\bibnamefont {Kakuyanagi}}, \bibinfo {author}
  {\bibfnamefont {S.}~\bibnamefont {Saito}}, \ and\ \bibinfo {author}
  {\bibfnamefont {K.}~\bibnamefont {Semba}},\ }\bibfield  {title} {\enquote
  {\bibinfo {title} {{Superconducting qubit-oscillator circuit beyond the
  ultrastrong-coupling regime}},}\ }\href {\doibase 10.1038/nphys3906}
  {\bibfield  {journal} {\bibinfo  {journal} {Nat. Phys.}\ }\textbf {\bibinfo
  {volume} {13}},\ \bibinfo {pages} {44} (\bibinfo {year} {2017})}\BibitemShut
  {NoStop}%
\bibitem [{\citenamefont {Yoshihara}\ \emph {et~al.}(2017)\citenamefont
  {Yoshihara}, \citenamefont {Fuse}, \citenamefont {Ashhab}, \citenamefont
  {Kakuyanagi}, \citenamefont {Saito},\ and\ \citenamefont
  {Semba}}]{Yoshihara2017PRA}%
  \BibitemOpen
  \bibfield  {author} {\bibinfo {author} {\bibfnamefont {F.}~\bibnamefont
  {Yoshihara}}, \bibinfo {author} {\bibfnamefont {T.}~\bibnamefont {Fuse}},
  \bibinfo {author} {\bibfnamefont {S.}~\bibnamefont {Ashhab}}, \bibinfo
  {author} {\bibfnamefont {K.}~\bibnamefont {Kakuyanagi}}, \bibinfo {author}
  {\bibfnamefont {S.}~\bibnamefont {Saito}}, \ and\ \bibinfo {author}
  {\bibfnamefont {K.}~\bibnamefont {Semba}},\ }\bibfield  {title} {\enquote
  {\bibinfo {title} {{Characteristic spectra of circuit quantum electrodynamics
  systems from the ultrastrong- to the deep-strong-coupling regime}},}\ }\href
  {\doibase 10.1103/PhysRevA.95.053824} {\bibfield  {journal} {\bibinfo
  {journal} {Phys. Rev. A}\ }\textbf {\bibinfo {volume} {95}},\ \bibinfo
  {pages} {053824} (\bibinfo {year} {2017})}\BibitemShut {NoStop}%
\bibitem [{\citenamefont {Forn-D{\'i}az}\ \emph {et~al.}(2019)\citenamefont
  {Forn-D{\'i}az}, \citenamefont {Lamata}, \citenamefont {Rico}, \citenamefont
  {Kono},\ and\ \citenamefont {Solano}}]{Forn-Diaz2018review}%
  \BibitemOpen
  \bibfield  {author} {\bibinfo {author} {\bibfnamefont {P.}~\bibnamefont
  {Forn-D{\'i}az}}, \bibinfo {author} {\bibfnamefont {L.}~\bibnamefont
  {Lamata}}, \bibinfo {author} {\bibfnamefont {E.}~\bibnamefont {Rico}},
  \bibinfo {author} {\bibfnamefont {J.}~\bibnamefont {Kono}}, \ and\ \bibinfo
  {author} {\bibfnamefont {E.}~\bibnamefont {Solano}},\ }\bibfield  {title}
  {\enquote {\bibinfo {title} {{Ultrastrong coupling regimes of light-matter
  interaction}},}\ }\href {\doibase 10.1103/RevModPhys.91.025005} {\bibfield
  {journal} {\bibinfo  {journal} {Rev. Mod. Phys.}\ }\textbf {\bibinfo {volume}
  {91}},\ \bibinfo {pages} {025005} (\bibinfo {year} {2019})}\BibitemShut
  {NoStop}%
\bibitem [{\citenamefont {Kockum}\ \emph {et~al.}(2019)\citenamefont {Kockum},
  \citenamefont {Miranowicz}, \citenamefont {{De Liberato}}, \citenamefont
  {Savasta},\ and\ \citenamefont {Nori}}]{Kockum2019}%
  \BibitemOpen
  \bibfield  {author} {\bibinfo {author} {\bibfnamefont {F.~A.}\ \bibnamefont
  {Kockum}}, \bibinfo {author} {\bibfnamefont {A.}~\bibnamefont {Miranowicz}},
  \bibinfo {author} {\bibfnamefont {S.}~\bibnamefont {{De Liberato}}}, \bibinfo
  {author} {\bibfnamefont {S.}~\bibnamefont {Savasta}}, \ and\ \bibinfo
  {author} {\bibfnamefont {F.}~\bibnamefont {Nori}},\ }\bibfield  {title}
  {\enquote {\bibinfo {title} {{Ultrastrong coupling between light and
  matter}},}\ }\href {\doibase 10.1038/s42254-018-0006-2} {\bibfield  {journal}
  {\bibinfo  {journal} {Nat. Rev. Phys.}\ }\textbf {\bibinfo {volume} {1}},\
  \bibinfo {pages} {19--40} (\bibinfo {year} {2019})}\BibitemShut {NoStop}%
\bibitem [{\citenamefont {Romero}\ \emph {et~al.}(2012)\citenamefont {Romero},
  \citenamefont {Ballester}, \citenamefont {Wang}, \citenamefont {Scarani},\
  and\ \citenamefont {Solano}}]{Romero2012}%
  \BibitemOpen
  \bibfield  {author} {\bibinfo {author} {\bibfnamefont {G.}~\bibnamefont
  {Romero}}, \bibinfo {author} {\bibfnamefont {D.}~\bibnamefont {Ballester}},
  \bibinfo {author} {\bibfnamefont {Y.~M.}\ \bibnamefont {Wang}}, \bibinfo
  {author} {\bibfnamefont {V.}~\bibnamefont {Scarani}}, \ and\ \bibinfo
  {author} {\bibfnamefont {E.}~\bibnamefont {Solano}},\ }\bibfield  {title}
  {\enquote {\bibinfo {title} {{Ultrafast Quantum Gates in Circuit QED}},}\
  }\href {\doibase 10.1103/PhysRevLett.108.120501} {\bibfield  {journal}
  {\bibinfo  {journal} {Phys. Rev. Lett.}\ }\textbf {\bibinfo {volume} {108}},\
  \bibinfo {pages} {120501} (\bibinfo {year} {2012})}\BibitemShut {NoStop}%
\bibitem [{\citenamefont {Devoret}\ and\ \citenamefont
  {Martinis}(2004)}]{Devoret2004}%
  \BibitemOpen
  \bibfield  {author} {\bibinfo {author} {\bibfnamefont {M.H.}\ \bibnamefont
  {Devoret}}\ and\ \bibinfo {author} {\bibfnamefont {J.M.}\ \bibnamefont
  {Martinis}},\ }\bibfield  {title} {\enquote {\bibinfo {title} {{Implementing
  qubits with superconducting integrated circuits}},}\ }\href@noop {}
  {\bibfield  {journal} {\bibinfo  {journal} {Quantum Inf. Process.}\ }\textbf
  {\bibinfo {volume} {3}},\ \bibinfo {pages} {163--203} (\bibinfo {year}
  {2004})}\BibitemShut {NoStop}%
\bibitem [{\citenamefont {Mooij}\ \emph {et~al.}(1999)\citenamefont {Mooij},
  \citenamefont {Orlando}, \citenamefont {Levitov}, \citenamefont {Tian},
  \citenamefont {van~der Wal},\ and\ \citenamefont {Lloyd}}]{Mooij1999}%
  \BibitemOpen
  \bibfield  {author} {\bibinfo {author} {\bibfnamefont {J.~E.}\ \bibnamefont
  {Mooij}}, \bibinfo {author} {\bibfnamefont {T.~P.}\ \bibnamefont {Orlando}},
  \bibinfo {author} {\bibfnamefont {L.}~\bibnamefont {Levitov}}, \bibinfo
  {author} {\bibfnamefont {L.}~\bibnamefont {Tian}}, \bibinfo {author}
  {\bibfnamefont {C.~H.}\ \bibnamefont {van~der Wal}}, \ and\ \bibinfo {author}
  {\bibfnamefont {S.}~\bibnamefont {Lloyd}},\ }\bibfield  {title} {\enquote
  {\bibinfo {title} {{Josephson Persistent-Current Qubit}},}\ }\href {\doibase
  10.1126/science.285.5430.1036} {\bibfield  {journal} {\bibinfo  {journal}
  {Science}\ }\textbf {\bibinfo {volume} {285}},\ \bibinfo {pages} {1036}
  (\bibinfo {year} {1999})}\BibitemShut {NoStop}%
\bibitem [{\citenamefont {Leggett}\ \emph {et~al.}(1987)\citenamefont
  {Leggett}, \citenamefont {Chakravarty}, \citenamefont {Dorsey}, \citenamefont
  {Fisher}, \citenamefont {Garg},\ and\ \citenamefont {Zwerger}}]{Leggett1987}%
  \BibitemOpen
  \bibfield  {author} {\bibinfo {author} {\bibfnamefont {A.~J.}\ \bibnamefont
  {Leggett}}, \bibinfo {author} {\bibfnamefont {S.}~\bibnamefont
  {Chakravarty}}, \bibinfo {author} {\bibfnamefont {A.~T.}\ \bibnamefont
  {Dorsey}}, \bibinfo {author} {\bibfnamefont {M.~P.~A.}\ \bibnamefont
  {Fisher}}, \bibinfo {author} {\bibfnamefont {A.}~\bibnamefont {Garg}}, \ and\
  \bibinfo {author} {\bibfnamefont {W.}~\bibnamefont {Zwerger}},\ }\bibfield
  {title} {\enquote {\bibinfo {title} {{Dynamics of the dissipative two-state
  system}},}\ }\href {\doibase 10.1103/RevModPhys.59.1} {\bibfield  {journal}
  {\bibinfo  {journal} {Rev. Mod. Phys.}\ }\textbf {\bibinfo {volume} {59}},\
  \bibinfo {pages} {1--85} (\bibinfo {year} {1987})}\BibitemShut {NoStop}%
\bibitem [{\citenamefont {Weiss}(2012, 4th ed.)}]{Weiss2012}%
  \BibitemOpen
  \bibfield  {author} {\bibinfo {author} {\bibfnamefont {U.}~\bibnamefont
  {Weiss}},\ }\href@noop {} {\emph {\bibinfo {title} {{Quantum Dissipative
  Systems}}}}\ (\bibinfo  {publisher} {World Scientific, Singapore},\ \bibinfo
  {year} {2012, 4th ed.})\BibitemShut {NoStop}%
\bibitem [{\citenamefont {Breuer}\ and\ \citenamefont
  {Petruccione}(2002)}]{Breuer2002}%
  \BibitemOpen
  \bibfield  {author} {\bibinfo {author} {\bibfnamefont {H.~P.}\ \bibnamefont
  {Breuer}}\ and\ \bibinfo {author} {\bibfnamefont {F.}~\bibnamefont
  {Petruccione}},\ }\href@noop {} {\emph {\bibinfo {title} {{The theory of open
  quantum systems}}}}\ (\bibinfo  {publisher} {Oxford University Press,
  Oxford},\ \bibinfo {year} {2002})\BibitemShut {NoStop}%
\bibitem [{\citenamefont {Peropadre}\ \emph {et~al.}(2013)\citenamefont
  {Peropadre}, \citenamefont {Zueco}, \citenamefont {Porras},\ and\
  \citenamefont {Garc{\'i}a-Ripoll}}]{Peropadre2013PRL}%
  \BibitemOpen
  \bibfield  {author} {\bibinfo {author} {\bibfnamefont {B.}~\bibnamefont
  {Peropadre}}, \bibinfo {author} {\bibfnamefont {D.}~\bibnamefont {Zueco}},
  \bibinfo {author} {\bibfnamefont {D.}~\bibnamefont {Porras}}, \ and\ \bibinfo
  {author} {\bibfnamefont {J.~J.}\ \bibnamefont {Garc{\'i}a-Ripoll}},\
  }\bibfield  {title} {\enquote {\bibinfo {title} {{Nonequilibrium and
  Nonperturbative Dynamics of Ultrastrong Coupling in Open Lines}},}\ }\href
  {\doibase 10.1103/PhysRevLett.111.243602} {\bibfield  {journal} {\bibinfo
  {journal} {Phys. Rev. Lett.}\ }\textbf {\bibinfo {volume} {111}},\ \bibinfo
  {pages} {243602} (\bibinfo {year} {2013})}\BibitemShut {NoStop}%
\bibitem [{\citenamefont {Forn-D{\'i}az}\ \emph
  {et~al.}(2017{\natexlab{a}})\citenamefont {Forn-D{\'i}az}, \citenamefont
  {Garc{\'i}a-Ripoll}, \citenamefont {Peropadre}, \citenamefont {Orgiazzi},
  \citenamefont {Yurtalan}, \citenamefont {Belyansky}, \citenamefont {Wilson},\
  and\ \citenamefont {Lupascu}}]{Forn-Diaz2017}%
  \BibitemOpen
  \bibfield  {author} {\bibinfo {author} {\bibfnamefont {P.}~\bibnamefont
  {Forn-D{\'i}az}}, \bibinfo {author} {\bibfnamefont {J.~J.}\ \bibnamefont
  {Garc{\'i}a-Ripoll}}, \bibinfo {author} {\bibfnamefont {B.}~\bibnamefont
  {Peropadre}}, \bibinfo {author} {\bibfnamefont {J.~L.}\ \bibnamefont
  {Orgiazzi}}, \bibinfo {author} {\bibfnamefont {M.~A.}\ \bibnamefont
  {Yurtalan}}, \bibinfo {author} {\bibfnamefont {R.}~\bibnamefont {Belyansky}},
  \bibinfo {author} {\bibfnamefont {C.~M.}\ \bibnamefont {Wilson}}, \ and\
  \bibinfo {author} {\bibfnamefont {A.}~\bibnamefont {Lupascu}},\ }\bibfield
  {title} {\enquote {\bibinfo {title} {{Ultrastrong coupling of a single
  artificial atom to an electromagnetic continuum in the nonperturbative
  regime}},}\ }\href {\doibase 10.1038/nphys3905} {\bibfield  {journal}
  {\bibinfo  {journal} {Nat. Phys.}\ }\textbf {\bibinfo {volume} {13}},\
  \bibinfo {pages} {39--43} (\bibinfo {year} {2017}{\natexlab{a}})}\BibitemShut
  {NoStop}%
\bibitem [{\citenamefont {Magazz{\`u}}\ \emph {et~al.}(2018)\citenamefont
  {Magazz{\`u}}, \citenamefont {Forn-D{\'i}az}, \citenamefont {Belyansky},
  \citenamefont {Orgiazzi}, \citenamefont {Yurtalan}, \citenamefont {Otto},
  \citenamefont {Lupascu}, \citenamefont {Wilson},\ and\ \citenamefont
  {Grifoni}}]{Magazzu2018}%
  \BibitemOpen
  \bibfield  {author} {\bibinfo {author} {\bibfnamefont {L.}~\bibnamefont
  {Magazz{\`u}}}, \bibinfo {author} {\bibfnamefont {P.}~\bibnamefont
  {Forn-D{\'i}az}}, \bibinfo {author} {\bibfnamefont {R.}~\bibnamefont
  {Belyansky}}, \bibinfo {author} {\bibfnamefont {J.-L.}\ \bibnamefont
  {Orgiazzi}}, \bibinfo {author} {\bibfnamefont {M.~A.}\ \bibnamefont
  {Yurtalan}}, \bibinfo {author} {\bibfnamefont {M.~R.}\ \bibnamefont {Otto}},
  \bibinfo {author} {\bibfnamefont {A.}~\bibnamefont {Lupascu}}, \bibinfo
  {author} {\bibfnamefont {C.~M.}\ \bibnamefont {Wilson}}, \ and\ \bibinfo
  {author} {\bibfnamefont {M.}~\bibnamefont {Grifoni}},\ }\bibfield  {title}
  {\enquote {\bibinfo {title} {{Probing the strongly driven spin-boson model in
  a superconducting quantum circuit}},}\ }\href {\doibase
  10.1038/s41467-018-03626-w} {\bibfield  {journal} {\bibinfo  {journal} {Nat.
  Commun.}\ }\textbf {\bibinfo {volume} {9}},\ \bibinfo {pages} {1403}
  (\bibinfo {year} {2018})}\BibitemShut {NoStop}%
\bibitem [{\citenamefont {Lepp{\"a}kangas}\ \emph {et~al.}(2018)\citenamefont
  {Lepp{\"a}kangas}, \citenamefont {Braum{\"u}ller}, \citenamefont {Hauck},
  \citenamefont {Reiner}, \citenamefont {Schwenk}, \citenamefont {Zanker},
  \citenamefont {Fritz}, \citenamefont {Ustinov}, \citenamefont {Weides},\ and\
  \citenamefont {Marthaler}}]{Ustinov2018}%
  \BibitemOpen
  \bibfield  {author} {\bibinfo {author} {\bibfnamefont {J.}~\bibnamefont
  {Lepp{\"a}kangas}}, \bibinfo {author} {\bibfnamefont {J.}~\bibnamefont
  {Braum{\"u}ller}}, \bibinfo {author} {\bibfnamefont {M.}~\bibnamefont
  {Hauck}}, \bibinfo {author} {\bibfnamefont {J.-M.}\ \bibnamefont {Reiner}},
  \bibinfo {author} {\bibfnamefont {I.}~\bibnamefont {Schwenk}}, \bibinfo
  {author} {\bibfnamefont {S.}~\bibnamefont {Zanker}}, \bibinfo {author}
  {\bibfnamefont {L.}~\bibnamefont {Fritz}}, \bibinfo {author} {\bibfnamefont
  {A.~V.}\ \bibnamefont {Ustinov}}, \bibinfo {author} {\bibfnamefont
  {M.}~\bibnamefont {Weides}}, \ and\ \bibinfo {author} {\bibfnamefont
  {M.}~\bibnamefont {Marthaler}},\ }\bibfield  {title} {\enquote {\bibinfo
  {title} {{Quantum simulation of the spin-boson model with a microwave
  circuit}},}\ }\href {\doibase 10.1103/PhysRevA.97.052321} {\bibfield
  {journal} {\bibinfo  {journal} {Phys. Rev. A}\ }\textbf {\bibinfo {volume}
  {97}},\ \bibinfo {pages} {052321} (\bibinfo {year} {2018})}\BibitemShut
  {NoStop}%
\bibitem [{\citenamefont {Javier}\ \emph {et~al.}(2019)\citenamefont {Javier},
  \citenamefont {S{\'e}bastien}, \citenamefont {Gheeraert}, \citenamefont
  {Dassonneville}, \citenamefont {Planat}, \citenamefont {Foroughi},
  \citenamefont {Krupko}, \citenamefont {Buisson}, \citenamefont {Naud},
  \citenamefont {Hasch-Guichard}, \citenamefont {Florens}, \citenamefont
  {Snyman},\ and\ \citenamefont {Roch}}]{Roch2019}%
  \BibitemOpen
  \bibfield  {author} {\bibinfo {author} {\bibfnamefont {P.~M.}\ \bibnamefont
  {Javier}}, \bibinfo {author} {\bibfnamefont {L.}~\bibnamefont
  {S{\'e}bastien}}, \bibinfo {author} {\bibfnamefont {N.}~\bibnamefont
  {Gheeraert}}, \bibinfo {author} {\bibfnamefont {R.}~\bibnamefont
  {Dassonneville}}, \bibinfo {author} {\bibfnamefont {L.}~\bibnamefont
  {Planat}}, \bibinfo {author} {\bibfnamefont {F.}~\bibnamefont {Foroughi}},
  \bibinfo {author} {\bibfnamefont {Y.}~\bibnamefont {Krupko}}, \bibinfo
  {author} {\bibfnamefont {O.}~\bibnamefont {Buisson}}, \bibinfo {author}
  {\bibfnamefont {C.}~\bibnamefont {Naud}}, \bibinfo {author} {\bibfnamefont
  {W.}~\bibnamefont {Hasch-Guichard}}, \bibinfo {author} {\bibfnamefont
  {S.}~\bibnamefont {Florens}}, \bibinfo {author} {\bibfnamefont
  {I.}~\bibnamefont {Snyman}}, \ and\ \bibinfo {author} {\bibfnamefont
  {N.}~\bibnamefont {Roch}},\ }\bibfield  {title} {\enquote {\bibinfo {title}
  {{A tunable Josephson platform to explore many-body quantum optics in
  circuit-QED}},}\ }\href{\doibase 10.1038/s41534-018-0104-0} {\bibfield  {journal} {\bibinfo  {journal}
  {npj Quantum Inf.}\ }\textbf {\bibinfo {volume} {5}},\ \bibinfo {pages} {19}
  (\bibinfo {year} {2019})}\BibitemShut {NoStop}%
\bibitem [{\citenamefont {Kuzmin}\ \emph {et~al.}(2019)\citenamefont {Kuzmin},
  \citenamefont {Mehta}, \citenamefont {Grabon}, \citenamefont {Mencia},\ and\
  \citenamefont {Manucharyan}}]{Kuzmin2019}%
  \BibitemOpen
  \bibfield  {author} {\bibinfo {author} {\bibfnamefont {R.}~\bibnamefont
  {Kuzmin}}, \bibinfo {author} {\bibfnamefont {N}~\bibnamefont {Mehta}},
  \bibinfo {author} {\bibfnamefont {N.}~\bibnamefont {Grabon}}, \bibinfo
  {author} {\bibfnamefont {R.}~\bibnamefont {Mencia}}, \ and\ \bibinfo {author}
  {\bibfnamefont {V.~E.}\ \bibnamefont {Manucharyan}},\ }\bibfield  {title}
  {\enquote {\bibinfo {title} {{Superstrong coupling in circuit quantum
  electrodynamics}},}\ }\href {\doibase 10.1038/s41534-019-0134-2} {\bibfield
  {journal} {\bibinfo  {journal} {npj Quantum Inf.}\ }\textbf {\bibinfo
  {volume} {5}},\ \bibinfo {pages} {20} (\bibinfo {year} {2019})}\BibitemShut
  {NoStop}%
\bibitem [{\citenamefont {Rossatto}\ \emph {et~al.}(2017)\citenamefont
  {Rossatto}, \citenamefont {Villas-B{\^o}as}, \citenamefont {Sanz},\ and\
  \citenamefont {Solano}}]{Rossatto2017}%
  \BibitemOpen
  \bibfield  {author} {\bibinfo {author} {\bibfnamefont {D.~Z.}\ \bibnamefont
  {Rossatto}}, \bibinfo {author} {\bibfnamefont {C.~J.}\ \bibnamefont
  {Villas-B{\^o}as}}, \bibinfo {author} {\bibfnamefont {M.}~\bibnamefont
  {Sanz}}, \ and\ \bibinfo {author} {\bibfnamefont {E.}~\bibnamefont
  {Solano}},\ }\bibfield  {title} {\enquote {\bibinfo {title} {{Spectral
  classification of coupling regimes in the quantum Rabi model}},}\ }\href
  {\doibase 10.1103/PhysRevA.96.013849} {\bibfield  {journal} {\bibinfo
  {journal} {Phys. Rev. A}\ }\textbf {\bibinfo {volume} {96}},\ \bibinfo
  {pages} {013849} (\bibinfo {year} {2017})}\BibitemShut {NoStop}%
\bibitem [{\citenamefont {D{\'i}az-Camacho}\ \emph {et~al.}(2016)\citenamefont
  {D{\'i}az-Camacho}, \citenamefont {Bermudez},\ and\ \citenamefont
  {Garc{\'i}a-Ripoll}}]{DiazCamacho2016}%
  \BibitemOpen
  \bibfield  {author} {\bibinfo {author} {\bibfnamefont {G.}~\bibnamefont
  {D{\'i}az-Camacho}}, \bibinfo {author} {\bibfnamefont {A.}~\bibnamefont
  {Bermudez}}, \ and\ \bibinfo {author} {\bibfnamefont {J.~J.}\ \bibnamefont
  {Garc{\'i}a-Ripoll}},\ }\bibfield  {title} {\enquote {\bibinfo {title}
  {{Dynamical polaron Ansatz: A theoretical tool for the ultrastrong-coupling
  regime of circuit QED}},}\ }\href {\doibase 10.1103/PhysRevA.93.043843}
  {\bibfield  {journal} {\bibinfo  {journal} {Phys. Rev. A}\ }\textbf
  {\bibinfo {volume} {93}},\ \bibinfo {pages} {043843} (\bibinfo {year}
  {2016})}\BibitemShut {NoStop}%
\bibitem [{\citenamefont {Armata}\ \emph {et~al.}(2017)\citenamefont {Armata},
  \citenamefont {Calajo}, \citenamefont {Jaako}, \citenamefont {Kim},\ and\
  \citenamefont {Rabl}}]{Armata2017}%
  \BibitemOpen
  \bibfield  {author} {\bibinfo {author} {\bibfnamefont {F.}~\bibnamefont
  {Armata}}, \bibinfo {author} {\bibfnamefont {G.}~\bibnamefont {Calajo}},
  \bibinfo {author} {\bibfnamefont {T.}~\bibnamefont {Jaako}}, \bibinfo
  {author} {\bibfnamefont {M.~S.}\ \bibnamefont {Kim}}, \ and\ \bibinfo
  {author} {\bibfnamefont {P.}~\bibnamefont {Rabl}},\ }\bibfield  {title}
  {\enquote {\bibinfo {title} {{Harvesting Multiqubit Entanglement from
  Ultrastrong Interactions in Circuit Quantum Electrodynamics}},}\ }\href
  {\doibase 10.1103/PhysRevLett.119.183602} {\bibfield  {journal} {\bibinfo
  {journal} {Phys. Rev. Lett.}\ }\textbf {\bibinfo {volume} {119}},\ \bibinfo
  {pages} {183602} (\bibinfo {year} {2017})}\BibitemShut {NoStop}%
\bibitem [{\citenamefont {{De Bernardis}}\ \emph {et~al.}(2018)\citenamefont
  {{De Bernardis}}, \citenamefont {Pilar}, \citenamefont {Jaako}, \citenamefont
  {{De Liberato}},\ and\ \citenamefont {Rabl}}]{DeBernardis2018}%
  \BibitemOpen
  \bibfield  {author} {\bibinfo {author} {\bibfnamefont {D.}~\bibnamefont {{De
  Bernardis}}}, \bibinfo {author} {\bibfnamefont {P.}~\bibnamefont {Pilar}},
  \bibinfo {author} {\bibfnamefont {T.}~\bibnamefont {Jaako}}, \bibinfo
  {author} {\bibfnamefont {S.}~\bibnamefont {{De Liberato}}}, \ and\ \bibinfo
  {author} {\bibfnamefont {P.}~\bibnamefont {Rabl}},\ }\bibfield  {title}
  {\enquote {\bibinfo {title} {{Breakdown of gauge invariance in
  ultrastrong-coupling cavity QED}},}\ }\href {\doibase 10.1103/PhysRevA.98.053819} {\bibfield  {journal}
  {\bibinfo  {journal} {Phys. Rev. A}\ }\textbf {\bibinfo {volume} {98}},\
  \bibinfo {pages} {053819} (\bibinfo {year} {2018})}\BibitemShut {NoStop}%
\bibitem [{\citenamefont {{Di Stefano}}\ \emph {et~al.}(2019)\citenamefont {{Di
  Stefano}}, \citenamefont {Settineri}, \citenamefont {Macr{\`\i}},
  \citenamefont {Garziano}, \citenamefont {Stassi}, \citenamefont {Savasta},\
  and\ \citenamefont {Nori}}]{DiStefano2019}%
  \BibitemOpen
  \bibfield  {author} {\bibinfo {author} {\bibfnamefont {O.}~\bibnamefont {{Di
  Stefano}}}, \bibinfo {author} {\bibfnamefont {A.}~\bibnamefont {Settineri}},
  \bibinfo {author} {\bibfnamefont {V.}~\bibnamefont {Macr{\`\i}}}, \bibinfo
  {author} {\bibfnamefont {L.}~\bibnamefont {Garziano}}, \bibinfo {author}
  {\bibfnamefont {R.}~\bibnamefont {Stassi}}, \bibinfo {author} {\bibfnamefont
  {S.}~\bibnamefont {Savasta}}, \ and\ \bibinfo {author} {\bibfnamefont
  {F.}~\bibnamefont {Nori}},\ }\bibfield  {title} {\enquote {\bibinfo {title}
  {{Resolution of gauge ambiguities in ultrastrong-coupling cavity quantum
  electrodynamics}},}\ }\href {\doibase 10.1038/s41567-019-0534-4} {\bibfield
  {journal} {\bibinfo  {journal} {Nat. Phys.}\ }\textbf {\bibinfo {volume}
  {15}},\ \bibinfo {pages} {803--808} (\bibinfo {year} {2019})}\BibitemShut
  {NoStop}%
\bibitem [{\citenamefont {Hausinger}\ and\ \citenamefont
  {Grifoni}(2011)}]{Hausinger2011}%
  \BibitemOpen
  \bibfield  {author} {\bibinfo {author} {\bibfnamefont {J.}~\bibnamefont
  {Hausinger}}\ and\ \bibinfo {author} {\bibfnamefont {M.}~\bibnamefont
  {Grifoni}},\ }\bibfield  {title} {\enquote {\bibinfo {title}
  {{Qubit-oscillator system under ultrastrong coupling and extreme driving}},}\
  }\href {\doibase 10.1103/PhysRevA.83.030301} {\bibfield  {journal} {\bibinfo
  {journal} {Phys. Rev. A}\ }\textbf {\bibinfo {volume} {83}},\ \bibinfo
  {pages} {030301(R)} (\bibinfo {year} {2011})}\BibitemShut {NoStop}%
\bibitem [{\citenamefont {Oliver}\ \emph {et~al.}(2005)\citenamefont {Oliver},
  \citenamefont {Yu}, \citenamefont {Lee}, \citenamefont {Berggren},
  \citenamefont {Levitov},\ and\ \citenamefont {Orlando}}]{Oliver2005}%
  \BibitemOpen
  \bibfield  {author} {\bibinfo {author} {\bibfnamefont {W.~D.}\ \bibnamefont
  {Oliver}}, \bibinfo {author} {\bibfnamefont {Y.}~\bibnamefont {Yu}}, \bibinfo
  {author} {\bibfnamefont {J.~C.}\ \bibnamefont {Lee}}, \bibinfo {author}
  {\bibfnamefont {K.~K.}\ \bibnamefont {Berggren}}, \bibinfo {author}
  {\bibfnamefont {L.~S.}\ \bibnamefont {Levitov}}, \ and\ \bibinfo {author}
  {\bibfnamefont {T.~P.}\ \bibnamefont {Orlando}},\ }\bibfield  {title}
  {\enquote {\bibinfo {title} {{Mach-Zehnder Interferometry in a Strongly
  Driven Superconducting Qubit}},}\ }\href {\doibase 10.1126/science.1119678}
  {\bibfield  {journal} {\bibinfo  {journal} {Science}\ }\textbf {\bibinfo
  {volume} {310}},\ \bibinfo {pages} {1653} (\bibinfo {year}
  {2005})}\BibitemShut {NoStop}%
\bibitem [{\citenamefont {Wilson}\ \emph {et~al.}(2007)\citenamefont {Wilson},
  \citenamefont {Duty}, \citenamefont {Persson}, \citenamefont {Sandberg},
  \citenamefont {Johansson},\ and\ \citenamefont {Delsing}}]{Wilson2007}%
  \BibitemOpen
  \bibfield  {author} {\bibinfo {author} {\bibfnamefont {C.~M.}\ \bibnamefont
  {Wilson}}, \bibinfo {author} {\bibfnamefont {T.}~\bibnamefont {Duty}},
  \bibinfo {author} {\bibfnamefont {F.}~\bibnamefont {Persson}}, \bibinfo
  {author} {\bibfnamefont {M.}~\bibnamefont {Sandberg}}, \bibinfo {author}
  {\bibfnamefont {G.}~\bibnamefont {Johansson}}, \ and\ \bibinfo {author}
  {\bibfnamefont {P.}~\bibnamefont {Delsing}},\ }\bibfield  {title} {\enquote
  {\bibinfo {title} {{Coherence Times of Dressed States of a Superconducting
  Qubit under Extreme Driving}},}\ }\href
  {https://link.aps.org/doi/10.1103/PhysRevLett.98.257003} {\bibfield
  {journal} {\bibinfo  {journal} {Phys. Rev. Lett.}\ }\textbf {\bibinfo
  {volume} {98}},\ \bibinfo {pages} {257003} (\bibinfo {year}
  {2007})}\BibitemShut {NoStop}%
\bibitem [{\citenamefont {Grifoni}\ and\ \citenamefont
  {H{\"a}nggi}(1998)}]{Grifoni1998}%
  \BibitemOpen
  \bibfield  {author} {\bibinfo {author} {\bibfnamefont {M.}~\bibnamefont
  {Grifoni}}\ and\ \bibinfo {author} {\bibfnamefont {P.}~\bibnamefont
  {H{\"a}nggi}},\ }\bibfield  {title} {\enquote {\bibinfo {title} {{Driven
  quantum tunneling}},}\ }\href {\doibase 10.1016/S0370-1573(98)00022-2}
  {\bibfield  {journal} {\bibinfo  {journal} {Phys. Rep.}\ }\textbf {\bibinfo
  {volume} {304}},\ \bibinfo {pages} {229--354} (\bibinfo {year}
  {1998})}\BibitemShut {NoStop}%
%
\bibitem [{\citenamefont {Hausinger}\ and\ \citenamefont
  {Grifoni}(2010)}]{Hausinger2010}%
  \BibitemOpen
  \bibfield  {author} {\bibinfo {author} {\bibfnamefont {J.}~\bibnamefont
  {Hausinger}}\ and\ \bibinfo {author} {\bibfnamefont {M.}~\bibnamefont
  {Grifoni}},\ }\bibfield  {title} {\enquote {\bibinfo {title}
  {{Dissipative two-level system under strong ac driving: A combination of Floquet and Van Vleck perturbation theory}},}\
  }\href {\doibase 10.1103/PhysRevA.81.022117} {\bibfield  {journal} {\bibinfo
  {journal} {Phys. Rev. A}\ }\textbf {\bibinfo {volume} {81}},\ \bibinfo
  {pages} {022117} (\bibinfo {year} {2010})}\BibitemShut {NoStop}
%
\bibitem [{\citenamefont {Kohler}(2017)}]{Kohler2017}%
  \BibitemOpen
  \bibfield  {author} {\bibinfo {author} {\bibfnamefont {S.}~\bibnamefont
  {Kohler}},\ }\bibfield  {title} {\enquote {\bibinfo {title} {{Dispersive
  Readout of Adiabatic Phases}},}\ }\href {\doibase
  10.1103/PhysRevLett.119.196802} {\bibfield  {journal} {\bibinfo  {journal}
  {Phys. Rev. Lett.}\ }\textbf {\bibinfo {volume} {119}},\ \bibinfo {pages}
  {196802} (\bibinfo {year} {2017})}\BibitemShut {NoStop}%
\bibitem [{\citenamefont {Engelhardt}\ and\ \citenamefont
  {Cao}(2021)}]{Engelhardt2021}%
  \BibitemOpen
  \bibfield  {author} {\bibinfo {author} {\bibfnamefont {G.}~\bibnamefont
  {Engelhardt}}\ and\ \bibinfo {author} {\bibfnamefont {J.}~\bibnamefont
  {Cao}},\ }\bibfield  {title} {\enquote {\bibinfo {title} {{Dynamical
  Symmetries and Symmetry-Protected Selection Rules in Periodically Driven
  Quantum Systems}},}\ }\href {\doibase 10.1103/PhysRevLett.126.090601}
  {\bibfield  {journal} {\bibinfo  {journal} {Phys. Rev. Lett.}\ }\textbf
  {\bibinfo {volume} {126}},\ \bibinfo {pages} {090601} (\bibinfo {year}
  {2021})}\BibitemShut {NoStop}%
\bibitem [{\citenamefont {Reimer}\ \emph {et~al.}(2018)\citenamefont {Reimer},
  \citenamefont {Pedersen}, \citenamefont {Tanger}, \citenamefont
  {Pletyukhov},\ and\ \citenamefont {Gritsev}}]{Reimer2018}%
  \BibitemOpen
  \bibfield  {author} {\bibinfo {author} {\bibfnamefont {V.}~\bibnamefont
  {Reimer}}, \bibinfo {author} {\bibfnamefont {K.~G.~L.}\ \bibnamefont
  {Pedersen}}, \bibinfo {author} {\bibfnamefont {N.}~\bibnamefont {Tanger}},
  \bibinfo {author} {\bibfnamefont {M.}~\bibnamefont {Pletyukhov}}, \ and\
  \bibinfo {author} {\bibfnamefont {V.}~\bibnamefont {Gritsev}},\ }\bibfield
  {title} {\enquote {\bibinfo {title} {{Nonadiabatic effects in periodically
  driven dissipative open quantum systems}},}\ }\href {\doibase
  10.1103/PhysRevA.97.043851} {\bibfield  {journal} {\bibinfo  {journal} {Phys.
  Rev. A}\ }\textbf {\bibinfo {volume} {97}},\ \bibinfo {pages} {043851}
  (\bibinfo {year} {2018})}\BibitemShut {NoStop}%
\bibitem [{\citenamefont {Yoshihara}\ \emph {et~al.}(2018)\citenamefont
  {Yoshihara}, \citenamefont {Fuse}, \citenamefont {Ao}, \citenamefont
  {Ashhab}, \citenamefont {Kakuyanagi}, \citenamefont {Saito}, \citenamefont
  {Aoki}, \citenamefont {Koshino},\ and\ \citenamefont
  {Semba}}]{Yoshihara2018}%
  \BibitemOpen
  \bibfield  {author} {\bibinfo {author} {\bibfnamefont {F.}~\bibnamefont
  {Yoshihara}}, \bibinfo {author} {\bibfnamefont {T.}~\bibnamefont {Fuse}},
  \bibinfo {author} {\bibfnamefont {Z.}~\bibnamefont {Ao}}, \bibinfo {author}
  {\bibfnamefont {S.}~\bibnamefont {Ashhab}}, \bibinfo {author} {\bibfnamefont
  {K.}~\bibnamefont {Kakuyanagi}}, \bibinfo {author} {\bibfnamefont
  {S.}~\bibnamefont {Saito}}, \bibinfo {author} {\bibfnamefont
  {T.}~\bibnamefont {Aoki}}, \bibinfo {author} {\bibfnamefont {K.}~\bibnamefont
  {Koshino}}, \ and\ \bibinfo {author} {\bibfnamefont {K.}~\bibnamefont
  {Semba}},\ }\bibfield  {title} {\enquote {\bibinfo {title} {{Inversion of
  Qubit Energy Levels in Qubit-Oscillator Circuits in the Deep-Strong-Coupling
  Regime}},}\ }\href {\doibase 10.1103/PhysRevLett.120.183601} {\bibfield
  {journal} {\bibinfo  {journal} {Phys. Rev. Lett.}\ }\textbf {\bibinfo
  {volume} {120}},\ \bibinfo {pages} {183601} (\bibinfo {year}
  {2018})}\BibitemShut {NoStop}%
\bibitem [{\citenamefont {Lolli}\ \emph {et~al.}(2015)\citenamefont {Lolli},
  \citenamefont {Baksic}, \citenamefont {Nagy}, \citenamefont {Manucharyan},\
  and\ \citenamefont {Ciuti}}]{Lolli2015}%
  \BibitemOpen
  \bibfield  {author} {\bibinfo {author} {\bibfnamefont {J.}~\bibnamefont
  {Lolli}}, \bibinfo {author} {\bibfnamefont {A.}~\bibnamefont {Baksic}},
  \bibinfo {author} {\bibfnamefont {D.}~\bibnamefont {Nagy}}, \bibinfo {author}
  {\bibfnamefont {V.~E.}\ \bibnamefont {Manucharyan}}, \ and\ \bibinfo {author}
  {\bibfnamefont {C.}~\bibnamefont {Ciuti}},\ }\bibfield  {title} {\enquote
  {\bibinfo {title} {{Ancillary Qubit Spectroscopy of Vacua in Cavity and
  Circuit Quantum Electrodynamics}},}\ }\href {\doibase
  10.1103/PhysRevLett.114.183601} {\bibfield  {journal} {\bibinfo  {journal}
  {Phys. Rev. Lett.}\ }\textbf {\bibinfo {volume} {114}},\ \bibinfo {pages}
  {183601} (\bibinfo {year} {2015})}\BibitemShut {NoStop}%
\bibitem [{\citenamefont {Falci}\ \emph {et~al.}(2019)\citenamefont {Falci},
  \citenamefont {Ridolfo}, \citenamefont {{Di Stefano}},\ and\ \citenamefont
  {Paladino}}]{Falci2019}%
  \BibitemOpen
  \bibfield  {author} {\bibinfo {author} {\bibfnamefont {G.}~\bibnamefont
  {Falci}}, \bibinfo {author} {\bibfnamefont {A.}~\bibnamefont {Ridolfo}},
  \bibinfo {author} {\bibfnamefont {P.~G.}\ \bibnamefont {{Di Stefano}}}, \
  and\ \bibinfo {author} {\bibfnamefont {E.}~\bibnamefont {Paladino}},\
  }\bibfield  {title} {\enquote {\bibinfo {title} {{Ultrastrong coupling probed
  by Coherent Population Transfer}},}\ }\href {\doibase
  10.1038/s41598-019-45187-y} {\bibfield  {journal} {\bibinfo  {journal} {Sci.
  Rep.}\ }\textbf {\bibinfo {volume} {9}},\ \bibinfo {pages} {9249} (\bibinfo
  {year} {2019})}\BibitemShut {NoStop}%
\bibitem [{\citenamefont {Ridolfo}\ \emph {et~al.}(2019)\citenamefont
  {Ridolfo}, \citenamefont {Falci}, \citenamefont {Pellegrino}, \citenamefont
  {Maccarrone},\ and\ \citenamefont {Paladino}}]{Ridolfo2019}%
  \BibitemOpen
  \bibfield  {author} {\bibinfo {author} {\bibfnamefont {A.}~\bibnamefont
  {Ridolfo}}, \bibinfo {author} {\bibfnamefont {G.}~\bibnamefont {Falci}},
  \bibinfo {author} {\bibfnamefont {F.~M.~D.}\ \bibnamefont {Pellegrino}},
  \bibinfo {author} {\bibfnamefont {G.~D.}\ \bibnamefont {Maccarrone}}, \ and\
  \bibinfo {author} {\bibfnamefont {E.}~\bibnamefont {Paladino}},\ }\bibfield
  {title} {\enquote {\bibinfo {title} {{Photon pair production by STIRAP in
  ultrastrongly coupled matter-radiation systems}},}\ }\href {\doibase
  10.1140/epjst/e2018-800076-1} {\bibfield  {journal} {\bibinfo  {journal}
  {Eur. Phys. J. Spec. Top.}\ }\textbf {\bibinfo {volume} {227}},\ \bibinfo
  {pages} {2183} (\bibinfo {year} {2019})}\BibitemShut {NoStop}%
\bibitem [{\citenamefont {Astafiev}\ \emph {et~al.}(2010)\citenamefont
  {Astafiev}, \citenamefont {Zagoskin}, \citenamefont {Abdumalikov},
  \citenamefont {Pashkin}, \citenamefont {Yamamoto}, \citenamefont {Inomata},
  \citenamefont {Nakamura},\ and\ \citenamefont {Tsai}}]{Astafiev2010}%
  \BibitemOpen
  \bibfield  {author} {\bibinfo {author} {\bibfnamefont {O.}~\bibnamefont
  {Astafiev}}, \bibinfo {author} {\bibfnamefont {A.~M.}\ \bibnamefont
  {Zagoskin}}, \bibinfo {author} {\bibfnamefont {A.~A.}\ \bibnamefont
  {Abdumalikov}}, \bibinfo {author} {\bibfnamefont {Y.~A.}\ \bibnamefont
  {Pashkin}}, \bibinfo {author} {\bibfnamefont {T.}~\bibnamefont {Yamamoto}},
  \bibinfo {author} {\bibfnamefont {K.}~\bibnamefont {Inomata}}, \bibinfo
  {author} {\bibfnamefont {Y.}~\bibnamefont {Nakamura}}, \ and\ \bibinfo
  {author} {\bibfnamefont {J.~S.}\ \bibnamefont {Tsai}},\ }\bibfield  {title}
  {\enquote {\bibinfo {title} {{Resonance Fluorescence of a Single Artificial
  Atom}},}\ }\href {\doibase 10.1126/science.1181918} {\bibfield  {journal}
  {\bibinfo  {journal} {Science}\ }\textbf {\bibinfo {volume} {327}},\ \bibinfo
  {pages} {840} (\bibinfo {year} {2010})}\BibitemShut {NoStop}%
\bibitem [{\citenamefont {Hoi}\ \emph {et~al.}(2011)\citenamefont {Hoi},
  \citenamefont {Wilson}, \citenamefont {Johansson}, \citenamefont {Palomaki},
  \citenamefont {Peropadre},\ and\ \citenamefont {Delsing}}]{Hoi2011}%
  \BibitemOpen
  \bibfield  {author} {\bibinfo {author} {\bibfnamefont {I.-C.}\ \bibnamefont
  {Hoi}}, \bibinfo {author} {\bibfnamefont {C.~M.}\ \bibnamefont {Wilson}},
  \bibinfo {author} {\bibfnamefont {G.}~\bibnamefont {Johansson}}, \bibinfo
  {author} {\bibfnamefont {T.}~\bibnamefont {Palomaki}}, \bibinfo {author}
  {\bibfnamefont {B.}~\bibnamefont {Peropadre}}, \ and\ \bibinfo {author}
  {\bibfnamefont {P.}~\bibnamefont {Delsing}},\ }\bibfield  {title} {\enquote
  {\bibinfo {title} {{Demonstration of a Single-Photon Router in the Microwave
  Regime}},}\ }\href {\doibase 10.1103/PhysRevLett.107.073601} {\bibfield
  {journal} {\bibinfo  {journal} {Phys. Rev. Lett.}\ }\textbf {\bibinfo
  {volume} {107}},\ \bibinfo {pages} {073601} (\bibinfo {year}
  {2011})}\BibitemShut {NoStop}%
\bibitem [{\citenamefont {Abdumalikov}\ \emph {et~al.}(2011)\citenamefont
  {Abdumalikov}, \citenamefont {Astafiev}, \citenamefont {Pashkin},
  \citenamefont {Nakamura},\ and\ \citenamefont {Tsai}}]{Abdumalikov2011}%
  \BibitemOpen
  \bibfield  {author} {\bibinfo {author} {\bibfnamefont {A.~A.}\ \bibnamefont
  {Abdumalikov}}, \bibinfo {author} {\bibfnamefont {O.~V.}\ \bibnamefont
  {Astafiev}}, \bibinfo {author} {\bibfnamefont {Y.~A.}\ \bibnamefont
  {Pashkin}}, \bibinfo {author} {\bibfnamefont {Y.}~\bibnamefont {Nakamura}}, \
  and\ \bibinfo {author} {\bibfnamefont {J.~S.}\ \bibnamefont {Tsai}},\
  }\bibfield  {title} {\enquote {\bibinfo {title} {{Dynamics of Coherent and
  Incoherent Emission from an Artificial Atom in a 1D Space}},}\ }\href
  {\doibase 10.1103/PhysRevLett.107.043604} {\bibfield  {journal} {\bibinfo
  {journal} {Phys. Rev. Lett.}\ }\textbf {\bibinfo {volume} {107}},\ \bibinfo
  {pages} {043604} (\bibinfo {year} {2011})}\BibitemShut {NoStop}%
\bibitem [{\citenamefont {{\emph{et al.}}}(2015)}]{Haeberlein2015}%
  \BibitemOpen
  \bibfield  {author} {\bibinfo {author} {\bibfnamefont {{M. Haeberlein}}\
  \bibnamefont {{\emph{et al.}}}},\ }\bibfield  {title} {\enquote {\bibinfo
  {title} {{Spin-boson model with an engineered reservoir in circuit quantum
  electrodynamics}},}\ }\href@noop {} {\bibfield  {journal} {\bibinfo
  {journal} {arXiv:1506.09114}\ } (\bibinfo {year} {2015})}\BibitemShut
  {NoStop}%
\bibitem [{\citenamefont {Forn-D{\'i}az}\ \emph
  {et~al.}(2017{\natexlab{b}})\citenamefont {Forn-D{\'i}az}, \citenamefont
  {Warren}, \citenamefont {Chang}, \citenamefont {Vadiraj},\ and\ \citenamefont
  {Wilson}}]{Forn-Diaz2017PRApp}%
  \BibitemOpen
  \bibfield  {author} {\bibinfo {author} {\bibfnamefont {P.}~\bibnamefont
  {Forn-D{\'i}az}}, \bibinfo {author} {\bibfnamefont {C.~W.}\ \bibnamefont
  {Warren}}, \bibinfo {author} {\bibfnamefont {C.~W.~S.}\ \bibnamefont
  {Chang}}, \bibinfo {author} {\bibfnamefont {A.~M.}\ \bibnamefont {Vadiraj}},
  \ and\ \bibinfo {author} {\bibfnamefont {C.~M.}\ \bibnamefont {Wilson}},\
  }\bibfield  {title} {\enquote {\bibinfo {title} {{On-Demand Microwave
  Generator of Shaped Single Photons}},}\ }\href {\doibase
  10.1103/PhysRevApplied.8.054015} {\bibfield  {journal} {\bibinfo  {journal}
  {Phys. Rev. Applied}\ }\textbf {\bibinfo {volume} {8}},\ \bibinfo {pages}
  {054015} (\bibinfo {year} {2017}{\natexlab{b}})}\BibitemShut {NoStop}%
\bibitem [{\citenamefont {Chiorescu}\ \emph {et~al.}(2004)\citenamefont
  {Chiorescu}, \citenamefont {Bertet}, \citenamefont {Semba}, \citenamefont
  {Nakamura}, \citenamefont {Harmans},\ and\ \citenamefont
  {Mooij}}]{Chiorescu2004}%
  \BibitemOpen
  \bibfield  {author} {\bibinfo {author} {\bibfnamefont {I.}~\bibnamefont
  {Chiorescu}}, \bibinfo {author} {\bibfnamefont {P.}~\bibnamefont {Bertet}},
  \bibinfo {author} {\bibfnamefont {K.}~\bibnamefont {Semba}}, \bibinfo
  {author} {\bibfnamefont {Y.}~\bibnamefont {Nakamura}}, \bibinfo {author}
  {\bibfnamefont {C.~J. P.~M.}\ \bibnamefont {Harmans}}, \ and\ \bibinfo
  {author} {\bibfnamefont {J.~E.}\ \bibnamefont {Mooij}},\ }\bibfield  {title}
  {\enquote {\bibinfo {title} {{Coherent dynamics of a flux qubit coupled to a
  harmonic oscillator}},}\ }
 \href {\doibase 10.1038/nature02831}    
   {\bibfield  {journal} {\bibinfo
  {journal} {Nature}\ }\textbf {\bibinfo {volume} {431}},\ \bibinfo {pages}
  {159--162} (\bibinfo {year} {2004})}\BibitemShut {NoStop}%
\bibitem [{\citenamefont {Thorwart}\ \emph {et~al.}(2004)\citenamefont
  {Thorwart}, \citenamefont {Paladino},\ and\ \citenamefont
  {Grifoni}}]{Thorwart2004}%
  \BibitemOpen
  \bibfield  {author} {\bibinfo {author} {\bibfnamefont {M.}~\bibnamefont
  {Thorwart}}, \bibinfo {author} {\bibfnamefont {E.}~\bibnamefont {Paladino}},
  \ and\ \bibinfo {author} {\bibfnamefont {M.}~\bibnamefont {Grifoni}},\
  }\bibfield  {title} {\enquote {\bibinfo {title} {{Dynamics of the spin-boson
  model with a structured environment}},}\ }\href {\doibase
  10.1016/j.chemphys.2003.10.007} {\bibfield  {journal} {\bibinfo  {journal}
  {Chem. Phys}\ }\textbf {\bibinfo {volume} {296}},\ \bibinfo {pages}
  {333--344} (\bibinfo {year} {2004})}\BibitemShut {NoStop}%
\bibitem [{\citenamefont {Johansson}\ \emph {et~al.}(2006)\citenamefont
  {Johansson}, \citenamefont {Saito}, \citenamefont {Meno}, \citenamefont
  {Nakano}, \citenamefont {Ueda}, \citenamefont {Semba},\ and\ \citenamefont
  {Takayanagi}}]{Johansson2006}%
  \BibitemOpen
  \bibfield  {author} {\bibinfo {author} {\bibfnamefont {J.}~\bibnamefont
  {Johansson}}, \bibinfo {author} {\bibfnamefont {S.}~\bibnamefont {Saito}},
  \bibinfo {author} {\bibfnamefont {T.}~\bibnamefont {Meno}}, \bibinfo {author}
  {\bibfnamefont {H.}~\bibnamefont {Nakano}}, \bibinfo {author} {\bibfnamefont
  {M.}~\bibnamefont {Ueda}}, \bibinfo {author} {\bibfnamefont {K.}~\bibnamefont
  {Semba}}, \ and\ \bibinfo {author} {\bibfnamefont {H.}~\bibnamefont
  {Takayanagi}},\ }\bibfield  {title} {\enquote {\bibinfo {title} {{Vacuum Rabi
  Oscillations in a Macroscopic Superconducting Qubit $LC$ Oscillator
  System}},}\ }\href {\doibase 10.1103/PhysRevLett.96.127006} {\bibfield
  {journal} {\bibinfo  {journal} {Phys. Rev. Lett.}\ }\textbf {\bibinfo
  {volume} {96}},\ \bibinfo {pages} {127006} (\bibinfo {year}
  {2006})}\BibitemShut {NoStop}%
\bibitem [{\citenamefont {{A. Ronzani}}\ \emph {et~al.}(2018)\citenamefont {{A.
  Ronzani}}, \citenamefont {{B. Karimi}}, \citenamefont {{J. Senior}},
  \citenamefont {{Y.-C. Chang}}, \citenamefont {{J. T. Peltonen}},
  \citenamefont {{C.D. Chen}},\ and\ \citenamefont {{J. P.
  Pekola}}}]{Pekola2018}%
  \BibitemOpen
  \bibfield  {author} {\bibinfo {author} {\bibnamefont {{A. Ronzani}}},
  \bibinfo {author} {\bibnamefont {{B. Karimi}}}, \bibinfo {author}
  {\bibnamefont {{J. Senior}}}, \bibinfo {author} {\bibnamefont {{Y.-C.
  Chang}}}, \bibinfo {author} {\bibnamefont {{J. T. Peltonen}}}, \bibinfo
  {author} {\bibnamefont {{C.D. Chen}}}, \ and\ \bibinfo {author} {\bibnamefont
  {{J. P. Pekola}}},\ }\bibfield  {title} {\enquote {\bibinfo {title} {{Tunable
  photonic heat transport in a quantum heat valve}},}\ }\href {\doibase
  10.1038/s41567-018-0199-4} {\bibfield  {journal} {\bibinfo  {journal} {Nat.
  Phys.}\ }\textbf {\bibinfo {volume} {14}},\ \bibinfo {pages} {991--995}
  (\bibinfo {year} {2018})}\BibitemShut {NoStop}%
\bibitem [{\citenamefont {Beaudoin}\ \emph
  {et~al.}(2011{\natexlab{a}})\citenamefont {Beaudoin}, \citenamefont
  {Gambetta},\ and\ \citenamefont {Blais}}]{Blais2011}%
  \BibitemOpen
  \bibfield  {author} {\bibinfo {author} {\bibfnamefont {F.}~\bibnamefont
  {Beaudoin}}, \bibinfo {author} {\bibfnamefont {J.~M.}\ \bibnamefont
  {Gambetta}}, \ and\ \bibinfo {author} {\bibfnamefont {A.}~\bibnamefont
  {Blais}},\ }\bibfield  {title} {\enquote {\bibinfo {title} {{Dissipation and
  ultrastrong coupling in circuit QED}},}\ }\href@noop {} {\bibfield  {journal}
  {\bibinfo  {journal} {Phys. Rev. A}\ }\textbf {\bibinfo {volume} {84}},\
  \bibinfo {pages} {043832} (\bibinfo {year} {2011}{\natexlab{a}})}\BibitemShut
  {NoStop}%
\bibitem [{\citenamefont {Garg}\ \emph {et~al.}(1985)\citenamefont {Garg},
  \citenamefont {Onuchic},\ and\ \citenamefont {Ambegaokar}}]{Garg1985}%
  \BibitemOpen
  \bibfield  {author} {\bibinfo {author} {\bibfnamefont {A.}~\bibnamefont
  {Garg}}, \bibinfo {author} {\bibfnamefont {J.~N.}\ \bibnamefont {Onuchic}}, \
  and\ \bibinfo {author} {\bibfnamefont {V.}~\bibnamefont {Ambegaokar}},\
  }\bibfield  {title} {\enquote {\bibinfo {title} {{Effect of friction on
  electron transfer in biomolecules}},}\ }\href {\doibase 10.1063/1.449017}
  {\bibfield  {journal} {\bibinfo  {journal} {J. Chem. Phys.}\ }\textbf
  {\bibinfo {volume} {83}},\ \bibinfo {pages} {4491--4503} (\bibinfo {year}
  {1985})}\BibitemShut {NoStop}%
\bibitem [{\citenamefont {Magazz{\`u}}\ and\ \citenamefont
  {Grifoni}(2019)}]{Magazzu2019}%
  \BibitemOpen
  \bibfield  {author} {\bibinfo {author} {\bibfnamefont {L.}~\bibnamefont
  {Magazz{\`u}}}\ and\ \bibinfo {author} {\bibfnamefont {M.}~\bibnamefont
  {Grifoni}},\ }\bibfield  {title} {\enquote {\bibinfo {title} {{Transmission
  spectra of an ultrastrongly coupled qubit-dissipative resonator system}},}\
  }\href {\doibase 10.1088/1742-5468/ab3da8} {\bibfield  {journal} {\bibinfo
  {journal} {J. Stat. Mech.}\ }\textbf {\bibinfo {volume} {2019}},\ \bibinfo
  {pages} {104002} (\bibinfo {year} {2019})}\BibitemShut {NoStop}%
 \bibitem [{\citenamefont {Kohler}\ (2018)}]{Kohler2018}%
  \BibitemOpen
  \bibfield  {author} {\bibinfo {author} {\bibfnamefont {S.}~\bibnamefont
  {Kohler}},\ }\bibfield  {title}
  {\enquote {\bibinfo {title} {{Dispersive readout: Universal theory beyond the rotating-wave approximation}},}\
  }\href {\doibase 10.1103/PhysRevA.98.023849} {\bibfield  {journal}
  {\bibinfo  {journal} {Phys. Rev. A}\ }\textbf {\bibinfo {volume} {98}},\
  \bibinfo {pages} {023849} (\bibinfo {year} {2018})}\BibitemShut {NoStop}%
\bibitem [{\citenamefont {{De Liberato}}\ \emph {et~al.}(2009)\citenamefont
  {{De Liberato}}, \citenamefont {Gerace}, \citenamefont {Carusotto},\ and\
  \citenamefont {Ciuti}}]{DeLiberato2009}%
  \BibitemOpen
  \bibfield  {author} {\bibinfo {author} {\bibfnamefont {S.}~\bibnamefont {{De
  Liberato}}}, \bibinfo {author} {\bibfnamefont {D.}~\bibnamefont {Gerace}},
  \bibinfo {author} {\bibfnamefont {I.}~\bibnamefont {Carusotto}}, \ and\
  \bibinfo {author} {\bibfnamefont {C.}~\bibnamefont {Ciuti}},\ }\bibfield
  {title} {\enquote {\bibinfo {title} {{Extracavity quantum vacuum radiation
  from a single qubit}},}\ }\href {\doibase 10.1103/PhysRevA.80.053810}
  {\bibfield  {journal} {\bibinfo  {journal} {Phys. Rev. A}\ }\textbf {\bibinfo
  {volume} {80}},\ \bibinfo {pages} {053810} (\bibinfo {year}
  {2009})}\BibitemShut {NoStop}%
\bibitem [{\citenamefont {Beaudoin}\ \emph
  {et~al.}(2011{\natexlab{b}})\citenamefont {Beaudoin}, \citenamefont
  {Gambetta},\ and\ \citenamefont {Blais}}]{Bourassa2011}%
  \BibitemOpen
  \bibfield  {author} {\bibinfo {author} {\bibfnamefont {F.}\
  \bibnamefont {Beaudoin}}, \bibinfo {author} {\bibfnamefont {J. M.}\
  \bibnamefont {Gambetta}}, \ and\ \bibinfo {author} {\bibfnamefont
  {A.}~\bibnamefont {Blais}},\ }\bibfield  {title} {\enquote {\bibinfo {title}
  {{Dissipation and ultrastrong coupling in circuit QED}},}\ }\href {\doibase
  10.1103/PhysRevA.84.043832} {\bibfield  {journal} {\bibinfo  {journal} {Phys.
  Rev. A}\ }\textbf {\bibinfo {volume} {84}},\ \bibinfo {pages} {043832}
  (\bibinfo {year} {2011}{\natexlab{b}})}\BibitemShut {NoStop}%
\bibitem [{\citenamefont {Caldeira}\ and\ \citenamefont
  {Leggett}(1981)}]{Caldeira1981}%
  \BibitemOpen
  \bibfield  {author} {\bibinfo {author} {\bibfnamefont {A.~O.}\ \bibnamefont
  {Caldeira}}\ and\ \bibinfo {author} {\bibfnamefont {A.~J.}\ \bibnamefont
  {Leggett}},\ }\bibfield  {title} {\enquote {\bibinfo {title} {{Influence of
  Dissipation on Quantum Tunneling in Macroscopic Systems}},}\ }\href {\doibase
  10.1103/PhysRevLett.46.211} {\bibfield  {journal} {\bibinfo  {journal} {Phys.
  Rev. Lett.}\ }\textbf {\bibinfo {volume} {46}},\ \bibinfo {pages} {211--214}
  (\bibinfo {year} {1981})}\BibitemShut {NoStop}%
\bibitem [{\citenamefont {Caldeira}\ and\ \citenamefont
  {Leggett}(1983)}]{Caldeira1983}%
  \BibitemOpen
  \bibfield  {author} {\bibinfo {author} {\bibfnamefont {A.~O}\ \bibnamefont
  {Caldeira}}\ and\ \bibinfo {author} {\bibfnamefont {A.~J}\ \bibnamefont
  {Leggett}},\ }\bibfield  {title} {\enquote {\bibinfo {title} {{Quantum
  tunnelling in a dissipative system}},}\ }\href {\doibase
  10.1016/0003-4916(83)90202-6} {\bibfield  {journal} {\bibinfo  {journal}
  {Ann. Phys.}\ }\textbf {\bibinfo {volume} {149}},\ \bibinfo {pages}
  {374--456} (\bibinfo {year} {1983})}\BibitemShut {NoStop}%
\bibitem [{\citenamefont {{\emph{et al.}}}(2018)}]{Maleeva2018}%
  \BibitemOpen
  \bibfield  {author} {\bibinfo {author} {\bibfnamefont {{N. Maleeva}}\
  \bibnamefont {{\emph{et al.}}}},\ }\bibfield  {title} {\enquote {\bibinfo
  {title} {{Circuit quantum electrodynamics of granular aluminum resonators}},}\ }\href {\doibase 10.1038/s41467-018-06386-9} {\bibfield  {journal}
  {\bibinfo  {journal} {Nat. Commun.}\ }\textbf {\bibinfo {volume} {9}},\
  \bibinfo {pages} {3889} (\bibinfo {year} {2018})}\BibitemShut {NoStop}%
\bibitem [{\citenamefont {{\emph{et al.}}}(2019)}]{Gruenhaupt2019}%
  \BibitemOpen
  \bibfield  {author} {\bibinfo {author} {\bibfnamefont {{L. Gr{\"u}nhaupt}}\
  \bibnamefont {{\emph{et al.}}}},\ }\bibfield  {title} {\enquote {\bibinfo
  {title} {{Granular aluminium as a superconducting material for high-impedance quantum circuits}},}\ }\href {\doibase 10.1038/s41563-019-0350-3} {\bibfield  {journal}
  {\bibinfo  {journal} {Nat. Mater.}\ }\textbf {\bibinfo {volume} {18}},\
  \bibinfo {pages} {816–819} (\bibinfo {year} {2019})}\BibitemShut {NoStop}
%
\bibitem [{\citenamefont {Goorden}\ \emph {et~al.}(2004)\citenamefont
  {Goorden}, \citenamefont {Thorwart},\ and\ \citenamefont
  {Grifoni}}]{Goorden2004}%
  \BibitemOpen
  \bibfield  {author} {\bibinfo {author} {\bibfnamefont {M.~C.}\ \bibnamefont
  {Goorden}}, \bibinfo {author} {\bibfnamefont {M.}~\bibnamefont {Thorwart}}, \
  and\ \bibinfo {author} {\bibfnamefont {M.}~\bibnamefont {Grifoni}},\
  }\bibfield  {title} {\enquote {\bibinfo {title} {{Entanglement spectroscopy
  of a driven solid-state qubit and its detector}},}\ }\href {\doibase
  10.1103/PhysRevLett.93.267005} {\bibfield  {journal} {\bibinfo  {journal}
  {Phys. Rev. Lett.}\ }\textbf {\bibinfo {volume} {93}},\ \bibinfo {pages}
  {267005} (\bibinfo {year} {2004})}\BibitemShut {NoStop}%
\bibitem [{\citenamefont {{F. Nesi}}\ \emph {et~al.}(2007)\citenamefont {{F.
  Nesi}}, \citenamefont {{M. Grifoni}},\ and\ \citenamefont {{E.
  Paladino}}}]{Nesi2007NJP}%
  \BibitemOpen
  \bibfield  {author} {\bibinfo {author} {\bibnamefont {{F. Nesi}}}, \bibinfo
  {author} {\bibnamefont {{M. Grifoni}}}, \ and\ \bibinfo {author}
  {\bibnamefont {{E. Paladino}}},\ }\bibfield  {title} {\enquote {\bibinfo
  {title} {{Dynamics of a qubit coupled to a broadened harmonic mode at finite
  detuning}},}\ }\href {\doibase 10.1088/1367-2630/9/9/316} {\bibfield
  {journal} {\bibinfo  {journal} {New J. Phys.}\ }\textbf {\bibinfo {volume}
  {9}},\ \bibinfo {pages} {316--316} (\bibinfo {year} {2007})}\BibitemShut
  {NoStop}%
\bibitem [{\citenamefont {Zueco}\ and\ \citenamefont
  {Garc{\'i}a-Ripoll}(2019)}]{Zueco2019}%
  \BibitemOpen
  \bibfield  {author} {\bibinfo {author} {\bibfnamefont {D.}~\bibnamefont
  {Zueco}}\ and\ \bibinfo {author} {\bibfnamefont {J.~J.}\ \bibnamefont
  {Garc{\'i}a-Ripoll}},\ }\bibfield  {title} {\enquote {\bibinfo {title}
  {{Ultrastrongly dissipative quantum Rabi model}},}\ }\href {\doibase
  10.1103/PhysRevA.99.013807} {\bibfield  {journal} {\bibinfo  {journal} {Phys.
  Rev. A}\ }\textbf {\bibinfo {volume} {99}},\ \bibinfo {pages} {013807}
  (\bibinfo {year} {2019})}\BibitemShut {NoStop}%
\bibitem [{\citenamefont {Irish}\ \emph {et~al.}(2005)\citenamefont {Irish},
  \citenamefont {Gea-Banacloche}, \citenamefont {Martin},\ and\ \citenamefont
  {Schwab}}]{Irish2005}%
  \BibitemOpen
  \bibfield  {author} {\bibinfo {author} {\bibfnamefont {E.~K.}\ \bibnamefont
  {Irish}}, \bibinfo {author} {\bibfnamefont {J.}~\bibnamefont
  {Gea-Banacloche}}, \bibinfo {author} {\bibfnamefont {I.}~\bibnamefont
  {Martin}}, \ and\ \bibinfo {author} {\bibfnamefont {K.~C.}\ \bibnamefont
  {Schwab}},\ }\bibfield  {title} {\enquote {\bibinfo {title} {{Dynamics of a
  two-level system strongly coupled to a high-frequency quantum oscillator}},}\
  }\href {\doibase 10.1103/PhysRevB.72.195410} {\bibfield  {journal} {\bibinfo
  {journal} {Phys. Rev. B}\ }\textbf {\bibinfo {volume} {72}},\ \bibinfo
  {pages} {195410} (\bibinfo {year} {2005})}\BibitemShut {NoStop}%
\bibitem [{\citenamefont {Irish}(2007)}]{Irish2007}%
  \BibitemOpen
  \bibfield  {author} {\bibinfo {author} {\bibfnamefont {E.~K.}\ \bibnamefont
  {Irish}},\ }\bibfield  {title} {\enquote {\bibinfo {title} {{Generalized
  Rotating-Wave Approximation for Arbitrarily Large Coupling}},}\ }\href
  {\doibase 10.1103/PhysRevLett.99.173601} {\bibfield  {journal} {\bibinfo
  {journal} {Phys. Rev. Lett.}\ }\textbf {\bibinfo {volume} {99}},\ \bibinfo
  {pages} {173601} (\bibinfo {year} {2007})}\BibitemShut {NoStop}%
\bibitem [{\citenamefont {{U. Vool}}\ and\ \citenamefont {{M.
  Devoret}}(2017)}]{Vool2017}%
  \BibitemOpen
  \bibfield  {author} {\bibinfo {author} {\bibnamefont {{U. Vool}}}\ and\
  \bibinfo {author} {\bibnamefont {{M. Devoret}}},\ }\bibfield  {title}
  {\enquote {\bibinfo {title} {{Introduction to quantum electromagnetic
  circuits}},}\ }\href {\doibase 10.1002/cta.2359} {\bibfield  {journal}
  {\bibinfo  {journal} {Int. J. Circ. Theor. Appl.}\ }\textbf {\bibinfo
  {volume} {45}},\ \bibinfo {pages} {897--934} (\bibinfo {year}
  {2017})}\BibitemShut {NoStop}%
\bibitem [{\citenamefont {Dekker}(1987)}]{Dekker1987}%
  \BibitemOpen
  \bibfield  {author} {\bibinfo {author} {\bibfnamefont {H.}~\bibnamefont
  {Dekker}},\ }\bibfield  {title} {\enquote {\bibinfo {title}
  {{Noninteracting-blip approximation for a two-level system coupled to a heat
  bath}},}\ }\href {\doibase 10.1103/PhysRevA.35.1436} {\bibfield  {journal}
  {\bibinfo  {journal} {Phys. Rev. A}\ }\textbf {\bibinfo {volume} {35}},\
  \bibinfo {pages} {1436--1437} (\bibinfo {year} {1987})}\BibitemShut {NoStop}%
\bibitem [{\citenamefont {Gradshteyn}\ and\ \citenamefont
  {Ryzhik}(1980)}]{Gradshteyn1980}%
  \BibitemOpen
  \bibfield  {author} {\bibinfo {author} {\bibfnamefont {I.~S.}\ \bibnamefont
  {Gradshteyn}}\ and\ \bibinfo {author} {\bibfnamefont {I.~M.}\ \bibnamefont
  {Ryzhik}},\ }\href@noop {} {\emph {\bibinfo {title} {{Table of Integrals,
  Series, and Products}}}}\ (\bibinfo  {publisher} {Academic Press, New York},\
  \bibinfo {year} {1980})\BibitemShut {NoStop}%
  \bibitem [{\citenamefont {Nicolin}\ and\ \citenamefont
  {Segal}(2011)}]{Nicolin2011}%
  \BibitemOpen
  \bibfield  {author} {\bibinfo {author} {\bibfnamefont {L.}~\bibnamefont
  {Nicolin}}\ and\ \bibinfo {author} {\bibfnamefont {D.}~\bibnamefont
  {Segal}},\ }\bibfield  {title} {\enquote {\bibinfo {title} {{Non-equilibrium
  spin-boson model: Counting statistics and the heat exchange fluctuation
  theorem}},}\ }\href {\doibase 10.1063/1.3655674} {\bibfield  {journal}
  {\bibinfo  {journal} {J. Chem. Phys.}\ }\textbf {\bibinfo {volume} {135}},\
  \bibinfo {pages} {164106} (\bibinfo {year} {2011})}\BibitemShut {NoStop}%
\bibitem [{\citenamefont {Boudjada}\ and\ \citenamefont
  {Segal}(2014)}]{Boudjada2014}%
  \BibitemOpen
  \bibfield  {author} {\bibinfo {author} {\bibfnamefont {N.}~\bibnamefont
  {Boudjada}}\ and\ \bibinfo {author} {\bibfnamefont {D.}~\bibnamefont
  {Segal}},\ }\bibfield  {title} {\enquote {\bibinfo {title} {{From Dissipative
  Dynamics to Studies of Heat Transfer at the Nanoscale: Analysis of the
  Spin-Boson Model}},}\ }\href {\doibase 10.1021/jp5091685} {\bibfield
  {journal} {\bibinfo  {journal} {J. Phys. Chem. A}\ }\textbf {\bibinfo
  {volume} {118}},\ \bibinfo {pages} {11323--11336} (\bibinfo {year}
  {2014})}\BibitemShut {NoStop}%
\bibitem [{\citenamefont {Segal}(2014)}]{Segal2014}%
  \BibitemOpen
  \bibfield  {author} {\bibinfo {author} {\bibfnamefont {D.}~\bibnamefont
  {Segal}},\ }\bibfield  {title} {\enquote {\bibinfo {title} {{Heat transfer in
  the spin-boson model: A comparative study in the incoherent tunneling
  regime}},}\ }\href {\doibase 10.1103/PhysRevE.90.012148} {\bibfield
  {journal} {\bibinfo  {journal} {Phys. Rev. E}\ }\textbf {\bibinfo {volume}
  {90}},\ \bibinfo {pages} {012148} (\bibinfo {year} {2014})}\BibitemShut
  {NoStop}%
\bibitem [{\citenamefont {Yamamoto}\ \emph {et~al.}(2018)\citenamefont
  {Yamamoto}, \citenamefont {Kato}, \citenamefont {Kato},\ and\ \citenamefont
  {Saito}}]{Yamamoto2018}%
  \BibitemOpen
  \bibfield  {author} {\bibinfo {author} {\bibfnamefont {T.}~\bibnamefont
  {Yamamoto}}, \bibinfo {author} {\bibfnamefont {M.}~\bibnamefont {Kato}},
  \bibinfo {author} {\bibfnamefont {T.}~\bibnamefont {Kato}}, \ and\ \bibinfo
  {author} {\bibfnamefont {K.}~\bibnamefont {Saito}},\ }\bibfield  {title}
  {\enquote {\bibinfo {title} {{Heat transport via a local two-state system
  near thermal equilibrium}},}\ }\href {\doibase 10.1088/1367-2630/aadf09}
  {\bibfield  {journal} {\bibinfo  {journal} {New J. Phys.}\ }\textbf {\bibinfo
  {volume} {20}},\ \bibinfo {pages} {093014} (\bibinfo {year}
  {2018})}\BibitemShut {NoStop}%
\bibitem [{\citenamefont {Nesi}\ (2007)}]{Nesi2007}%
  \BibitemOpen
  \bibfield  {author} {\bibinfo {author} {\bibfnamefont {F.}~\bibnamefont
  {Nesi}}, \bibinfo {author} {\bibfnamefont {M.}~\bibnamefont
  {Grifoni}},\ and\ \bibinfo {author} {\bibfnamefont {E.}~\bibnamefont
  {Paladino}},\ }\bibfield  {title}
  {\enquote {\bibinfo {title} {{Dynamics of a qubit coupled to a broadened harmonic mode at finite detuning}},}\
  }\href {\doibase 10.1088/1367-2630/9/9/316} {\bibfield  {journal}
  {\bibinfo  {journal} {New J. Phys.}\ }\textbf {\bibinfo {volume} {9}},\
  \bibinfo {pages} {316} (\bibinfo {year} {2007})}\BibitemShut {NoStop}%
 \bibitem [{\citenamefont {Imamoglu}\ (1997)}]{Imamoglu1997}%
  \BibitemOpen
  \bibfield  {author} {\bibinfo {author} {\bibfnamefont {A.}~\bibnamefont
  {Imamo\={g}lu}}, \bibinfo {author} {\bibfnamefont {H.}~\bibnamefont
  {Schmidt}}, \bibinfo {author} {\bibfnamefont {G.}~\bibnamefont
  {Woods}},\ and\ \bibinfo {author} {\bibfnamefont {M.}~\bibnamefont
  {Deutsch}},\ }\bibfield  {title}
  {\enquote {\bibinfo {title} {{Strongly Interacting Photons in a Nonlinear Cavity}},}\
  }\href {\doibase 10.1103/PhysRevLett.79.1467} {\bibfield  {journal}
  {\bibinfo  {journal} {Phys. Rev. Lett.}\ }\textbf {\bibinfo {volume} {79}},\
  \bibinfo {pages} {1467} (\bibinfo {year} {1997})}\BibitemShut {NoStop}%
\bibitem [{\citenamefont {Birnbaum}\ (2005)}]{Birnbaum2005}%
  \BibitemOpen
  \bibfield  {author} {\bibinfo {author} {\bibfnamefont {K. M.}~\bibnamefont
  {Birnbaum}}, \bibinfo {author} {\bibfnamefont {A.}~\bibnamefont
  {Boca}}, \bibinfo {author} {\bibfnamefont {R.}~\bibnamefont
  {Miller}}, \bibinfo {author} {\bibfnamefont {A. D.}~\bibnamefont
  {Boozer}}, \bibinfo {author} {\bibfnamefont {T. E.}~\bibnamefont
  {Northup}},\ and\ \bibinfo {author} {\bibfnamefont {H. J.}~\bibnamefont
  {Kimble}},\ }\bibfield  {title}
  {\enquote {\bibinfo {title} {{Photon blockade in an optical cavity with one trapped atom}},}\
  }\href {\doibase 10.1038/nature03804} {\bibfield  {journal}
  {\bibinfo  {journal} {Nature}\ }\textbf {\bibinfo {volume} {436}},\
  \bibinfo {pages} {87-90} (\bibinfo {year} {2005})}\BibitemShut {NoStop}%
\bibitem [{\citenamefont {Hoffman}\ (2011)}]{Hoffman2011}%
  \BibitemOpen
  \bibfield  {author} {\bibinfo {author} {\bibfnamefont {A. J.}~\bibnamefont
  {Hoffman}}, \bibinfo {author} {\bibfnamefont {S. J.}~\bibnamefont
  {Srinivasan}}, \bibinfo {author} {\bibfnamefont {S.}~\bibnamefont
  {Schmidt}}, \bibinfo {author} {\bibfnamefont {L.}~\bibnamefont
  {Spietz}}, \bibinfo {author} {\bibfnamefont {J.}~\bibnamefont
  {Aumentado}}, \bibinfo {author} {\bibfnamefont {H. E.}~\bibnamefont
  {T\"ureci}},\ and\ \bibinfo {author} {\bibfnamefont {A. A.}~\bibnamefont
  {Houck}},\ }\bibfield  {title}
  {\enquote {\bibinfo {title} {{Dispersive Photon Blockade in a Superconducting Circuit}},}\
  }\href {\doibase 10.1103/PhysRevLett.107.053602} {\bibfield  {journal}
  {\bibinfo  {journal} {Phys. Rev. Lett.}\ }\textbf {\bibinfo {volume} {107}},\
  \bibinfo {pages} {053602} (\bibinfo {year} {2011})}\BibitemShut {NoStop}
\bibitem [{\citenamefont {Ridolfo}\ (2012)}]{Ridolfo2012}%
  \BibitemOpen
  \bibfield  {author} {\bibinfo {author} {\bibfnamefont {A.}~\bibnamefont
  {Ridolfo}}, \bibinfo {author} {\bibfnamefont {M.}~\bibnamefont
  {Leib}}, \bibinfo {author} {\bibfnamefont {S.}~\bibnamefont
  {Savasta}},\ and\ \bibinfo {author} {\bibfnamefont {M. J.}~\bibnamefont
  {Hartmann}},\ }\bibfield  {title}
  {\enquote {\bibinfo {title} {{Photon Blockade in the Ultrastrong Coupling Regime}},}\
  }\href {\doibase 10.1103/PhysRevLett.109.193602} {\bibfield  {journal}
  {\bibinfo  {journal} {Phys. Rev. Lett.}\ }\textbf {\bibinfo {volume} {109}},\
  \bibinfo {pages} {193602} (\bibinfo {year} {2012})}\BibitemShut {NoStop}%
\bibitem [{\citenamefont {Carmichael}\ (2012)}]{Carmichael2015}%
  \BibitemOpen
  \bibfield  {author} {\bibinfo {author} {\bibfnamefont {H. J.}~\bibnamefont
  {Carmichael}},\ }\bibfield  {title}
  {\enquote {\bibinfo {title} {{Breakdown of Photon Blockade: A Dissipative Quantum Phase Transition in Zero Dimensions}},}\
  }\href {\doibase 10.1103/PhysRevX.5.031028} {\bibfield  {journal}
  {\bibinfo  {journal} {Phys. Rev. X}\ }\textbf {\bibinfo {volume} {5}},\
  \bibinfo {pages} { 031028} (\bibinfo {year} {2015})}\BibitemShut {NoStop}%
\bibitem [{\citenamefont {Curtis}\ (2021)}]{Curtis2021}%
  \BibitemOpen
  \bibfield  {author} {\bibinfo {author} {\bibfnamefont {J. B.}~\bibnamefont
  {Curtis}}, \bibinfo {author} {\bibfnamefont {I.}~\bibnamefont
  {Boettcher}}, \bibinfo {author} {\bibfnamefont {J. T.}~\bibnamefont
  {Young}}, \bibinfo {author} {\bibfnamefont {M. F.}~\bibnamefont
  {Maghrebi}}, \bibinfo {author} {\bibfnamefont {H.}~\bibnamefont
  {Carmichael}}, \bibinfo {author} {\bibfnamefont {A. V.}~\bibnamefont
  {Gorshkov}},\ and\ \bibinfo {author} {\bibfnamefont {M.}~\bibnamefont
  {Foss-Feig}},\ }\bibfield  {title}
  {\enquote {\bibinfo {title} {{Critical theory for the breakdown of photon blockade}},}\
  }\href {\doibase 10.1103/PhysRevResearch.3.023062} {\bibfield  {journal}
  {\bibinfo  {journal} {Phys. Rev. Research}\ }\textbf {\bibinfo {volume} {3}},\
  \bibinfo {pages} {023062} (\bibinfo {year} {2021})}\BibitemShut {NoStop}%
\bibitem [{\citenamefont {LeBoite}\ (2016)}]{LeBoite2016}%
  \BibitemOpen
  \bibfield  {author} {\bibinfo {author} {\bibfnamefont {A.}~\bibnamefont
  {Le Boit\'{e}}}, \bibinfo {author} {\bibfnamefont {M.-J.}~\bibnamefont
  {Hwang}}, \bibinfo {author} {\bibfnamefont {H.}~\bibnamefont
  {Nha}},\ and\ \bibinfo {author} {\bibfnamefont {M. B.}~\bibnamefont
  {Plenio}},\ }\bibfield  {title}
  {\enquote {\bibinfo {title} {{Fate of photon blockade in the deep strong-coupling regime}},}\
  }\href {\doibase 10.1103/PhysRevA.94.033827} {\bibfield  {journal}
  {\bibinfo  {journal} {Phys. Rev. A}\ }\textbf {\bibinfo {volume} {94}},\
  \bibinfo {pages} {033827} (\bibinfo {year} {2016})}\BibitemShut {NoStop}%
\bibitem [{\citenamefont {Hoi}\ (2012)}]{Hoi2012}%
  \BibitemOpen
  \bibfield  {author} {\bibinfo {author} {\bibfnamefont {I.-C.}~\bibnamefont
  {Hoi}}, \bibinfo {author} {\bibfnamefont {T.}~\bibnamefont
  {Palomaki}}, \bibinfo {author} {\bibfnamefont {J.}~\bibnamefont
  {Lindkvist}}, \bibinfo {author} {\bibfnamefont {G.}~\bibnamefont
  {Johansson}}, \bibinfo {author} {\bibfnamefont {P.}~\bibnamefont
  {Delsing}},\ and\ \bibinfo {author} {\bibfnamefont {C. M.}~\bibnamefont
  {Wilson}},\ }\bibfield  {title}
  {\enquote {\bibinfo {title} {{Generation of Nonclassical Microwave States Using an Artificial Atom in 1D Open Space}},}\
  }\href {\doibase 10.1103/PhysRevLett.108.263601} {\bibfield  {journal}
  {\bibinfo  {journal} {Phys. Rev. Lett.}\ }\textbf {\bibinfo {volume} {108}},\
  \bibinfo {pages} {263601} (\bibinfo {year} {2012})}\BibitemShut {NoStop}%
 \bibitem [{\citenamefont {Carrega}\ \emph {et~al.}(2015)\citenamefont
  {Carrega}, \citenamefont {Solinas}, \citenamefont {Braggio}, \citenamefont
  {Sassetti},\ and\ \citenamefont {Weiss}}]{Carrega2015}%
  \BibitemOpen
  \bibfield  {author} {\bibinfo {author} {\bibfnamefont {M.}~\bibnamefont
  {Carrega}}, \bibinfo {author} {\bibfnamefont {P.}~\bibnamefont {Solinas}},
  \bibinfo {author} {\bibfnamefont {A.}~\bibnamefont {Braggio}}, \bibinfo
  {author} {\bibfnamefont {M.}~\bibnamefont {Sassetti}}, \ and\ \bibinfo
  {author} {\bibfnamefont {U.}~\bibnamefont {Weiss}},\ }\bibfield  {title}
  {\enquote {\bibinfo {title} {{Functional integral approach to time-dependent
  heat exchange in open quantum systems: general method and applications}},}\
  }\href {\doibase 10.1088/1367-2630/17/4/045030} {\bibfield  {journal}
  {\bibinfo  {journal} {New J. Phys.}\ }\textbf {\bibinfo {volume} {17}},\
  \bibinfo {pages} {045030} (\bibinfo {year} {2015})}\BibitemShut {NoStop}%
\bibitem [{\citenamefont {Carrega}\ \emph {et~al.}(2016)\citenamefont
  {Carrega}, \citenamefont {Solinas}, \citenamefont {Sassetti},\ and\
  \citenamefont {Weiss}}]{Carrega2016}%
  \BibitemOpen
  \bibfield  {author} {\bibinfo {author} {\bibfnamefont {M.}~\bibnamefont
  {Carrega}}, \bibinfo {author} {\bibfnamefont {P.}~\bibnamefont {Solinas}},
  \bibinfo {author} {\bibfnamefont {M.}~\bibnamefont {Sassetti}}, \ and\
  \bibinfo {author} {\bibfnamefont {U.}~\bibnamefont {Weiss}},\ }\bibfield
  {title} {\enquote {\bibinfo {title} {{Energy Exchange in Driven Open Quantum
  Systems at Strong Coupling}},}\ }\href {\doibase
  10.1103/PhysRevLett.116.240403} {\bibfield  {journal} {\bibinfo  {journal}
  {Phys. Rev. Lett.}\ }\textbf {\bibinfo {volume} {116}},\ \bibinfo {pages}
  {240403} (\bibinfo {year} {2016})}\BibitemShut {NoStop}%
\bibitem [{\citenamefont {{M. Grifoni}}\ \emph {et~al.}(1995)\citenamefont {{M.
  Grifoni}}, \citenamefont {{M. Sassetti}}, \citenamefont {{P. H{\"a}nggi}},\
  and\ \citenamefont {{U. Weiss}}}]{Grifoni1995}%
  \BibitemOpen
  \bibfield  {author} {\bibinfo {author} {\bibnamefont {{M. Grifoni}}},
  \bibinfo {author} {\bibnamefont {{M. Sassetti}}}, \bibinfo {author}
  {\bibnamefont {{P. H{\"a}nggi}}}, \ and\ \bibinfo {author} {\bibnamefont {{U.
  Weiss}}},\ }\bibfield  {title} {\enquote {\bibinfo {title} {{Cooperative
  effects in the nonlinearly driven spin-boson system}},}\ }\href {\doibase
  10.1103/PhysRevE.52.3596} {\bibfield  {journal} {\bibinfo  {journal} {Phys.
  Rev. E}\ }\textbf {\bibinfo {volume} {52}},\ \bibinfo {pages} {3596--3607}
  (\bibinfo {year} {1995})}\BibitemShut {NoStop}%
\end{thebibliography}
%

\end{document}